\documentclass[twocolumn]{aastex631}

\pdfoutput=1
\usepackage[utf8]{inputenc}
\usepackage{array}
\usepackage{xspace}
\usepackage{xstring}
\usepackage{acronym}
\usepackage{tabularx}
\usepackage{comment}
\usepackage{multirow}
\usepackage{rotating}
\usepackage{booktabs}
\usepackage{amsmath,amssymb,graphicx}
\usepackage{comment}
\usepackage{dcolumn}
\usepackage{makecell}
\usepackage{scrextend}
\usepackage{xcolor}
\usepackage{textgreek}
\usepackage{graphicx}
\usepackage[caption=false]{subfig}
\usepackage{makerobust}
\usepackage{ulem}
\usepackage{physics}
\usepackage{verbatim, ifthen} 

\newcommand{\aoneminus}[1]{\IfEqCase{#1}{{GW231123cg_combined}{0.19}{GW231123cg_nrsur}{0.19}{GW231123cg_xphm}{0.20}{GW231123cg_xo4a}{0.06}{GW231123cg_tphm}{0.14}{GW231123cg_seob}{0.16}}}
\newcommand{\aonemed}[1]{\IfEqCase{#1}{{GW231123cg_combined}{0.90}{GW231123cg_nrsur}{0.90}{GW231123cg_xphm}{0.79}{GW231123cg_xo4a}{0.92}{GW231123cg_tphm}{0.92}{GW231123cg_seob}{0.91}}}
\newcommand{\aoneplus}[1]{\IfEqCase{#1}{{GW231123cg_combined}{0.10}{GW231123cg_nrsur}{0.10}{GW231123cg_xphm}{0.21}{GW231123cg_xo4a}{0.07}{GW231123cg_tphm}{0.08}{GW231123cg_seob}{0.09}}}
\newcommand{\aonezeropercent}[1]{\IfEqCase{#1}{{GW231123cg_combined}{0.04}{GW231123cg_nrsur}{0.04}{GW231123cg_xphm}{0.10}{GW231123cg_xo4a}{0.51}{GW231123cg_tphm}{0.15}{GW231123cg_seob}{0.06}}}
\newcommand{\aoneonepercent}[1]{\IfEqCase{#1}{{GW231123cg_combined}{0.48}{GW231123cg_nrsur}{0.43}{GW231123cg_xphm}{0.42}{GW231123cg_xo4a}{0.81}{GW231123cg_tphm}{0.61}{GW231123cg_seob}{0.55}}}
\newcommand{\aoneninetyninepercent}[1]{\IfEqCase{#1}{{GW231123cg_combined}{0.99}{GW231123cg_nrsur}{0.99}{GW231123cg_xphm}{0.99}{GW231123cg_xo4a}{0.99}{GW231123cg_tphm}{0.99}{GW231123cg_seob}{0.99}}}
\newcommand{\aonefivepercent}[1]{\IfEqCase{#1}{{GW231123cg_combined}{0.64}{GW231123cg_nrsur}{0.63}{GW231123cg_xphm}{0.53}{GW231123cg_xo4a}{0.84}{GW231123cg_tphm}{0.72}{GW231123cg_seob}{0.70}}}
\newcommand{\aoneninetyfivepercent}[1]{\IfEqCase{#1}{{GW231123cg_combined}{0.98}{GW231123cg_nrsur}{0.98}{GW231123cg_xphm}{0.96}{GW231123cg_xo4a}{0.98}{GW231123cg_tphm}{0.98}{GW231123cg_seob}{0.98}}}
\newcommand{\aoneninetypercent}[1]{\IfEqCase{#1}{{GW231123cg_combined}{0.97}{GW231123cg_nrsur}{0.98}{GW231123cg_xphm}{0.94}{GW231123cg_xo4a}{0.97}{GW231123cg_tphm}{0.98}{GW231123cg_seob}{0.98}}}
\newcommand{\atwominus}[1]{\IfEqCase{#1}{{GW231123cg_combined}{0.52}{GW231123cg_nrsur}{0.22}{GW231123cg_xphm}{0.47}{GW231123cg_xo4a}{0.47}{GW231123cg_tphm}{0.25}{GW231123cg_seob}{0.35}}}
\newcommand{\atwomed}[1]{\IfEqCase{#1}{{GW231123cg_combined}{0.80}{GW231123cg_nrsur}{0.91}{GW231123cg_xphm}{0.67}{GW231123cg_xo4a}{0.47}{GW231123cg_tphm}{0.87}{GW231123cg_seob}{0.81}}}
\newcommand{\atwoplus}[1]{\IfEqCase{#1}{{GW231123cg_combined}{0.20}{GW231123cg_nrsur}{0.09}{GW231123cg_xphm}{0.33}{GW231123cg_xo4a}{0.41}{GW231123cg_tphm}{0.13}{GW231123cg_seob}{0.19}}}
\newcommand{\atwozeropercent}[1]{\IfEqCase{#1}{{GW231123cg_combined}{0.00}{GW231123cg_nrsur}{0.00}{GW231123cg_xphm}{0.00}{GW231123cg_xo4a}{0.00}{GW231123cg_tphm}{0.01}{GW231123cg_seob}{0.00}}}
\newcommand{\atwoonepercent}[1]{\IfEqCase{#1}{{GW231123cg_combined}{0.03}{GW231123cg_nrsur}{0.29}{GW231123cg_xphm}{0.02}{GW231123cg_xo4a}{0.01}{GW231123cg_tphm}{0.25}{GW231123cg_seob}{0.10}}}
\newcommand{\atwoninetyninepercent}[1]{\IfEqCase{#1}{{GW231123cg_combined}{0.99}{GW231123cg_nrsur}{0.99}{GW231123cg_xphm}{0.98}{GW231123cg_xo4a}{0.98}{GW231123cg_tphm}{0.99}{GW231123cg_seob}{0.99}}}
\newcommand{\atwofivepercent}[1]{\IfEqCase{#1}{{GW231123cg_combined}{0.14}{GW231123cg_nrsur}{0.59}{GW231123cg_xphm}{0.11}{GW231123cg_xo4a}{0.05}{GW231123cg_tphm}{0.51}{GW231123cg_seob}{0.33}}}
\newcommand{\atwoninetyfivepercent}[1]{\IfEqCase{#1}{{GW231123cg_combined}{0.98}{GW231123cg_nrsur}{0.98}{GW231123cg_xphm}{0.96}{GW231123cg_xo4a}{0.93}{GW231123cg_tphm}{0.98}{GW231123cg_seob}{0.98}}}
\newcommand{\atwoninetypercent}[1]{\IfEqCase{#1}{{GW231123cg_combined}{0.96}{GW231123cg_nrsur}{0.98}{GW231123cg_xphm}{0.94}{GW231123cg_xo4a}{0.88}{GW231123cg_tphm}{0.97}{GW231123cg_seob}{0.96}}}
\newcommand{\chieffminus}[1]{\IfEqCase{#1}{{GW231123cg_combined}{0.41}{GW231123cg_nrsur}{0.35}{GW231123cg_xphm}{0.25}{GW231123cg_xo4a}{0.19}{GW231123cg_tphm}{0.19}{GW231123cg_seob}{0.25}}}
\newcommand{\chieffmed}[1]{\IfEqCase{#1}{{GW231123cg_combined}{0.32}{GW231123cg_nrsur}{0.27}{GW231123cg_xphm}{0.03}{GW231123cg_xo4a}{0.31}{GW231123cg_tphm}{0.43}{GW231123cg_seob}{0.43}}}
\newcommand{\chieffplus}[1]{\IfEqCase{#1}{{GW231123cg_combined}{0.25}{GW231123cg_nrsur}{0.24}{GW231123cg_xphm}{0.17}{GW231123cg_xo4a}{0.19}{GW231123cg_tphm}{0.16}{GW231123cg_seob}{0.20}}}
\newcommand{\chieffzeropercent}[1]{\IfEqCase{#1}{{GW231123cg_combined}{-0.73}{GW231123cg_nrsur}{-0.54}{GW231123cg_xphm}{-0.73}{GW231123cg_xo4a}{-0.33}{GW231123cg_tphm}{-0.33}{GW231123cg_seob}{-0.36}}}
\newcommand{\chieffonepercent}[1]{\IfEqCase{#1}{{GW231123cg_combined}{-0.24}{GW231123cg_nrsur}{-0.21}{GW231123cg_xphm}{-0.38}{GW231123cg_xo4a}{0.03}{GW231123cg_tphm}{0.12}{GW231123cg_seob}{0.06}}}
\newcommand{\chieffninetyninepercent}[1]{\IfEqCase{#1}{{GW231123cg_combined}{0.64}{GW231123cg_nrsur}{0.60}{GW231123cg_xphm}{0.27}{GW231123cg_xo4a}{0.55}{GW231123cg_tphm}{0.64}{GW231123cg_seob}{0.69}}}
\newcommand{\chiefffivepercent}[1]{\IfEqCase{#1}{{GW231123cg_combined}{-0.09}{GW231123cg_nrsur}{-0.08}{GW231123cg_xphm}{-0.23}{GW231123cg_xo4a}{0.12}{GW231123cg_tphm}{0.23}{GW231123cg_seob}{0.19}}}
\newcommand{\chieffninetyfivepercent}[1]{\IfEqCase{#1}{{GW231123cg_combined}{0.56}{GW231123cg_nrsur}{0.51}{GW231123cg_xphm}{0.20}{GW231123cg_xo4a}{0.51}{GW231123cg_tphm}{0.58}{GW231123cg_seob}{0.63}}}
\newcommand{\chieffninetypercent}[1]{\IfEqCase{#1}{{GW231123cg_combined}{0.52}{GW231123cg_nrsur}{0.46}{GW231123cg_xphm}{0.16}{GW231123cg_xo4a}{0.48}{GW231123cg_tphm}{0.55}{GW231123cg_seob}{0.59}}}
\newcommand{\chipminus}[1]{\IfEqCase{#1}{{GW231123cg_combined}{0.19}{GW231123cg_nrsur}{0.17}{GW231123cg_xphm}{0.21}{GW231123cg_xo4a}{0.12}{GW231123cg_tphm}{0.17}{GW231123cg_seob}{0.19}}}
\newcommand{\chipmed}[1]{\IfEqCase{#1}{{GW231123cg_combined}{0.77}{GW231123cg_nrsur}{0.76}{GW231123cg_xphm}{0.74}{GW231123cg_xo4a}{0.82}{GW231123cg_tphm}{0.76}{GW231123cg_seob}{0.74}}}
\newcommand{\chipplus}[1]{\IfEqCase{#1}{{GW231123cg_combined}{0.18}{GW231123cg_nrsur}{0.19}{GW231123cg_xphm}{0.21}{GW231123cg_xo4a}{0.10}{GW231123cg_tphm}{0.17}{GW231123cg_seob}{0.20}}}
\newcommand{\chipzeropercent}[1]{\IfEqCase{#1}{{GW231123cg_combined}{0.21}{GW231123cg_nrsur}{0.21}{GW231123cg_xphm}{0.21}{GW231123cg_xo4a}{0.51}{GW231123cg_tphm}{0.31}{GW231123cg_seob}{0.22}}}
\newcommand{\chiponepercent}[1]{\IfEqCase{#1}{{GW231123cg_combined}{0.45}{GW231123cg_nrsur}{0.46}{GW231123cg_xphm}{0.42}{GW231123cg_xo4a}{0.67}{GW231123cg_tphm}{0.48}{GW231123cg_seob}{0.44}}}
\newcommand{\chipninetyninepercent}[1]{\IfEqCase{#1}{{GW231123cg_combined}{0.96}{GW231123cg_nrsur}{0.97}{GW231123cg_xphm}{0.97}{GW231123cg_xo4a}{0.96}{GW231123cg_tphm}{0.95}{GW231123cg_seob}{0.96}}}
\newcommand{\chipfivepercent}[1]{\IfEqCase{#1}{{GW231123cg_combined}{0.55}{GW231123cg_nrsur}{0.56}{GW231123cg_xphm}{0.50}{GW231123cg_xo4a}{0.70}{GW231123cg_tphm}{0.57}{GW231123cg_seob}{0.52}}}
\newcommand{\chipninetyfivepercent}[1]{\IfEqCase{#1}{{GW231123cg_combined}{0.92}{GW231123cg_nrsur}{0.93}{GW231123cg_xphm}{0.93}{GW231123cg_xo4a}{0.92}{GW231123cg_tphm}{0.92}{GW231123cg_seob}{0.92}}}
\newcommand{\chipninetypercent}[1]{\IfEqCase{#1}{{GW231123cg_combined}{0.89}{GW231123cg_nrsur}{0.90}{GW231123cg_xphm}{0.90}{GW231123cg_xo4a}{0.90}{GW231123cg_tphm}{0.89}{GW231123cg_seob}{0.89}}}
\newcommand{\chirpmassminus}[1]{\IfEqCase{#1}{{GW231123cg_combined}{27}{GW231123cg_nrsur}{13}{GW231123cg_xphm}{15}{GW231123cg_xo4a}{19}{GW231123cg_tphm}{11}{GW231123cg_seob}{9}}}
\newcommand{\chirpmassmed}[1]{\IfEqCase{#1}{{GW231123cg_combined}{137}{GW231123cg_nrsur}{138}{GW231123cg_xphm}{119}{GW231123cg_xo4a}{120}{GW231123cg_tphm}{154}{GW231123cg_seob}{147}}}
\newcommand{\chirpmassplus}[1]{\IfEqCase{#1}{{GW231123cg_combined}{21}{GW231123cg_nrsur}{7}{GW231123cg_xphm}{13}{GW231123cg_xo4a}{7}{GW231123cg_tphm}{9}{GW231123cg_seob}{9}}}
\newcommand{\chirpmasszeropercent}[1]{\IfEqCase{#1}{{GW231123cg_combined}{78}{GW231123cg_nrsur}{106}{GW231123cg_xphm}{77}{GW231123cg_xo4a}{89}{GW231123cg_tphm}{120}{GW231123cg_seob}{114}}}
\newcommand{\chirpmassonepercent}[1]{\IfEqCase{#1}{{GW231123cg_combined}{99}{GW231123cg_nrsur}{119}{GW231123cg_xphm}{98}{GW231123cg_xo4a}{96}{GW231123cg_tphm}{138}{GW231123cg_seob}{134}}}
\newcommand{\chirpmassninetyninepercent}[1]{\IfEqCase{#1}{{GW231123cg_combined}{163}{GW231123cg_nrsur}{147}{GW231123cg_xphm}{142}{GW231123cg_xo4a}{130}{GW231123cg_tphm}{166}{GW231123cg_seob}{159}}}
\newcommand{\chirpmassfivepercent}[1]{\IfEqCase{#1}{{GW231123cg_combined}{110}{GW231123cg_nrsur}{124}{GW231123cg_xphm}{104}{GW231123cg_xo4a}{101}{GW231123cg_tphm}{143}{GW231123cg_seob}{138}}}
\newcommand{\chirpmassninetyfivepercent}[1]{\IfEqCase{#1}{{GW231123cg_combined}{158}{GW231123cg_nrsur}{144}{GW231123cg_xphm}{132}{GW231123cg_xo4a}{127}{GW231123cg_tphm}{163}{GW231123cg_seob}{156}}}
\newcommand{\chirpmassninetypercent}[1]{\IfEqCase{#1}{{GW231123cg_combined}{155}{GW231123cg_nrsur}{143}{GW231123cg_xphm}{128}{GW231123cg_xo4a}{125}{GW231123cg_tphm}{161}{GW231123cg_seob}{154}}}
\newcommand{\chirpmasssourceminus}[1]{\IfEqCase{#1}{{GW231123cg_combined}{30}{GW231123cg_nrsur}{14}{GW231123cg_xphm}{15}{GW231123cg_xo4a}{11}{GW231123cg_tphm}{9}{GW231123cg_seob}{11}}}
\newcommand{\chirpmasssourcemed}[1]{\IfEqCase{#1}{{GW231123cg_combined}{101}{GW231123cg_nrsur}{102}{GW231123cg_xphm}{101}{GW231123cg_xo4a}{75}{GW231123cg_tphm}{105}{GW231123cg_seob}{104}}}
\newcommand{\chirpmasssourceplus}[1]{\IfEqCase{#1}{{GW231123cg_combined}{13}{GW231123cg_nrsur}{10}{GW231123cg_xphm}{14}{GW231123cg_xo4a}{12}{GW231123cg_tphm}{12}{GW231123cg_seob}{12}}}
\newcommand{\chirpmasssourcezeropercent}[1]{\IfEqCase{#1}{{GW231123cg_combined}{55}{GW231123cg_nrsur}{61}{GW231123cg_xphm}{64}{GW231123cg_xo4a}{55}{GW231123cg_tphm}{81}{GW231123cg_seob}{76}}}
\newcommand{\chirpmasssourceonepercent}[1]{\IfEqCase{#1}{{GW231123cg_combined}{64}{GW231123cg_nrsur}{79}{GW231123cg_xphm}{80}{GW231123cg_xo4a}{60}{GW231123cg_tphm}{92}{GW231123cg_seob}{87}}}
\newcommand{\chirpmasssourceninetyninepercent}[1]{\IfEqCase{#1}{{GW231123cg_combined}{121}{GW231123cg_nrsur}{119}{GW231123cg_xphm}{123}{GW231123cg_xo4a}{92}{GW231123cg_tphm}{123}{GW231123cg_seob}{123}}}
\newcommand{\chirpmasssourcefivepercent}[1]{\IfEqCase{#1}{{GW231123cg_combined}{71}{GW231123cg_nrsur}{88}{GW231123cg_xphm}{86}{GW231123cg_xo4a}{64}{GW231123cg_tphm}{96}{GW231123cg_seob}{93}}}
\newcommand{\chirpmasssourceninetyfivepercent}[1]{\IfEqCase{#1}{{GW231123cg_combined}{114}{GW231123cg_nrsur}{112}{GW231123cg_xphm}{116}{GW231123cg_xo4a}{87}{GW231123cg_tphm}{117}{GW231123cg_seob}{116}}}
\newcommand{\chirpmasssourceninetypercent}[1]{\IfEqCase{#1}{{GW231123cg_combined}{111}{GW231123cg_nrsur}{110}{GW231123cg_xphm}{112}{GW231123cg_xo4a}{85}{GW231123cg_tphm}{113}{GW231123cg_seob}{113}}}
\newcommand{\comovingdistanceminus}[1]{\IfEqCase{#1}{{GW231123cg_combined}{971}{GW231123cg_nrsur}{685}{GW231123cg_xphm}{244}{GW231123cg_xo4a}{693}{GW231123cg_tphm}{587}{GW231123cg_seob}{584}}}
\newcommand{\comovingdistancemed}[1]{\IfEqCase{#1}{{GW231123cg_combined}{1599}{GW231123cg_nrsur}{1394}{GW231123cg_xphm}{746}{GW231123cg_xo4a}{2214}{GW231123cg_tphm}{1816}{GW231123cg_seob}{1616}}}
\newcommand{\comovingdistanceplus}[1]{\IfEqCase{#1}{{GW231123cg_combined}{900}{GW231123cg_nrsur}{836}{GW231123cg_xphm}{277}{GW231123cg_xo4a}{502}{GW231123cg_tphm}{529}{GW231123cg_seob}{660}}}
\newcommand{\comovingdistancezeropercent}[1]{\IfEqCase{#1}{{GW231123cg_combined}{101}{GW231123cg_nrsur}{207}{GW231123cg_xphm}{101}{GW231123cg_xo4a}{639}{GW231123cg_tphm}{468}{GW231123cg_seob}{368}}}
\newcommand{\comovingdistanceonepercent}[1]{\IfEqCase{#1}{{GW231123cg_combined}{488}{GW231123cg_nrsur}{537}{GW231123cg_xphm}{313}{GW231123cg_xo4a}{1306}{GW231123cg_tphm}{943}{GW231123cg_seob}{744}}}
\newcommand{\comovingdistanceninetyninepercent}[1]{\IfEqCase{#1}{{GW231123cg_combined}{2765}{GW231123cg_nrsur}{2789}{GW231123cg_xphm}{1142}{GW231123cg_xo4a}{2894}{GW231123cg_tphm}{2557}{GW231123cg_seob}{2569}}}
\newcommand{\comovingdistancefivepercent}[1]{\IfEqCase{#1}{{GW231123cg_combined}{627}{GW231123cg_nrsur}{708}{GW231123cg_xphm}{502}{GW231123cg_xo4a}{1521}{GW231123cg_tphm}{1229}{GW231123cg_seob}{1033}}}
\newcommand{\comovingdistanceninetyfivepercent}[1]{\IfEqCase{#1}{{GW231123cg_combined}{2499}{GW231123cg_nrsur}{2230}{GW231123cg_xphm}{1023}{GW231123cg_xo4a}{2717}{GW231123cg_tphm}{2345}{GW231123cg_seob}{2276}}}
\newcommand{\comovingdistanceninetypercent}[1]{\IfEqCase{#1}{{GW231123cg_combined}{2335}{GW231123cg_nrsur}{2007}{GW231123cg_xphm}{959}{GW231123cg_xo4a}{2616}{GW231123cg_tphm}{2234}{GW231123cg_seob}{2119}}}
\newcommand{\cosiotaminus}[1]{\IfEqCase{#1}{{GW231123cg_combined}{1.19}{GW231123cg_nrsur}{1.25}{GW231123cg_xphm}{0.70}{GW231123cg_xo4a}{1.43}{GW231123cg_tphm}{0.47}{GW231123cg_seob}{1.22}}}
\newcommand{\cosiotamed}[1]{\IfEqCase{#1}{{GW231123cg_combined}{0.44}{GW231123cg_nrsur}{0.51}{GW231123cg_xphm}{0.09}{GW231123cg_xo4a}{0.80}{GW231123cg_tphm}{-0.30}{GW231123cg_seob}{0.43}}}
\newcommand{\cosiotaplus}[1]{\IfEqCase{#1}{{GW231123cg_combined}{0.46}{GW231123cg_nrsur}{0.30}{GW231123cg_xphm}{0.56}{GW231123cg_xo4a}{0.16}{GW231123cg_tphm}{1.08}{GW231123cg_seob}{0.40}}}
\newcommand{\cosiotazeropercent}[1]{\IfEqCase{#1}{{GW231123cg_combined}{-1.00}{GW231123cg_nrsur}{-0.99}{GW231123cg_xphm}{-0.94}{GW231123cg_xo4a}{-1.00}{GW231123cg_tphm}{-0.98}{GW231123cg_seob}{-0.99}}}
\newcommand{\cosiotaonepercent}[1]{\IfEqCase{#1}{{GW231123cg_combined}{-0.88}{GW231123cg_nrsur}{-0.82}{GW231123cg_xphm}{-0.72}{GW231123cg_xo4a}{-0.97}{GW231123cg_tphm}{-0.88}{GW231123cg_seob}{-0.88}}}
\newcommand{\cosiotaninetyninepercent}[1]{\IfEqCase{#1}{{GW231123cg_combined}{0.97}{GW231123cg_nrsur}{0.91}{GW231123cg_xphm}{0.73}{GW231123cg_xo4a}{0.99}{GW231123cg_tphm}{0.88}{GW231123cg_seob}{0.91}}}
\newcommand{\cosiotafivepercent}[1]{\IfEqCase{#1}{{GW231123cg_combined}{-0.74}{GW231123cg_nrsur}{-0.74}{GW231123cg_xphm}{-0.61}{GW231123cg_xo4a}{-0.63}{GW231123cg_tphm}{-0.77}{GW231123cg_seob}{-0.79}}}
\newcommand{\cosiotaninetyfivepercent}[1]{\IfEqCase{#1}{{GW231123cg_combined}{0.90}{GW231123cg_nrsur}{0.82}{GW231123cg_xphm}{0.65}{GW231123cg_xo4a}{0.97}{GW231123cg_tphm}{0.78}{GW231123cg_seob}{0.83}}}
\newcommand{\cosiotaninetypercent}[1]{\IfEqCase{#1}{{GW231123cg_combined}{0.84}{GW231123cg_nrsur}{0.77}{GW231123cg_xphm}{0.59}{GW231123cg_xo4a}{0.95}{GW231123cg_tphm}{0.70}{GW231123cg_seob}{0.78}}}
\newcommand{\costhetajnminus}[1]{\IfEqCase{#1}{{GW231123cg_combined}{0.88}{GW231123cg_nrsur}{0.77}{GW231123cg_xphm}{0.41}{GW231123cg_xo4a}{1.74}{GW231123cg_tphm}{0.27}{GW231123cg_seob}{0.93}}}
\newcommand{\costhetajnmed}[1]{\IfEqCase{#1}{{GW231123cg_combined}{0.30}{GW231123cg_nrsur}{0.29}{GW231123cg_xphm}{-0.00}{GW231123cg_xo4a}{0.88}{GW231123cg_tphm}{-0.35}{GW231123cg_seob}{0.33}}}
\newcommand{\costhetajnplus}[1]{\IfEqCase{#1}{{GW231123cg_combined}{0.64}{GW231123cg_nrsur}{0.32}{GW231123cg_xphm}{0.41}{GW231123cg_xo4a}{0.10}{GW231123cg_tphm}{0.98}{GW231123cg_seob}{0.32}}}
\newcommand{\costhetajnzeropercent}[1]{\IfEqCase{#1}{{GW231123cg_combined}{-1.00}{GW231123cg_nrsur}{-0.98}{GW231123cg_xphm}{-0.77}{GW231123cg_xo4a}{-1.00}{GW231123cg_tphm}{-0.94}{GW231123cg_seob}{-0.94}}}
\newcommand{\costhetajnonepercent}[1]{\IfEqCase{#1}{{GW231123cg_combined}{-0.86}{GW231123cg_nrsur}{-0.62}{GW231123cg_xphm}{-0.53}{GW231123cg_xo4a}{-0.99}{GW231123cg_tphm}{-0.70}{GW231123cg_seob}{-0.72}}}
\newcommand{\costhetajnninetyninepercent}[1]{\IfEqCase{#1}{{GW231123cg_combined}{0.98}{GW231123cg_nrsur}{0.83}{GW231123cg_xphm}{0.50}{GW231123cg_xo4a}{1.00}{GW231123cg_tphm}{0.71}{GW231123cg_seob}{0.75}}}
\newcommand{\costhetajnfivepercent}[1]{\IfEqCase{#1}{{GW231123cg_combined}{-0.58}{GW231123cg_nrsur}{-0.48}{GW231123cg_xphm}{-0.42}{GW231123cg_xo4a}{-0.86}{GW231123cg_tphm}{-0.62}{GW231123cg_seob}{-0.60}}}
\newcommand{\costhetajnninetyfivepercent}[1]{\IfEqCase{#1}{{GW231123cg_combined}{0.94}{GW231123cg_nrsur}{0.61}{GW231123cg_xphm}{0.40}{GW231123cg_xo4a}{0.98}{GW231123cg_tphm}{0.63}{GW231123cg_seob}{0.65}}}
\newcommand{\costhetajnninetypercent}[1]{\IfEqCase{#1}{{GW231123cg_combined}{0.88}{GW231123cg_nrsur}{0.53}{GW231123cg_xphm}{0.33}{GW231123cg_xo4a}{0.97}{GW231123cg_tphm}{0.58}{GW231123cg_seob}{0.59}}}
\newcommand{\costiltoneminus}[1]{\IfEqCase{#1}{{GW231123cg_combined}{0.78}{GW231123cg_nrsur}{0.66}{GW231123cg_xphm}{0.39}{GW231123cg_xo4a}{0.27}{GW231123cg_tphm}{0.29}{GW231123cg_seob}{0.35}}}
\newcommand{\costiltonemed}[1]{\IfEqCase{#1}{{GW231123cg_combined}{0.48}{GW231123cg_nrsur}{0.67}{GW231123cg_xphm}{-0.25}{GW231123cg_xo4a}{0.44}{GW231123cg_tphm}{0.57}{GW231123cg_seob}{0.60}}}
\newcommand{\costiltoneplus}[1]{\IfEqCase{#1}{{GW231123cg_combined}{0.49}{GW231123cg_nrsur}{0.33}{GW231123cg_xphm}{0.37}{GW231123cg_xo4a}{0.25}{GW231123cg_tphm}{0.29}{GW231123cg_seob}{0.33}}}
\newcommand{\costiltonezeropercent}[1]{\IfEqCase{#1}{{GW231123cg_combined}{-0.93}{GW231123cg_nrsur}{-0.85}{GW231123cg_xphm}{-0.93}{GW231123cg_xo4a}{-0.78}{GW231123cg_tphm}{-0.64}{GW231123cg_seob}{-0.83}}}
\newcommand{\costiltoneonepercent}[1]{\IfEqCase{#1}{{GW231123cg_combined}{-0.62}{GW231123cg_nrsur}{-0.45}{GW231123cg_xphm}{-0.75}{GW231123cg_xo4a}{-0.07}{GW231123cg_tphm}{0.08}{GW231123cg_seob}{-0.07}}}
\newcommand{\costiltoneninetyninepercent}[1]{\IfEqCase{#1}{{GW231123cg_combined}{0.97}{GW231123cg_nrsur}{0.99}{GW231123cg_xphm}{0.62}{GW231123cg_xo4a}{0.69}{GW231123cg_tphm}{0.92}{GW231123cg_seob}{0.96}}}
\newcommand{\costiltonefivepercent}[1]{\IfEqCase{#1}{{GW231123cg_combined}{-0.41}{GW231123cg_nrsur}{-0.17}{GW231123cg_xphm}{-0.61}{GW231123cg_xo4a}{0.11}{GW231123cg_tphm}{0.25}{GW231123cg_seob}{0.19}}}
\newcommand{\costiltoneninetyfivepercent}[1]{\IfEqCase{#1}{{GW231123cg_combined}{0.88}{GW231123cg_nrsur}{0.96}{GW231123cg_xphm}{0.15}{GW231123cg_xo4a}{0.65}{GW231123cg_tphm}{0.84}{GW231123cg_seob}{0.89}}}
\newcommand{\costiltoneninetypercent}[1]{\IfEqCase{#1}{{GW231123cg_combined}{0.81}{GW231123cg_nrsur}{0.93}{GW231123cg_xphm}{0.04}{GW231123cg_xo4a}{0.62}{GW231123cg_tphm}{0.78}{GW231123cg_seob}{0.84}}}
\newcommand{\costilttwominus}[1]{\IfEqCase{#1}{{GW231123cg_combined}{0.64}{GW231123cg_nrsur}{0.49}{GW231123cg_xphm}{0.72}{GW231123cg_xo4a}{0.84}{GW231123cg_tphm}{0.43}{GW231123cg_seob}{0.49}}}
\newcommand{\costilttwomed}[1]{\IfEqCase{#1}{{GW231123cg_combined}{0.34}{GW231123cg_nrsur}{-0.07}{GW231123cg_xphm}{0.61}{GW231123cg_xo4a}{0.31}{GW231123cg_tphm}{0.41}{GW231123cg_seob}{0.44}}}
\newcommand{\costilttwoplus}[1]{\IfEqCase{#1}{{GW231123cg_combined}{0.66}{GW231123cg_nrsur}{0.48}{GW231123cg_xphm}{0.39}{GW231123cg_xo4a}{0.69}{GW231123cg_tphm}{0.43}{GW231123cg_seob}{0.56}}}
\newcommand{\costilttwozeropercent}[1]{\IfEqCase{#1}{{GW231123cg_combined}{-1.00}{GW231123cg_nrsur}{-1.00}{GW231123cg_xphm}{-1.00}{GW231123cg_xo4a}{-1.00}{GW231123cg_tphm}{-0.97}{GW231123cg_seob}{-1.00}}}
\newcommand{\costilttwoonepercent}[1]{\IfEqCase{#1}{{GW231123cg_combined}{-0.81}{GW231123cg_nrsur}{-0.72}{GW231123cg_xphm}{-0.80}{GW231123cg_xo4a}{-0.94}{GW231123cg_tphm}{-0.29}{GW231123cg_seob}{-0.51}}}
\newcommand{\costilttwoninetyninepercent}[1]{\IfEqCase{#1}{{GW231123cg_combined}{0.98}{GW231123cg_nrsur}{0.66}{GW231123cg_xphm}{0.99}{GW231123cg_xo4a}{0.99}{GW231123cg_tphm}{0.91}{GW231123cg_seob}{0.97}}}
\newcommand{\costilttwofivepercent}[1]{\IfEqCase{#1}{{GW231123cg_combined}{-0.48}{GW231123cg_nrsur}{-0.55}{GW231123cg_xphm}{-0.39}{GW231123cg_xo4a}{-0.73}{GW231123cg_tphm}{-0.06}{GW231123cg_seob}{-0.20}}}
\newcommand{\costilttwoninetyfivepercent}[1]{\IfEqCase{#1}{{GW231123cg_combined}{0.91}{GW231123cg_nrsur}{0.42}{GW231123cg_xphm}{0.97}{GW231123cg_xo4a}{0.93}{GW231123cg_tphm}{0.80}{GW231123cg_seob}{0.89}}}
\newcommand{\costilttwoninetypercent}[1]{\IfEqCase{#1}{{GW231123cg_combined}{0.83}{GW231123cg_nrsur}{0.31}{GW231123cg_xphm}{0.93}{GW231123cg_xo4a}{0.87}{GW231123cg_tphm}{0.72}{GW231123cg_seob}{0.82}}}
\newcommand{\decminus}[1]{\IfEqCase{#1}{{GW231123cg_combined}{0.67493}{GW231123cg_nrsur}{0.67564}{GW231123cg_xphm}{0.31025}{GW231123cg_xo4a}{0.79192}{GW231123cg_tphm}{0.57655}{GW231123cg_seob}{0.72002}}}
\newcommand{\decmed}[1]{\IfEqCase{#1}{{GW231123cg_combined}{0.38870}{GW231123cg_nrsur}{0.40098}{GW231123cg_xphm}{0.35264}{GW231123cg_xo4a}{0.62850}{GW231123cg_tphm}{0.26541}{GW231123cg_seob}{0.37019}}}
\newcommand{\decplus}[1]{\IfEqCase{#1}{{GW231123cg_combined}{0.38023}{GW231123cg_nrsur}{0.38260}{GW231123cg_xphm}{0.19326}{GW231123cg_xo4a}{0.14385}{GW231123cg_tphm}{0.50842}{GW231123cg_seob}{0.40310}}}
\newcommand{\deczeropercent}[1]{\IfEqCase{#1}{{GW231123cg_combined}{-1.50323}{GW231123cg_nrsur}{-1.43925}{GW231123cg_xphm}{-1.50323}{GW231123cg_xo4a}{-1.39159}{GW231123cg_tphm}{-1.48402}{GW231123cg_seob}{-1.44368}}}
\newcommand{\deconepercent}[1]{\IfEqCase{#1}{{GW231123cg_combined}{-0.59271}{GW231123cg_nrsur}{-0.55345}{GW231123cg_xphm}{-0.77134}{GW231123cg_xo4a}{-0.51813}{GW231123cg_tphm}{-0.49575}{GW231123cg_seob}{-0.52483}}}
\newcommand{\decninetyninepercent}[1]{\IfEqCase{#1}{{GW231123cg_combined}{0.80697}{GW231123cg_nrsur}{0.82148}{GW231123cg_xphm}{0.64128}{GW231123cg_xo4a}{0.79774}{GW231123cg_tphm}{0.81386}{GW231123cg_seob}{0.80728}}}
\newcommand{\decfivepercent}[1]{\IfEqCase{#1}{{GW231123cg_combined}{-0.28624}{GW231123cg_nrsur}{-0.27467}{GW231123cg_xphm}{0.04239}{GW231123cg_xo4a}{-0.16343}{GW231123cg_tphm}{-0.31114}{GW231123cg_seob}{-0.34983}}}
\newcommand{\decninetyfivepercent}[1]{\IfEqCase{#1}{{GW231123cg_combined}{0.76893}{GW231123cg_nrsur}{0.78358}{GW231123cg_xphm}{0.54589}{GW231123cg_xo4a}{0.77235}{GW231123cg_tphm}{0.77383}{GW231123cg_seob}{0.77330}}}
\newcommand{\decninetypercent}[1]{\IfEqCase{#1}{{GW231123cg_combined}{0.74257}{GW231123cg_nrsur}{0.75759}{GW231123cg_xphm}{0.49897}{GW231123cg_xo4a}{0.75610}{GW231123cg_tphm}{0.74506}{GW231123cg_seob}{0.74755}}}
\newcommand{\finalmasssourceminus}[1]{\IfEqCase{#1}{{GW231123cg_combined}{42}{GW231123cg_nrsur}{32}{GW231123cg_xphm}{25}{GW231123cg_xo4a}{17}{GW231123cg_tphm}{17}{GW231123cg_seob}{22}}}
\newcommand{\finalmasssourcemed}[1]{\IfEqCase{#1}{{GW231123cg_combined}{222}{GW231123cg_nrsur}{222}{GW231123cg_xphm}{231}{GW231123cg_xo4a}{190}{GW231123cg_tphm}{227}{GW231123cg_seob}{226}}}
\newcommand{\finalmasssourceplus}[1]{\IfEqCase{#1}{{GW231123cg_combined}{28}{GW231123cg_nrsur}{22}{GW231123cg_xphm}{27}{GW231123cg_xo4a}{29}{GW231123cg_tphm}{26}{GW231123cg_seob}{25}}}
\newcommand{\finalmasssourcezeropercent}[1]{\IfEqCase{#1}{{GW231123cg_combined}{144}{GW231123cg_nrsur}{144}{GW231123cg_xphm}{174}{GW231123cg_xo4a}{156}{GW231123cg_tphm}{176}{GW231123cg_seob}{177}}}
\newcommand{\finalmasssourceonepercent}[1]{\IfEqCase{#1}{{GW231123cg_combined}{172}{GW231123cg_nrsur}{172}{GW231123cg_xphm}{196}{GW231123cg_xo4a}{167}{GW231123cg_tphm}{203}{GW231123cg_seob}{196}}}
\newcommand{\finalmasssourceninetyninepercent}[1]{\IfEqCase{#1}{{GW231123cg_combined}{265}{GW231123cg_nrsur}{258}{GW231123cg_xphm}{272}{GW231123cg_xo4a}{228}{GW231123cg_tphm}{266}{GW231123cg_seob}{266}}}
\newcommand{\finalmasssourcefivepercent}[1]{\IfEqCase{#1}{{GW231123cg_combined}{181}{GW231123cg_nrsur}{191}{GW231123cg_xphm}{205}{GW231123cg_xo4a}{173}{GW231123cg_tphm}{210}{GW231123cg_seob}{205}}}
\newcommand{\finalmasssourceninetyfivepercent}[1]{\IfEqCase{#1}{{GW231123cg_combined}{250}{GW231123cg_nrsur}{244}{GW231123cg_xphm}{258}{GW231123cg_xo4a}{218}{GW231123cg_tphm}{253}{GW231123cg_seob}{252}}}
\newcommand{\finalmasssourceninetypercent}[1]{\IfEqCase{#1}{{GW231123cg_combined}{243}{GW231123cg_nrsur}{239}{GW231123cg_xphm}{251}{GW231123cg_xo4a}{212}{GW231123cg_tphm}{246}{GW231123cg_seob}{245}}}
\newcommand{\finalspinminus}[1]{\IfEqCase{#1}{{GW231123cg_combined}{0.16}{GW231123cg_nrsur}{0.11}{GW231123cg_xphm}{0.11}{GW231123cg_xo4a}{0.07}{GW231123cg_tphm}{0.04}{GW231123cg_seob}{0.06}}}
\newcommand{\finalspinmed}[1]{\IfEqCase{#1}{{GW231123cg_combined}{0.84}{GW231123cg_nrsur}{0.82}{GW231123cg_xphm}{0.70}{GW231123cg_xo4a}{0.85}{GW231123cg_tphm}{0.88}{GW231123cg_seob}{0.88}}}
\newcommand{\finalspinplus}[1]{\IfEqCase{#1}{{GW231123cg_combined}{0.08}{GW231123cg_nrsur}{0.06}{GW231123cg_xphm}{0.08}{GW231123cg_xo4a}{0.06}{GW231123cg_tphm}{0.04}{GW231123cg_seob}{0.05}}}
\newcommand{\finalspinzeropercent}[1]{\IfEqCase{#1}{{GW231123cg_combined}{0.24}{GW231123cg_nrsur}{0.52}{GW231123cg_xphm}{0.24}{GW231123cg_xo4a}{0.54}{GW231123cg_tphm}{0.69}{GW231123cg_seob}{0.65}}}
\newcommand{\finalspinonepercent}[1]{\IfEqCase{#1}{{GW231123cg_combined}{0.58}{GW231123cg_nrsur}{0.64}{GW231123cg_xphm}{0.51}{GW231123cg_xo4a}{0.74}{GW231123cg_tphm}{0.80}{GW231123cg_seob}{0.77}}}
\newcommand{\finalspinninetyninepercent}[1]{\IfEqCase{#1}{{GW231123cg_combined}{0.92}{GW231123cg_nrsur}{0.89}{GW231123cg_xphm}{0.84}{GW231123cg_xo4a}{0.92}{GW231123cg_tphm}{0.92}{GW231123cg_seob}{0.93}}}
\newcommand{\finalspinfivepercent}[1]{\IfEqCase{#1}{{GW231123cg_combined}{0.66}{GW231123cg_nrsur}{0.69}{GW231123cg_xphm}{0.58}{GW231123cg_xo4a}{0.77}{GW231123cg_tphm}{0.83}{GW231123cg_seob}{0.81}}}
\newcommand{\finalspinninetyfivepercent}[1]{\IfEqCase{#1}{{GW231123cg_combined}{0.91}{GW231123cg_nrsur}{0.87}{GW231123cg_xphm}{0.78}{GW231123cg_xo4a}{0.91}{GW231123cg_tphm}{0.92}{GW231123cg_seob}{0.92}}}
\newcommand{\finalspinninetypercent}[1]{\IfEqCase{#1}{{GW231123cg_combined}{0.90}{GW231123cg_nrsur}{0.86}{GW231123cg_xphm}{0.76}{GW231123cg_xo4a}{0.90}{GW231123cg_tphm}{0.91}{GW231123cg_seob}{0.91}}}
\newcommand{\geocenttimeminus}[1]{\IfEqCase{#1}{{GW231123cg_combined}{0.0}{GW231123cg_nrsur}{0.0}{GW231123cg_xphm}{0.0}{GW231123cg_xo4a}{0.0}{GW231123cg_tphm}{0.0}{GW231123cg_seob}{0.0}}}
\newcommand{\geocenttimemed}[1]{\IfEqCase{#1}{{GW231123cg_combined}{1384782888.6}{GW231123cg_nrsur}{1384782888.6}{GW231123cg_xphm}{1384782888.6}{GW231123cg_xo4a}{1384782888.6}{GW231123cg_tphm}{1384782888.6}{GW231123cg_seob}{1384782888.6}}}
\newcommand{\geocenttimeplus}[1]{\IfEqCase{#1}{{GW231123cg_combined}{0.0}{GW231123cg_nrsur}{0.0}{GW231123cg_xphm}{0.0}{GW231123cg_xo4a}{0.0}{GW231123cg_tphm}{0.0}{GW231123cg_seob}{0.0}}}
\newcommand{\geocenttimezeropercent}[1]{\IfEqCase{#1}{{GW231123cg_combined}{1384782888.6}{GW231123cg_nrsur}{1384782888.6}{GW231123cg_xphm}{1384782888.6}{GW231123cg_xo4a}{1384782888.6}{GW231123cg_tphm}{1384782888.6}{GW231123cg_seob}{1384782888.6}}}
\newcommand{\geocenttimeonepercent}[1]{\IfEqCase{#1}{{GW231123cg_combined}{1384782888.6}{GW231123cg_nrsur}{1384782888.6}{GW231123cg_xphm}{1384782888.6}{GW231123cg_xo4a}{1384782888.6}{GW231123cg_tphm}{1384782888.6}{GW231123cg_seob}{1384782888.6}}}
\newcommand{\geocenttimeninetyninepercent}[1]{\IfEqCase{#1}{{GW231123cg_combined}{1384782888.6}{GW231123cg_nrsur}{1384782888.6}{GW231123cg_xphm}{1384782888.6}{GW231123cg_xo4a}{1384782888.6}{GW231123cg_tphm}{1384782888.6}{GW231123cg_seob}{1384782888.6}}}
\newcommand{\geocenttimefivepercent}[1]{\IfEqCase{#1}{{GW231123cg_combined}{1384782888.6}{GW231123cg_nrsur}{1384782888.6}{GW231123cg_xphm}{1384782888.6}{GW231123cg_xo4a}{1384782888.6}{GW231123cg_tphm}{1384782888.6}{GW231123cg_seob}{1384782888.6}}}
\newcommand{\geocenttimeninetyfivepercent}[1]{\IfEqCase{#1}{{GW231123cg_combined}{1384782888.6}{GW231123cg_nrsur}{1384782888.6}{GW231123cg_xphm}{1384782888.6}{GW231123cg_xo4a}{1384782888.6}{GW231123cg_tphm}{1384782888.6}{GW231123cg_seob}{1384782888.6}}}
\newcommand{\geocenttimeninetypercent}[1]{\IfEqCase{#1}{{GW231123cg_combined}{1384782888.6}{GW231123cg_nrsur}{1384782888.6}{GW231123cg_xphm}{1384782888.6}{GW231123cg_xo4a}{1384782888.6}{GW231123cg_tphm}{1384782888.6}{GW231123cg_seob}{1384782888.6}}}
\newcommand{\iotaminus}[1]{\IfEqCase{#1}{{GW231123cg_combined}{0.66}{GW231123cg_nrsur}{0.42}{GW231123cg_xphm}{0.62}{GW231123cg_xo4a}{0.38}{GW231123cg_tphm}{1.20}{GW231123cg_seob}{0.54}}}
\newcommand{\iotamed}[1]{\IfEqCase{#1}{{GW231123cg_combined}{1.11}{GW231123cg_nrsur}{1.03}{GW231123cg_xphm}{1.48}{GW231123cg_xo4a}{0.64}{GW231123cg_tphm}{1.88}{GW231123cg_seob}{1.13}}}
\newcommand{\iotaplus}[1]{\IfEqCase{#1}{{GW231123cg_combined}{1.29}{GW231123cg_nrsur}{1.37}{GW231123cg_xphm}{0.74}{GW231123cg_xo4a}{1.61}{GW231123cg_tphm}{0.57}{GW231123cg_seob}{1.36}}}
\newcommand{\iotazeropercent}[1]{\IfEqCase{#1}{{GW231123cg_combined}{0.01}{GW231123cg_nrsur}{0.10}{GW231123cg_xphm}{0.16}{GW231123cg_xo4a}{0.01}{GW231123cg_tphm}{0.20}{GW231123cg_seob}{0.06}}}
\newcommand{\iotaonepercent}[1]{\IfEqCase{#1}{{GW231123cg_combined}{0.25}{GW231123cg_nrsur}{0.43}{GW231123cg_xphm}{0.75}{GW231123cg_xo4a}{0.13}{GW231123cg_tphm}{0.49}{GW231123cg_seob}{0.43}}}
\newcommand{\iotaninetyninepercent}[1]{\IfEqCase{#1}{{GW231123cg_combined}{2.65}{GW231123cg_nrsur}{2.53}{GW231123cg_xphm}{2.37}{GW231123cg_xo4a}{2.89}{GW231123cg_tphm}{2.64}{GW231123cg_seob}{2.65}}}
\newcommand{\iotafivepercent}[1]{\IfEqCase{#1}{{GW231123cg_combined}{0.45}{GW231123cg_nrsur}{0.62}{GW231123cg_xphm}{0.87}{GW231123cg_xo4a}{0.25}{GW231123cg_tphm}{0.68}{GW231123cg_seob}{0.59}}}
\newcommand{\iotaninetyfivepercent}[1]{\IfEqCase{#1}{{GW231123cg_combined}{2.41}{GW231123cg_nrsur}{2.40}{GW231123cg_xphm}{2.23}{GW231123cg_xo4a}{2.25}{GW231123cg_tphm}{2.45}{GW231123cg_seob}{2.48}}}
\newcommand{\iotaninetypercent}[1]{\IfEqCase{#1}{{GW231123cg_combined}{2.30}{GW231123cg_nrsur}{2.33}{GW231123cg_xphm}{2.11}{GW231123cg_xo4a}{1.15}{GW231123cg_tphm}{2.35}{GW231123cg_seob}{2.39}}}
\newcommand{\loglikelihoodminus}[1]{\IfEqCase{#1}{{GW231123cg_combined}{8.5}{GW231123cg_nrsur}{5.8}{GW231123cg_xphm}{5.5}{GW231123cg_xo4a}{5.1}{GW231123cg_tphm}{5.4}{GW231123cg_seob}{5.3}}}
\newcommand{\loglikelihoodmed}[1]{\IfEqCase{#1}{{GW231123cg_combined}{210.4}{GW231123cg_nrsur}{208.8}{GW231123cg_xphm}{204.4}{GW231123cg_xo4a}{212.9}{GW231123cg_tphm}{212.8}{GW231123cg_seob}{211.2}}}
\newcommand{\loglikelihoodplus}[1]{\IfEqCase{#1}{{GW231123cg_combined}{5.1}{GW231123cg_nrsur}{4.0}{GW231123cg_xphm}{3.7}{GW231123cg_xo4a}{3.2}{GW231123cg_tphm}{3.5}{GW231123cg_seob}{3.3}}}
\newcommand{\loglikelihoodzeropercent}[1]{\IfEqCase{#1}{{GW231123cg_combined}{188.7}{GW231123cg_nrsur}{188.0}{GW231123cg_xphm}{188.7}{GW231123cg_xo4a}{197.2}{GW231123cg_tphm}{195.2}{GW231123cg_seob}{194.3}}}
\newcommand{\loglikelihoodonepercent}[1]{\IfEqCase{#1}{{GW231123cg_combined}{198.6}{GW231123cg_nrsur}{199.9}{GW231123cg_xphm}{196.0}{GW231123cg_xo4a}{205.2}{GW231123cg_tphm}{204.4}{GW231123cg_seob}{203.1}}}
\newcommand{\loglikelihoodninetyninepercent}[1]{\IfEqCase{#1}{{GW231123cg_combined}{216.7}{GW231123cg_nrsur}{214.5}{GW231123cg_xphm}{209.1}{GW231123cg_xo4a}{217.1}{GW231123cg_tphm}{217.3}{GW231123cg_seob}{215.4}}}
\newcommand{\loglikelihoodfivepercent}[1]{\IfEqCase{#1}{{GW231123cg_combined}{201.9}{GW231123cg_nrsur}{203.0}{GW231123cg_xphm}{198.9}{GW231123cg_xo4a}{207.8}{GW231123cg_tphm}{207.4}{GW231123cg_seob}{205.9}}}
\newcommand{\loglikelihoodninetyfivepercent}[1]{\IfEqCase{#1}{{GW231123cg_combined}{215.4}{GW231123cg_nrsur}{212.8}{GW231123cg_xphm}{208.1}{GW231123cg_xo4a}{216.2}{GW231123cg_tphm}{216.3}{GW231123cg_seob}{214.5}}}
\newcommand{\loglikelihoodninetypercent}[1]{\IfEqCase{#1}{{GW231123cg_combined}{214.6}{GW231123cg_nrsur}{212.0}{GW231123cg_xphm}{207.4}{GW231123cg_xo4a}{215.6}{GW231123cg_tphm}{215.6}{GW231123cg_seob}{214.0}}}
\newcommand{\logpriorminus}[1]{\IfEqCase{#1}{{GW231123cg_combined}{8.7}{GW231123cg_nrsur}{8.8}{GW231123cg_xphm}{8.7}{GW231123cg_xo4a}{8.7}{GW231123cg_tphm}{8.7}{GW231123cg_seob}{8.6}}}
\newcommand{\logpriormed}[1]{\IfEqCase{#1}{{GW231123cg_combined}{106.3}{GW231123cg_nrsur}{106.9}{GW231123cg_xphm}{106.3}{GW231123cg_xo4a}{106.0}{GW231123cg_tphm}{106.3}{GW231123cg_seob}{106.0}}}
\newcommand{\logpriorplus}[1]{\IfEqCase{#1}{{GW231123cg_combined}{6.8}{GW231123cg_nrsur}{6.7}{GW231123cg_xphm}{6.8}{GW231123cg_xo4a}{6.8}{GW231123cg_tphm}{6.7}{GW231123cg_seob}{6.8}}}
\newcommand{\logpriorzeropercent}[1]{\IfEqCase{#1}{{GW231123cg_combined}{77.7}{GW231123cg_nrsur}{80.2}{GW231123cg_xphm}{80.5}{GW231123cg_xo4a}{82.0}{GW231123cg_tphm}{77.7}{GW231123cg_seob}{84.4}}}
\newcommand{\logprioronepercent}[1]{\IfEqCase{#1}{{GW231123cg_combined}{93.3}{GW231123cg_nrsur}{93.8}{GW231123cg_xphm}{93.3}{GW231123cg_xo4a}{92.9}{GW231123cg_tphm}{93.3}{GW231123cg_seob}{93.1}}}
\newcommand{\logpriorninetyninepercent}[1]{\IfEqCase{#1}{{GW231123cg_combined}{115.3}{GW231123cg_nrsur}{115.8}{GW231123cg_xphm}{115.6}{GW231123cg_xo4a}{115.1}{GW231123cg_tphm}{115.2}{GW231123cg_seob}{115.0}}}
\newcommand{\logpriorfivepercent}[1]{\IfEqCase{#1}{{GW231123cg_combined}{97.5}{GW231123cg_nrsur}{98.1}{GW231123cg_xphm}{97.5}{GW231123cg_xo4a}{97.2}{GW231123cg_tphm}{97.6}{GW231123cg_seob}{97.4}}}
\newcommand{\logpriorninetyfivepercent}[1]{\IfEqCase{#1}{{GW231123cg_combined}{113.1}{GW231123cg_nrsur}{113.6}{GW231123cg_xphm}{113.1}{GW231123cg_xo4a}{112.8}{GW231123cg_tphm}{113.0}{GW231123cg_seob}{112.8}}}
\newcommand{\logpriorninetypercent}[1]{\IfEqCase{#1}{{GW231123cg_combined}{111.7}{GW231123cg_nrsur}{112.3}{GW231123cg_xphm}{111.7}{GW231123cg_xo4a}{111.4}{GW231123cg_tphm}{111.6}{GW231123cg_seob}{111.5}}}
\newcommand{\luminositydistanceminus}[1]{\IfEqCase{#1}{{GW231123cg_combined}{1500}{GW231123cg_nrsur}{1000}{GW231123cg_xphm}{300}{GW231123cg_xo4a}{1400}{GW231123cg_tphm}{1100}{GW231123cg_seob}{1000}}}
\newcommand{\luminositydistancemed}[1]{\IfEqCase{#1}{{GW231123cg_combined}{2200}{GW231123cg_nrsur}{1900}{GW231123cg_xphm}{900}{GW231123cg_xo4a}{3500}{GW231123cg_tphm}{2700}{GW231123cg_seob}{2300}}}
\newcommand{\luminositydistanceplus}[1]{\IfEqCase{#1}{{GW231123cg_combined}{1900}{GW231123cg_nrsur}{1700}{GW231123cg_xphm}{400}{GW231123cg_xo4a}{1200}{GW231123cg_tphm}{1200}{GW231123cg_seob}{1400}}}
\newcommand{\luminositydistancezeropercent}[1]{\IfEqCase{#1}{{GW231123cg_combined}{100}{GW231123cg_nrsur}{200}{GW231123cg_xphm}{100}{GW231123cg_xo4a}{700}{GW231123cg_tphm}{500}{GW231123cg_seob}{400}}}
\newcommand{\luminositydistanceonepercent}[1]{\IfEqCase{#1}{{GW231123cg_combined}{500}{GW231123cg_nrsur}{600}{GW231123cg_xphm}{300}{GW231123cg_xo4a}{1700}{GW231123cg_tphm}{1200}{GW231123cg_seob}{900}}}
\newcommand{\luminositydistanceninetyninepercent}[1]{\IfEqCase{#1}{{GW231123cg_combined}{4900}{GW231123cg_nrsur}{4900}{GW231123cg_xphm}{1500}{GW231123cg_xo4a}{5200}{GW231123cg_tphm}{4300}{GW231123cg_seob}{4400}}}
\newcommand{\luminositydistancefivepercent}[1]{\IfEqCase{#1}{{GW231123cg_combined}{700}{GW231123cg_nrsur}{800}{GW231123cg_xphm}{600}{GW231123cg_xo4a}{2100}{GW231123cg_tphm}{1600}{GW231123cg_seob}{1300}}}
\newcommand{\luminositydistanceninetyfivepercent}[1]{\IfEqCase{#1}{{GW231123cg_combined}{4200}{GW231123cg_nrsur}{3500}{GW231123cg_xphm}{1300}{GW231123cg_xo4a}{4700}{GW231123cg_tphm}{3800}{GW231123cg_seob}{3600}}}
\newcommand{\luminositydistanceninetypercent}[1]{\IfEqCase{#1}{{GW231123cg_combined}{3800}{GW231123cg_nrsur}{3000}{GW231123cg_xphm}{1200}{GW231123cg_xo4a}{4500}{GW231123cg_tphm}{3500}{GW231123cg_seob}{3300}}}
\newcommand{\massoneminus}[1]{\IfEqCase{#1}{{GW231123cg_combined}{24}{GW231123cg_nrsur}{22}{GW231123cg_xphm}{10}{GW231123cg_xo4a}{20}{GW231123cg_tphm}{20}{GW231123cg_seob}{21}}}
\newcommand{\massonemed}[1]{\IfEqCase{#1}{{GW231123cg_combined}{186}{GW231123cg_nrsur}{173}{GW231123cg_xphm}{176}{GW231123cg_xo4a}{226}{GW231123cg_tphm}{195}{GW231123cg_seob}{188}}}
\newcommand{\massoneplus}[1]{\IfEqCase{#1}{{GW231123cg_combined}{49}{GW231123cg_nrsur}{19}{GW231123cg_xphm}{12}{GW231123cg_xo4a}{31}{GW231123cg_tphm}{25}{GW231123cg_seob}{28}}}
\newcommand{\massonezeropercent}[1]{\IfEqCase{#1}{{GW231123cg_combined}{122}{GW231123cg_nrsur}{136}{GW231123cg_xphm}{144}{GW231123cg_xo4a}{122}{GW231123cg_tphm}{154}{GW231123cg_seob}{146}}}
\newcommand{\massoneonepercent}[1]{\IfEqCase{#1}{{GW231123cg_combined}{151}{GW231123cg_nrsur}{145}{GW231123cg_xphm}{161}{GW231123cg_xo4a}{191}{GW231123cg_tphm}{168}{GW231123cg_seob}{160}}}
\newcommand{\massoneninetyninepercent}[1]{\IfEqCase{#1}{{GW231123cg_combined}{257}{GW231123cg_nrsur}{205}{GW231123cg_xphm}{194}{GW231123cg_xo4a}{270}{GW231123cg_tphm}{233}{GW231123cg_seob}{233}}}
\newcommand{\massonefivepercent}[1]{\IfEqCase{#1}{{GW231123cg_combined}{162}{GW231123cg_nrsur}{151}{GW231123cg_xphm}{165}{GW231123cg_xo4a}{207}{GW231123cg_tphm}{175}{GW231123cg_seob}{167}}}
\newcommand{\massoneninetyfivepercent}[1]{\IfEqCase{#1}{{GW231123cg_combined}{235}{GW231123cg_nrsur}{192}{GW231123cg_xphm}{187}{GW231123cg_xo4a}{257}{GW231123cg_tphm}{220}{GW231123cg_seob}{216}}}
\newcommand{\massoneninetypercent}[1]{\IfEqCase{#1}{{GW231123cg_combined}{228}{GW231123cg_nrsur}{187}{GW231123cg_xphm}{185}{GW231123cg_xo4a}{245}{GW231123cg_tphm}{214}{GW231123cg_seob}{208}}}
\newcommand{\massonesourceminus}[1]{\IfEqCase{#1}{{GW231123cg_combined}{18}{GW231123cg_nrsur}{16}{GW231123cg_xphm}{13}{GW231123cg_xo4a}{16}{GW231123cg_tphm}{13}{GW231123cg_seob}{15}}}
\newcommand{\massonesourcemed}[1]{\IfEqCase{#1}{{GW231123cg_combined}{137}{GW231123cg_nrsur}{128}{GW231123cg_xphm}{149}{GW231123cg_xo4a}{143}{GW231123cg_tphm}{133}{GW231123cg_seob}{133}}}
\newcommand{\massonesourceplus}[1]{\IfEqCase{#1}{{GW231123cg_combined}{23}{GW231123cg_nrsur}{16}{GW231123cg_xphm}{14}{GW231123cg_xo4a}{26}{GW231123cg_tphm}{19}{GW231123cg_seob}{19}}}
\newcommand{\massonesourcezeropercent}[1]{\IfEqCase{#1}{{GW231123cg_combined}{83}{GW231123cg_nrsur}{83}{GW231123cg_xphm}{115}{GW231123cg_xo4a}{95}{GW231123cg_tphm}{107}{GW231123cg_seob}{101}}}
\newcommand{\massonesourceonepercent}[1]{\IfEqCase{#1}{{GW231123cg_combined}{111}{GW231123cg_nrsur}{103}{GW231123cg_xphm}{132}{GW231123cg_xo4a}{122}{GW231123cg_tphm}{116}{GW231123cg_seob}{114}}}
\newcommand{\massonesourceninetyninepercent}[1]{\IfEqCase{#1}{{GW231123cg_combined}{172}{GW231123cg_nrsur}{153}{GW231123cg_xphm}{170}{GW231123cg_xo4a}{185}{GW231123cg_tphm}{162}{GW231123cg_seob}{163}}}
\newcommand{\massonesourcefivepercent}[1]{\IfEqCase{#1}{{GW231123cg_combined}{119}{GW231123cg_nrsur}{112}{GW231123cg_xphm}{137}{GW231123cg_xo4a}{127}{GW231123cg_tphm}{121}{GW231123cg_seob}{119}}}
\newcommand{\massonesourceninetyfivepercent}[1]{\IfEqCase{#1}{{GW231123cg_combined}{161}{GW231123cg_nrsur}{144}{GW231123cg_xphm}{163}{GW231123cg_xo4a}{169}{GW231123cg_tphm}{152}{GW231123cg_seob}{152}}}
\newcommand{\massonesourceninetypercent}[1]{\IfEqCase{#1}{{GW231123cg_combined}{155}{GW231123cg_nrsur}{140}{GW231123cg_xphm}{160}{GW231123cg_xo4a}{163}{GW231123cg_tphm}{147}{GW231123cg_seob}{148}}}
\newcommand{\masstwominus}[1]{\IfEqCase{#1}{{GW231123cg_combined}{59}{GW231123cg_nrsur}{18}{GW231123cg_xphm}{24}{GW231123cg_xo4a}{30}{GW231123cg_tphm}{20}{GW231123cg_seob}{23}}}
\newcommand{\masstwomed}[1]{\IfEqCase{#1}{{GW231123cg_combined}{139}{GW231123cg_nrsur}{144}{GW231123cg_xphm}{108}{GW231123cg_xo4a}{88}{GW231123cg_tphm}{160}{GW231123cg_seob}{153}}}
\newcommand{\masstwoplus}[1]{\IfEqCase{#1}{{GW231123cg_combined}{30}{GW231123cg_nrsur}{14}{GW231123cg_xphm}{22}{GW231123cg_xo4a}{11}{GW231123cg_tphm}{17}{GW231123cg_seob}{17}}}
\newcommand{\masstwozeropercent}[1]{\IfEqCase{#1}{{GW231123cg_combined}{47}{GW231123cg_nrsur}{68}{GW231123cg_xphm}{48}{GW231123cg_xo4a}{47}{GW231123cg_tphm}{98}{GW231123cg_seob}{82}}}
\newcommand{\masstwoonepercent}[1]{\IfEqCase{#1}{{GW231123cg_combined}{58}{GW231123cg_nrsur}{116}{GW231123cg_xphm}{75}{GW231123cg_xo4a}{52}{GW231123cg_tphm}{130}{GW231123cg_seob}{116}}}
\newcommand{\masstwoninetyninepercent}[1]{\IfEqCase{#1}{{GW231123cg_combined}{177}{GW231123cg_nrsur}{163}{GW231123cg_xphm}{143}{GW231123cg_xo4a}{106}{GW231123cg_tphm}{183}{GW231123cg_seob}{176}}}
\newcommand{\masstwofivepercent}[1]{\IfEqCase{#1}{{GW231123cg_combined}{81}{GW231123cg_nrsur}{126}{GW231123cg_xphm}{84}{GW231123cg_xo4a}{58}{GW231123cg_tphm}{140}{GW231123cg_seob}{130}}}
\newcommand{\masstwoninetyfivepercent}[1]{\IfEqCase{#1}{{GW231123cg_combined}{169}{GW231123cg_nrsur}{158}{GW231123cg_xphm}{129}{GW231123cg_xo4a}{99}{GW231123cg_tphm}{177}{GW231123cg_seob}{170}}}
\newcommand{\masstwoninetypercent}[1]{\IfEqCase{#1}{{GW231123cg_combined}{164}{GW231123cg_nrsur}{156}{GW231123cg_xphm}{124}{GW231123cg_xo4a}{96}{GW231123cg_tphm}{173}{GW231123cg_seob}{166}}}
\newcommand{\masstwosourceminus}[1]{\IfEqCase{#1}{{GW231123cg_combined}{50}{GW231123cg_nrsur}{20}{GW231123cg_xphm}{22}{GW231123cg_xo4a}{18}{GW231123cg_tphm}{17}{GW231123cg_seob}{22}}}
\newcommand{\masstwosourcemed}[1]{\IfEqCase{#1}{{GW231123cg_combined}{101}{GW231123cg_nrsur}{108}{GW231123cg_xphm}{92}{GW231123cg_xo4a}{55}{GW231123cg_tphm}{110}{GW231123cg_seob}{109}}}
\newcommand{\masstwosourceplus}[1]{\IfEqCase{#1}{{GW231123cg_combined}{22}{GW231123cg_nrsur}{16}{GW231123cg_xphm}{21}{GW231123cg_xo4a}{12}{GW231123cg_tphm}{16}{GW231123cg_seob}{17}}}
\newcommand{\masstwosourcezeropercent}[1]{\IfEqCase{#1}{{GW231123cg_combined}{29}{GW231123cg_nrsur}{42}{GW231123cg_xphm}{40}{GW231123cg_xo4a}{29}{GW231123cg_tphm}{60}{GW231123cg_seob}{52}}}
\newcommand{\masstwosourceonepercent}[1]{\IfEqCase{#1}{{GW231123cg_combined}{37}{GW231123cg_nrsur}{73}{GW231123cg_xphm}{62}{GW231123cg_xo4a}{33}{GW231123cg_tphm}{84}{GW231123cg_seob}{75}}}
\newcommand{\masstwosourceninetyninepercent}[1]{\IfEqCase{#1}{{GW231123cg_combined}{131}{GW231123cg_nrsur}{130}{GW231123cg_xphm}{124}{GW231123cg_xo4a}{74}{GW231123cg_tphm}{133}{GW231123cg_seob}{134}}}
\newcommand{\masstwosourcefivepercent}[1]{\IfEqCase{#1}{{GW231123cg_combined}{51}{GW231123cg_nrsur}{88}{GW231123cg_xphm}{70}{GW231123cg_xo4a}{37}{GW231123cg_tphm}{92}{GW231123cg_seob}{87}}}
\newcommand{\masstwosourceninetyfivepercent}[1]{\IfEqCase{#1}{{GW231123cg_combined}{123}{GW231123cg_nrsur}{123}{GW231123cg_xphm}{113}{GW231123cg_xo4a}{67}{GW231123cg_tphm}{126}{GW231123cg_seob}{126}}}
\newcommand{\masstwosourceninetypercent}[1]{\IfEqCase{#1}{{GW231123cg_combined}{119}{GW231123cg_nrsur}{120}{GW231123cg_xphm}{108}{GW231123cg_xo4a}{64}{GW231123cg_tphm}{122}{GW231123cg_seob}{122}}}
\newcommand{\massratiominus}[1]{\IfEqCase{#1}{{GW231123cg_combined}{0.37}{GW231123cg_nrsur}{0.12}{GW231123cg_xphm}{0.14}{GW231123cg_xo4a}{0.16}{GW231123cg_tphm}{0.14}{GW231123cg_seob}{0.15}}}
\newcommand{\massratiomed}[1]{\IfEqCase{#1}{{GW231123cg_combined}{0.74}{GW231123cg_nrsur}{0.85}{GW231123cg_xphm}{0.61}{GW231123cg_xo4a}{0.39}{GW231123cg_tphm}{0.82}{GW231123cg_seob}{0.82}}}
\newcommand{\massratioplus}[1]{\IfEqCase{#1}{{GW231123cg_combined}{0.23}{GW231123cg_nrsur}{0.15}{GW231123cg_xphm}{0.13}{GW231123cg_xo4a}{0.07}{GW231123cg_tphm}{0.16}{GW231123cg_seob}{0.18}}}
\newcommand{\massratiozeropercent}[1]{\IfEqCase{#1}{{GW231123cg_combined}{0.17}{GW231123cg_nrsur}{0.27}{GW231123cg_xphm}{0.27}{GW231123cg_xo4a}{0.17}{GW231123cg_tphm}{0.40}{GW231123cg_seob}{0.28}}}
\newcommand{\massratioonepercent}[1]{\IfEqCase{#1}{{GW231123cg_combined}{0.22}{GW231123cg_nrsur}{0.59}{GW231123cg_xphm}{0.42}{GW231123cg_xo4a}{0.20}{GW231123cg_tphm}{0.58}{GW231123cg_seob}{0.51}}}
\newcommand{\massrationinetyninepercent}[1]{\IfEqCase{#1}{{GW231123cg_combined}{0.99}{GW231123cg_nrsur}{0.99}{GW231123cg_xphm}{0.81}{GW231123cg_xo4a}{0.55}{GW231123cg_tphm}{0.99}{GW231123cg_seob}{0.99}}}
\newcommand{\massratiofivepercent}[1]{\IfEqCase{#1}{{GW231123cg_combined}{0.36}{GW231123cg_nrsur}{0.68}{GW231123cg_xphm}{0.47}{GW231123cg_xo4a}{0.22}{GW231123cg_tphm}{0.66}{GW231123cg_seob}{0.62}}}
\newcommand{\massrationinetyfivepercent}[1]{\IfEqCase{#1}{{GW231123cg_combined}{0.96}{GW231123cg_nrsur}{0.97}{GW231123cg_xphm}{0.74}{GW231123cg_xo4a}{0.46}{GW231123cg_tphm}{0.96}{GW231123cg_seob}{0.97}}}
\newcommand{\massrationinetypercent}[1]{\IfEqCase{#1}{{GW231123cg_combined}{0.92}{GW231123cg_nrsur}{0.95}{GW231123cg_xphm}{0.71}{GW231123cg_xo4a}{0.44}{GW231123cg_tphm}{0.94}{GW231123cg_seob}{0.95}}}
\newcommand{\networkmatchedfiltersnrminus}[1]{\IfEqCase{#1}{{GW231123cg_combined}{0.3}{GW231123cg_nrsur}{0.3}{GW231123cg_xphm}{0.3}{GW231123cg_xo4a}{0.2}{GW231123cg_tphm}{0.3}{GW231123cg_seob}{0.3}}}
\newcommand{\networkmatchedfiltersnrmed}[1]{\IfEqCase{#1}{{GW231123cg_combined}{20.7}{GW231123cg_nrsur}{20.6}{GW231123cg_xphm}{20.5}{GW231123cg_xo4a}{20.8}{GW231123cg_tphm}{20.8}{GW231123cg_seob}{20.7}}}
\newcommand{\networkmatchedfiltersnrplus}[1]{\IfEqCase{#1}{{GW231123cg_combined}{0.2}{GW231123cg_nrsur}{0.2}{GW231123cg_xphm}{0.2}{GW231123cg_xo4a}{0.2}{GW231123cg_tphm}{0.2}{GW231123cg_seob}{0.2}}}
\newcommand{\networkmatchedfiltersnrzeropercent}[1]{\IfEqCase{#1}{{GW231123cg_combined}{19.6}{GW231123cg_nrsur}{19.6}{GW231123cg_xphm}{19.6}{GW231123cg_xo4a}{20.1}{GW231123cg_tphm}{20.0}{GW231123cg_seob}{19.9}}}
\newcommand{\networkmatchedfiltersnronepercent}[1]{\IfEqCase{#1}{{GW231123cg_combined}{20.2}{GW231123cg_nrsur}{20.2}{GW231123cg_xphm}{20.1}{GW231123cg_xo4a}{20.4}{GW231123cg_tphm}{20.4}{GW231123cg_seob}{20.3}}}
\newcommand{\networkmatchedfiltersnrninetyninepercent}[1]{\IfEqCase{#1}{{GW231123cg_combined}{21.0}{GW231123cg_nrsur}{21.0}{GW231123cg_xphm}{20.8}{GW231123cg_xo4a}{21.0}{GW231123cg_tphm}{21.0}{GW231123cg_seob}{21.0}}}
\newcommand{\networkmatchedfiltersnrfivepercent}[1]{\IfEqCase{#1}{{GW231123cg_combined}{20.4}{GW231123cg_nrsur}{20.4}{GW231123cg_xphm}{20.2}{GW231123cg_xo4a}{20.5}{GW231123cg_tphm}{20.5}{GW231123cg_seob}{20.5}}}
\newcommand{\networkmatchedfiltersnrninetyfivepercent}[1]{\IfEqCase{#1}{{GW231123cg_combined}{20.9}{GW231123cg_nrsur}{20.9}{GW231123cg_xphm}{20.7}{GW231123cg_xo4a}{20.9}{GW231123cg_tphm}{21.0}{GW231123cg_seob}{20.9}}}
\newcommand{\networkmatchedfiltersnrninetypercent}[1]{\IfEqCase{#1}{{GW231123cg_combined}{20.9}{GW231123cg_nrsur}{20.8}{GW231123cg_xphm}{20.7}{GW231123cg_xo4a}{20.9}{GW231123cg_tphm}{20.9}{GW231123cg_seob}{20.9}}}
\newcommand{\networkoptimalsnrminus}[1]{\IfEqCase{#1}{{GW231123cg_combined}{1.7}{GW231123cg_nrsur}{1.7}{GW231123cg_xphm}{1.7}{GW231123cg_xo4a}{1.7}{GW231123cg_tphm}{1.7}{GW231123cg_seob}{1.7}}}
\newcommand{\networkoptimalsnrmed}[1]{\IfEqCase{#1}{{GW231123cg_combined}{20.5}{GW231123cg_nrsur}{20.5}{GW231123cg_xphm}{20.3}{GW231123cg_xo4a}{20.6}{GW231123cg_tphm}{20.7}{GW231123cg_seob}{20.6}}}
\newcommand{\networkoptimalsnrplus}[1]{\IfEqCase{#1}{{GW231123cg_combined}{1.7}{GW231123cg_nrsur}{1.6}{GW231123cg_xphm}{1.7}{GW231123cg_xo4a}{1.7}{GW231123cg_tphm}{1.6}{GW231123cg_seob}{1.7}}}
\newcommand{\networkoptimalsnrzeropercent}[1]{\IfEqCase{#1}{{GW231123cg_combined}{16.4}{GW231123cg_nrsur}{16.5}{GW231123cg_xphm}{16.3}{GW231123cg_xo4a}{16.5}{GW231123cg_tphm}{16.4}{GW231123cg_seob}{16.8}}}
\newcommand{\networkoptimalsnronepercent}[1]{\IfEqCase{#1}{{GW231123cg_combined}{18.2}{GW231123cg_nrsur}{18.1}{GW231123cg_xphm}{18.0}{GW231123cg_xo4a}{18.3}{GW231123cg_tphm}{18.3}{GW231123cg_seob}{18.2}}}
\newcommand{\networkoptimalsnrninetyninepercent}[1]{\IfEqCase{#1}{{GW231123cg_combined}{22.9}{GW231123cg_nrsur}{22.8}{GW231123cg_xphm}{22.7}{GW231123cg_xo4a}{23.0}{GW231123cg_tphm}{23.0}{GW231123cg_seob}{22.9}}}
\newcommand{\networkoptimalsnrfivepercent}[1]{\IfEqCase{#1}{{GW231123cg_combined}{18.9}{GW231123cg_nrsur}{18.8}{GW231123cg_xphm}{18.6}{GW231123cg_xo4a}{19.0}{GW231123cg_tphm}{19.0}{GW231123cg_seob}{18.9}}}
\newcommand{\networkoptimalsnrninetyfivepercent}[1]{\IfEqCase{#1}{{GW231123cg_combined}{22.2}{GW231123cg_nrsur}{22.1}{GW231123cg_xphm}{22.0}{GW231123cg_xo4a}{22.3}{GW231123cg_tphm}{22.3}{GW231123cg_seob}{22.3}}}
\newcommand{\networkoptimalsnrninetypercent}[1]{\IfEqCase{#1}{{GW231123cg_combined}{21.8}{GW231123cg_nrsur}{21.8}{GW231123cg_xphm}{21.6}{GW231123cg_xo4a}{21.9}{GW231123cg_tphm}{21.9}{GW231123cg_seob}{21.9}}}
\newcommand{\phaseminus}[1]{\IfEqCase{#1}{{GW231123cg_combined}{3.03}{GW231123cg_nrsur}{2.28}{GW231123cg_xphm}{0.66}{GW231123cg_xo4a}{1.03}{GW231123cg_tphm}{3.12}{GW231123cg_seob}{2.61}}}
\newcommand{\phasemed}[1]{\IfEqCase{#1}{{GW231123cg_combined}{3.94}{GW231123cg_nrsur}{3.38}{GW231123cg_xphm}{5.04}{GW231123cg_xo4a}{1.63}{GW231123cg_tphm}{4.31}{GW231123cg_seob}{3.87}}}
\newcommand{\phaseplus}[1]{\IfEqCase{#1}{{GW231123cg_combined}{1.72}{GW231123cg_nrsur}{2.11}{GW231123cg_xphm}{0.72}{GW231123cg_xo4a}{2.45}{GW231123cg_tphm}{1.48}{GW231123cg_seob}{1.82}}}
\newcommand{\phasezeropercent}[1]{\IfEqCase{#1}{{GW231123cg_combined}{0.00}{GW231123cg_nrsur}{0.00}{GW231123cg_xphm}{0.00}{GW231123cg_xo4a}{0.00}{GW231123cg_tphm}{0.00}{GW231123cg_seob}{0.00}}}
\newcommand{\phaseonepercent}[1]{\IfEqCase{#1}{{GW231123cg_combined}{0.22}{GW231123cg_nrsur}{0.24}{GW231123cg_xphm}{3.93}{GW231123cg_xo4a}{0.19}{GW231123cg_tphm}{0.16}{GW231123cg_seob}{0.19}}}
\newcommand{\phaseninetyninepercent}[1]{\IfEqCase{#1}{{GW231123cg_combined}{6.10}{GW231123cg_nrsur}{6.04}{GW231123cg_xphm}{6.07}{GW231123cg_xo4a}{6.06}{GW231123cg_tphm}{6.15}{GW231123cg_seob}{6.11}}}
\newcommand{\phasefivepercent}[1]{\IfEqCase{#1}{{GW231123cg_combined}{0.91}{GW231123cg_nrsur}{1.10}{GW231123cg_xphm}{4.38}{GW231123cg_xo4a}{0.59}{GW231123cg_tphm}{1.19}{GW231123cg_seob}{1.26}}}
\newcommand{\phaseninetyfivepercent}[1]{\IfEqCase{#1}{{GW231123cg_combined}{5.66}{GW231123cg_nrsur}{5.49}{GW231123cg_xphm}{5.76}{GW231123cg_xo4a}{4.08}{GW231123cg_tphm}{5.79}{GW231123cg_seob}{5.69}}}
\newcommand{\phaseninetypercent}[1]{\IfEqCase{#1}{{GW231123cg_combined}{5.38}{GW231123cg_nrsur}{5.09}{GW231123cg_xphm}{5.56}{GW231123cg_xo4a}{2.78}{GW231123cg_tphm}{5.50}{GW231123cg_seob}{5.32}}}
\newcommand{\phioneminus}[1]{\IfEqCase{#1}{{GW231123cg_combined}{2.33}{GW231123cg_nrsur}{2.33}{GW231123cg_xphm}{4.02}{GW231123cg_xo4a}{0.80}{GW231123cg_tphm}{2.45}{GW231123cg_seob}{2.57}}}
\newcommand{\phionemed}[1]{\IfEqCase{#1}{{GW231123cg_combined}{2.56}{GW231123cg_nrsur}{2.59}{GW231123cg_xphm}{4.24}{GW231123cg_xo4a}{0.94}{GW231123cg_tphm}{3.00}{GW231123cg_seob}{2.94}}}
\newcommand{\phioneplus}[1]{\IfEqCase{#1}{{GW231123cg_combined}{3.46}{GW231123cg_nrsur}{3.38}{GW231123cg_xphm}{1.88}{GW231123cg_xo4a}{5.13}{GW231123cg_tphm}{2.85}{GW231123cg_seob}{3.01}}}
\newcommand{\phionezeropercent}[1]{\IfEqCase{#1}{{GW231123cg_combined}{0.00}{GW231123cg_nrsur}{0.00}{GW231123cg_xphm}{0.00}{GW231123cg_xo4a}{0.00}{GW231123cg_tphm}{0.00}{GW231123cg_seob}{0.00}}}
\newcommand{\phioneonepercent}[1]{\IfEqCase{#1}{{GW231123cg_combined}{0.05}{GW231123cg_nrsur}{0.05}{GW231123cg_xphm}{0.03}{GW231123cg_xo4a}{0.03}{GW231123cg_tphm}{0.09}{GW231123cg_seob}{0.08}}}
\newcommand{\phioneninetyninepercent}[1]{\IfEqCase{#1}{{GW231123cg_combined}{6.24}{GW231123cg_nrsur}{6.23}{GW231123cg_xphm}{6.25}{GW231123cg_xo4a}{6.25}{GW231123cg_tphm}{6.18}{GW231123cg_seob}{6.22}}}
\newcommand{\phionefivepercent}[1]{\IfEqCase{#1}{{GW231123cg_combined}{0.23}{GW231123cg_nrsur}{0.25}{GW231123cg_xphm}{0.22}{GW231123cg_xo4a}{0.14}{GW231123cg_tphm}{0.55}{GW231123cg_seob}{0.37}}}
\newcommand{\phioneninetyfivepercent}[1]{\IfEqCase{#1}{{GW231123cg_combined}{6.02}{GW231123cg_nrsur}{5.97}{GW231123cg_xphm}{6.12}{GW231123cg_xo4a}{6.07}{GW231123cg_tphm}{5.85}{GW231123cg_seob}{5.95}}}
\newcommand{\phioneninetypercent}[1]{\IfEqCase{#1}{{GW231123cg_combined}{5.73}{GW231123cg_nrsur}{5.51}{GW231123cg_xphm}{5.97}{GW231123cg_xo4a}{4.49}{GW231123cg_tphm}{5.53}{GW231123cg_seob}{5.62}}}
\newcommand{\phionetwominus}[1]{\IfEqCase{#1}{{GW231123cg_combined}{3.77}{GW231123cg_nrsur}{1.14}{GW231123cg_xphm}{4.06}{GW231123cg_xo4a}{2.80}{GW231123cg_tphm}{5.08}{GW231123cg_seob}{4.60}}}
\newcommand{\phionetwomed}[1]{\IfEqCase{#1}{{GW231123cg_combined}{3.92}{GW231123cg_nrsur}{1.27}{GW231123cg_xphm}{4.36}{GW231123cg_xo4a}{3.09}{GW231123cg_tphm}{5.18}{GW231123cg_seob}{4.71}}}
\newcommand{\phionetwoplus}[1]{\IfEqCase{#1}{{GW231123cg_combined}{2.22}{GW231123cg_nrsur}{4.88}{GW231123cg_xphm}{1.66}{GW231123cg_xo4a}{2.89}{GW231123cg_tphm}{1.01}{GW231123cg_seob}{1.46}}}
\newcommand{\phionetwozeropercent}[1]{\IfEqCase{#1}{{GW231123cg_combined}{0.00}{GW231123cg_nrsur}{0.00}{GW231123cg_xphm}{0.00}{GW231123cg_xo4a}{0.00}{GW231123cg_tphm}{0.00}{GW231123cg_seob}{0.00}}}
\newcommand{\phionetwoonepercent}[1]{\IfEqCase{#1}{{GW231123cg_combined}{0.03}{GW231123cg_nrsur}{0.03}{GW231123cg_xphm}{0.06}{GW231123cg_xo4a}{0.05}{GW231123cg_tphm}{0.02}{GW231123cg_seob}{0.02}}}
\newcommand{\phionetwoninetyninepercent}[1]{\IfEqCase{#1}{{GW231123cg_combined}{6.25}{GW231123cg_nrsur}{6.25}{GW231123cg_xphm}{6.23}{GW231123cg_xo4a}{6.23}{GW231123cg_tphm}{6.27}{GW231123cg_seob}{6.26}}}
\newcommand{\phionetwofivepercent}[1]{\IfEqCase{#1}{{GW231123cg_combined}{0.15}{GW231123cg_nrsur}{0.13}{GW231123cg_xphm}{0.30}{GW231123cg_xo4a}{0.29}{GW231123cg_tphm}{0.10}{GW231123cg_seob}{0.11}}}
\newcommand{\phionetwoninetyfivepercent}[1]{\IfEqCase{#1}{{GW231123cg_combined}{6.14}{GW231123cg_nrsur}{6.14}{GW231123cg_xphm}{6.02}{GW231123cg_xo4a}{5.98}{GW231123cg_tphm}{6.19}{GW231123cg_seob}{6.17}}}
\newcommand{\phionetwoninetypercent}[1]{\IfEqCase{#1}{{GW231123cg_combined}{5.98}{GW231123cg_nrsur}{5.96}{GW231123cg_xphm}{5.80}{GW231123cg_xo4a}{5.68}{GW231123cg_tphm}{6.09}{GW231123cg_seob}{6.07}}}
\newcommand{\phitwominus}[1]{\IfEqCase{#1}{{GW231123cg_combined}{2.60}{GW231123cg_nrsur}{1.94}{GW231123cg_xphm}{3.17}{GW231123cg_xo4a}{2.75}{GW231123cg_tphm}{2.27}{GW231123cg_seob}{2.50}}}
\newcommand{\phitwomed}[1]{\IfEqCase{#1}{{GW231123cg_combined}{2.97}{GW231123cg_nrsur}{2.28}{GW231123cg_xphm}{3.54}{GW231123cg_xo4a}{3.02}{GW231123cg_tphm}{2.90}{GW231123cg_seob}{2.88}}}
\newcommand{\phitwoplus}[1]{\IfEqCase{#1}{{GW231123cg_combined}{2.91}{GW231123cg_nrsur}{3.53}{GW231123cg_xphm}{2.38}{GW231123cg_xo4a}{2.96}{GW231123cg_tphm}{2.87}{GW231123cg_seob}{3.00}}}
\newcommand{\phitwozeropercent}[1]{\IfEqCase{#1}{{GW231123cg_combined}{0.00}{GW231123cg_nrsur}{0.00}{GW231123cg_xphm}{0.00}{GW231123cg_xo4a}{0.00}{GW231123cg_tphm}{0.00}{GW231123cg_seob}{0.00}}}
\newcommand{\phitwoonepercent}[1]{\IfEqCase{#1}{{GW231123cg_combined}{0.08}{GW231123cg_nrsur}{0.08}{GW231123cg_xphm}{0.07}{GW231123cg_xo4a}{0.06}{GW231123cg_tphm}{0.15}{GW231123cg_seob}{0.09}}}
\newcommand{\phitwoninetyninepercent}[1]{\IfEqCase{#1}{{GW231123cg_combined}{6.20}{GW231123cg_nrsur}{6.19}{GW231123cg_xphm}{6.20}{GW231123cg_xo4a}{6.22}{GW231123cg_tphm}{6.15}{GW231123cg_seob}{6.19}}}
\newcommand{\phitwofivepercent}[1]{\IfEqCase{#1}{{GW231123cg_combined}{0.37}{GW231123cg_nrsur}{0.34}{GW231123cg_xphm}{0.36}{GW231123cg_xo4a}{0.28}{GW231123cg_tphm}{0.63}{GW231123cg_seob}{0.38}}}
\newcommand{\phitwoninetyfivepercent}[1]{\IfEqCase{#1}{{GW231123cg_combined}{5.89}{GW231123cg_nrsur}{5.81}{GW231123cg_xphm}{5.92}{GW231123cg_xo4a}{5.99}{GW231123cg_tphm}{5.77}{GW231123cg_seob}{5.87}}}
\newcommand{\phitwoninetypercent}[1]{\IfEqCase{#1}{{GW231123cg_combined}{5.53}{GW231123cg_nrsur}{5.34}{GW231123cg_xphm}{5.61}{GW231123cg_xo4a}{5.69}{GW231123cg_tphm}{5.44}{GW231123cg_seob}{5.50}}}
\newcommand{\phijlminus}[1]{\IfEqCase{#1}{{GW231123cg_combined}{2.28}{GW231123cg_nrsur}{1.01}{GW231123cg_xphm}{1.29}{GW231123cg_xo4a}{1.91}{GW231123cg_tphm}{3.02}{GW231123cg_seob}{0.89}}}
\newcommand{\phijlmed}[1]{\IfEqCase{#1}{{GW231123cg_combined}{2.52}{GW231123cg_nrsur}{1.99}{GW231123cg_xphm}{1.66}{GW231123cg_xo4a}{1.99}{GW231123cg_tphm}{3.20}{GW231123cg_seob}{2.72}}}
\newcommand{\phijlplus}[1]{\IfEqCase{#1}{{GW231123cg_combined}{3.50}{GW231123cg_nrsur}{3.31}{GW231123cg_xphm}{3.38}{GW231123cg_xo4a}{4.21}{GW231123cg_tphm}{2.86}{GW231123cg_seob}{3.15}}}
\newcommand{\phijlzeropercent}[1]{\IfEqCase{#1}{{GW231123cg_combined}{0.00}{GW231123cg_nrsur}{0.01}{GW231123cg_xphm}{0.00}{GW231123cg_xo4a}{0.00}{GW231123cg_tphm}{0.00}{GW231123cg_seob}{0.00}}}
\newcommand{\phijlonepercent}[1]{\IfEqCase{#1}{{GW231123cg_combined}{0.05}{GW231123cg_nrsur}{0.63}{GW231123cg_xphm}{0.11}{GW231123cg_xo4a}{0.02}{GW231123cg_tphm}{0.04}{GW231123cg_seob}{1.54}}}
\newcommand{\phijlninetyninepercent}[1]{\IfEqCase{#1}{{GW231123cg_combined}{6.23}{GW231123cg_nrsur}{5.75}{GW231123cg_xphm}{6.09}{GW231123cg_xo4a}{6.27}{GW231123cg_tphm}{6.25}{GW231123cg_seob}{6.10}}}
\newcommand{\phijlfivepercent}[1]{\IfEqCase{#1}{{GW231123cg_combined}{0.24}{GW231123cg_nrsur}{0.98}{GW231123cg_xphm}{0.37}{GW231123cg_xo4a}{0.08}{GW231123cg_tphm}{0.18}{GW231123cg_seob}{1.84}}}
\newcommand{\phijlninetyfivepercent}[1]{\IfEqCase{#1}{{GW231123cg_combined}{6.02}{GW231123cg_nrsur}{5.30}{GW231123cg_xphm}{5.03}{GW231123cg_xo4a}{6.20}{GW231123cg_tphm}{6.06}{GW231123cg_seob}{5.88}}}
\newcommand{\phijlninetypercent}[1]{\IfEqCase{#1}{{GW231123cg_combined}{5.73}{GW231123cg_nrsur}{5.07}{GW231123cg_xphm}{4.69}{GW231123cg_xo4a}{6.12}{GW231123cg_tphm}{5.69}{GW231123cg_seob}{5.73}}}
\newcommand{\psiminus}[1]{\IfEqCase{#1}{{GW231123cg_combined}{1.67}{GW231123cg_nrsur}{1.61}{GW231123cg_xphm}{2.59}{GW231123cg_xo4a}{1.08}{GW231123cg_tphm}{1.22}{GW231123cg_seob}{1.07}}}
\newcommand{\psimed}[1]{\IfEqCase{#1}{{GW231123cg_combined}{1.78}{GW231123cg_nrsur}{1.93}{GW231123cg_xphm}{2.65}{GW231123cg_xo4a}{1.35}{GW231123cg_tphm}{1.27}{GW231123cg_seob}{1.60}}}
\newcommand{\psiplus}[1]{\IfEqCase{#1}{{GW231123cg_combined}{1.24}{GW231123cg_nrsur}{0.85}{GW231123cg_xphm}{0.44}{GW231123cg_xo4a}{1.49}{GW231123cg_tphm}{1.81}{GW231123cg_seob}{0.92}}}
\newcommand{\psizeropercent}[1]{\IfEqCase{#1}{{GW231123cg_combined}{0.00}{GW231123cg_nrsur}{0.00}{GW231123cg_xphm}{0.00}{GW231123cg_xo4a}{0.00}{GW231123cg_tphm}{0.00}{GW231123cg_seob}{0.00}}}
\newcommand{\psionepercent}[1]{\IfEqCase{#1}{{GW231123cg_combined}{0.02}{GW231123cg_nrsur}{0.10}{GW231123cg_xphm}{0.01}{GW231123cg_xo4a}{0.06}{GW231123cg_tphm}{0.01}{GW231123cg_seob}{0.22}}}
\newcommand{\psininetyninepercent}[1]{\IfEqCase{#1}{{GW231123cg_combined}{3.12}{GW231123cg_nrsur}{3.06}{GW231123cg_xphm}{3.13}{GW231123cg_xo4a}{3.08}{GW231123cg_tphm}{3.13}{GW231123cg_seob}{2.88}}}
\newcommand{\psifivepercent}[1]{\IfEqCase{#1}{{GW231123cg_combined}{0.11}{GW231123cg_nrsur}{0.31}{GW231123cg_xphm}{0.06}{GW231123cg_xo4a}{0.27}{GW231123cg_tphm}{0.05}{GW231123cg_seob}{0.52}}}
\newcommand{\psininetyfivepercent}[1]{\IfEqCase{#1}{{GW231123cg_combined}{3.01}{GW231123cg_nrsur}{2.78}{GW231123cg_xphm}{3.09}{GW231123cg_xo4a}{2.84}{GW231123cg_tphm}{3.08}{GW231123cg_seob}{2.52}}}
\newcommand{\psininetypercent}[1]{\IfEqCase{#1}{{GW231123cg_combined}{2.86}{GW231123cg_nrsur}{2.58}{GW231123cg_xphm}{3.03}{GW231123cg_xo4a}{2.53}{GW231123cg_tphm}{3.02}{GW231123cg_seob}{2.33}}}
\newcommand{\raminus}[1]{\IfEqCase{#1}{{GW231123cg_combined}{0.61372}{GW231123cg_nrsur}{0.64287}{GW231123cg_xphm}{0.27833}{GW231123cg_xo4a}{0.87899}{GW231123cg_tphm}{0.48998}{GW231123cg_seob}{0.64301}}}
\newcommand{\ramed}[1]{\IfEqCase{#1}{{GW231123cg_combined}{3.38809}{GW231123cg_nrsur}{3.40656}{GW231123cg_xphm}{3.32536}{GW231123cg_xo4a}{3.82604}{GW231123cg_tphm}{3.22792}{GW231123cg_seob}{3.37055}}}
\newcommand{\raplus}[1]{\IfEqCase{#1}{{GW231123cg_combined}{1.28417}{GW231123cg_nrsur}{1.19932}{GW231123cg_xphm}{0.26207}{GW231123cg_xo4a}{1.26189}{GW231123cg_tphm}{1.36760}{GW231123cg_seob}{0.94739}}}
\newcommand{\razeropercent}[1]{\IfEqCase{#1}{{GW231123cg_combined}{0.04569}{GW231123cg_nrsur}{0.47435}{GW231123cg_xphm}{0.08340}{GW231123cg_xo4a}{2.61325}{GW231123cg_tphm}{2.34311}{GW231123cg_seob}{0.04569}}}
\newcommand{\raonepercent}[1]{\IfEqCase{#1}{{GW231123cg_combined}{2.66448}{GW231123cg_nrsur}{2.64655}{GW231123cg_xphm}{2.71328}{GW231123cg_xo4a}{2.75541}{GW231123cg_tphm}{2.65355}{GW231123cg_seob}{2.63413}}}
\newcommand{\raninetyninepercent}[1]{\IfEqCase{#1}{{GW231123cg_combined}{5.24702}{GW231123cg_nrsur}{5.28267}{GW231123cg_xphm}{5.42530}{GW231123cg_xo4a}{5.21853}{GW231123cg_tphm}{5.26580}{GW231123cg_seob}{5.19422}}}
\newcommand{\rafivepercent}[1]{\IfEqCase{#1}{{GW231123cg_combined}{2.77437}{GW231123cg_nrsur}{2.76369}{GW231123cg_xphm}{3.04702}{GW231123cg_xo4a}{2.94705}{GW231123cg_tphm}{2.73794}{GW231123cg_seob}{2.72754}}}
\newcommand{\raninetyfivepercent}[1]{\IfEqCase{#1}{{GW231123cg_combined}{4.67226}{GW231123cg_nrsur}{4.60588}{GW231123cg_xphm}{3.58743}{GW231123cg_xo4a}{5.08793}{GW231123cg_tphm}{4.59552}{GW231123cg_seob}{4.31794}}}
\newcommand{\raninetypercent}[1]{\IfEqCase{#1}{{GW231123cg_combined}{4.22724}{GW231123cg_nrsur}{4.18941}{GW231123cg_xphm}{3.52066}{GW231123cg_xo4a}{4.81750}{GW231123cg_tphm}{4.15176}{GW231123cg_seob}{4.08598}}}
\newcommand{\radiatedenergyminus}[1]{\IfEqCase{#1}{{GW231123cg_combined}{6.2}{GW231123cg_nrsur}{2.8}{GW231123cg_xphm}{3.1}{GW231123cg_xo4a}{2.2}{GW231123cg_tphm}{2.6}{GW231123cg_seob}{2.8}}}
\newcommand{\radiatedenergymed}[1]{\IfEqCase{#1}{{GW231123cg_combined}{13.4}{GW231123cg_nrsur}{14.6}{GW231123cg_xphm}{10.3}{GW231123cg_xo4a}{8.1}{GW231123cg_tphm}{15.2}{GW231123cg_seob}{15.1}}}
\newcommand{\radiatedenergyplus}[1]{\IfEqCase{#1}{{GW231123cg_combined}{4.3}{GW231123cg_nrsur}{3.0}{GW231123cg_xphm}{3.1}{GW231123cg_xo4a}{2.1}{GW231123cg_tphm}{3.1}{GW231123cg_seob}{3.7}}}
\newcommand{\radiatedenergyzeropercent}[1]{\IfEqCase{#1}{{GW231123cg_combined}{3.5}{GW231123cg_nrsur}{8.1}{GW231123cg_xphm}{3.5}{GW231123cg_xo4a}{4.4}{GW231123cg_tphm}{8.2}{GW231123cg_seob}{8.3}}}
\newcommand{\radiatedenergyonepercent}[1]{\IfEqCase{#1}{{GW231123cg_combined}{5.8}{GW231123cg_nrsur}{10.6}{GW231123cg_xphm}{6.1}{GW231123cg_xo4a}{5.2}{GW231123cg_tphm}{11.5}{GW231123cg_seob}{11.2}}}
\newcommand{\radiatedenergyninetyninepercent}[1]{\IfEqCase{#1}{{GW231123cg_combined}{19.4}{GW231123cg_nrsur}{18.9}{GW231123cg_xphm}{15.5}{GW231123cg_xo4a}{11.1}{GW231123cg_tphm}{19.8}{GW231123cg_seob}{20.7}}}
\newcommand{\radiatedenergyfivepercent}[1]{\IfEqCase{#1}{{GW231123cg_combined}{7.1}{GW231123cg_nrsur}{11.8}{GW231123cg_xphm}{7.2}{GW231123cg_xo4a}{5.9}{GW231123cg_tphm}{12.6}{GW231123cg_seob}{12.2}}}
\newcommand{\radiatedenergyninetyfivepercent}[1]{\IfEqCase{#1}{{GW231123cg_combined}{17.6}{GW231123cg_nrsur}{17.5}{GW231123cg_xphm}{13.4}{GW231123cg_xo4a}{10.2}{GW231123cg_tphm}{18.3}{GW231123cg_seob}{18.8}}}
\newcommand{\radiatedenergyninetypercent}[1]{\IfEqCase{#1}{{GW231123cg_combined}{16.8}{GW231123cg_nrsur}{16.8}{GW231123cg_xphm}{12.5}{GW231123cg_xo4a}{9.7}{GW231123cg_tphm}{17.5}{GW231123cg_seob}{17.9}}}
\newcommand{\redshiftminus}[1]{\IfEqCase{#1}{{GW231123cg_combined}{0.25}{GW231123cg_nrsur}{0.18}{GW231123cg_xphm}{0.06}{GW231123cg_xo4a}{0.20}{GW231123cg_tphm}{0.16}{GW231123cg_seob}{0.16}}}
\newcommand{\redshiftmed}[1]{\IfEqCase{#1}{{GW231123cg_combined}{0.40}{GW231123cg_nrsur}{0.34}{GW231123cg_xphm}{0.18}{GW231123cg_xo4a}{0.58}{GW231123cg_tphm}{0.46}{GW231123cg_seob}{0.40}}}
\newcommand{\redshiftplus}[1]{\IfEqCase{#1}{{GW231123cg_combined}{0.27}{GW231123cg_nrsur}{0.24}{GW231123cg_xphm}{0.07}{GW231123cg_xo4a}{0.17}{GW231123cg_tphm}{0.16}{GW231123cg_seob}{0.20}}}
\newcommand{\redshiftzeropercent}[1]{\IfEqCase{#1}{{GW231123cg_combined}{0.02}{GW231123cg_nrsur}{0.05}{GW231123cg_xphm}{0.02}{GW231123cg_xo4a}{0.15}{GW231123cg_tphm}{0.11}{GW231123cg_seob}{0.08}}}
\newcommand{\redshiftonepercent}[1]{\IfEqCase{#1}{{GW231123cg_combined}{0.11}{GW231123cg_nrsur}{0.13}{GW231123cg_xphm}{0.07}{GW231123cg_xo4a}{0.32}{GW231123cg_tphm}{0.23}{GW231123cg_seob}{0.18}}}
\newcommand{\redshiftninetyninepercent}[1]{\IfEqCase{#1}{{GW231123cg_combined}{0.76}{GW231123cg_nrsur}{0.77}{GW231123cg_xphm}{0.28}{GW231123cg_xo4a}{0.81}{GW231123cg_tphm}{0.69}{GW231123cg_seob}{0.70}}}
\newcommand{\redshiftfivepercent}[1]{\IfEqCase{#1}{{GW231123cg_combined}{0.15}{GW231123cg_nrsur}{0.17}{GW231123cg_xphm}{0.12}{GW231123cg_xo4a}{0.38}{GW231123cg_tphm}{0.30}{GW231123cg_seob}{0.25}}}
\newcommand{\redshiftninetyfivepercent}[1]{\IfEqCase{#1}{{GW231123cg_combined}{0.67}{GW231123cg_nrsur}{0.59}{GW231123cg_xphm}{0.25}{GW231123cg_xo4a}{0.75}{GW231123cg_tphm}{0.62}{GW231123cg_seob}{0.60}}}
\newcommand{\redshiftninetypercent}[1]{\IfEqCase{#1}{{GW231123cg_combined}{0.62}{GW231123cg_nrsur}{0.52}{GW231123cg_xphm}{0.23}{GW231123cg_xo4a}{0.71}{GW231123cg_tphm}{0.59}{GW231123cg_seob}{0.55}}}
\newcommand{\spinonexminus}[1]{\IfEqCase{#1}{{GW231123cg_combined}{0.91}{GW231123cg_nrsur}{0.80}{GW231123cg_xphm}{1.04}{GW231123cg_xo4a}{1.21}{GW231123cg_tphm}{0.68}{GW231123cg_seob}{0.76}}}
\newcommand{\spinonexmed}[1]{\IfEqCase{#1}{{GW231123cg_combined}{0.17}{GW231123cg_nrsur}{0.07}{GW231123cg_xphm}{0.25}{GW231123cg_xo4a}{0.61}{GW231123cg_tphm}{-0.06}{GW231123cg_seob}{0.03}}}
\newcommand{\spinonexplus}[1]{\IfEqCase{#1}{{GW231123cg_combined}{0.65}{GW231123cg_nrsur}{0.69}{GW231123cg_xphm}{0.60}{GW231123cg_xo4a}{0.25}{GW231123cg_tphm}{0.77}{GW231123cg_seob}{0.72}}}
\newcommand{\spinonexzeropercent}[1]{\IfEqCase{#1}{{GW231123cg_combined}{-0.99}{GW231123cg_nrsur}{-0.98}{GW231123cg_xphm}{-0.99}{GW231123cg_xo4a}{-0.98}{GW231123cg_tphm}{-0.96}{GW231123cg_seob}{-0.99}}}
\newcommand{\spinonexonepercent}[1]{\IfEqCase{#1}{{GW231123cg_combined}{-0.87}{GW231123cg_nrsur}{-0.88}{GW231123cg_xphm}{-0.90}{GW231123cg_xo4a}{-0.84}{GW231123cg_tphm}{-0.85}{GW231123cg_seob}{-0.87}}}
\newcommand{\spinonexninetyninepercent}[1]{\IfEqCase{#1}{{GW231123cg_combined}{0.90}{GW231123cg_nrsur}{0.87}{GW231123cg_xphm}{0.93}{GW231123cg_xo4a}{0.92}{GW231123cg_tphm}{0.84}{GW231123cg_seob}{0.88}}}
\newcommand{\spinonexfivepercent}[1]{\IfEqCase{#1}{{GW231123cg_combined}{-0.74}{GW231123cg_nrsur}{-0.73}{GW231123cg_xphm}{-0.80}{GW231123cg_xo4a}{-0.60}{GW231123cg_tphm}{-0.74}{GW231123cg_seob}{-0.73}}}
\newcommand{\spinonexninetyfivepercent}[1]{\IfEqCase{#1}{{GW231123cg_combined}{0.82}{GW231123cg_nrsur}{0.76}{GW231123cg_xphm}{0.84}{GW231123cg_xo4a}{0.86}{GW231123cg_tphm}{0.71}{GW231123cg_seob}{0.75}}}
\newcommand{\spinonexninetypercent}[1]{\IfEqCase{#1}{{GW231123cg_combined}{0.77}{GW231123cg_nrsur}{0.67}{GW231123cg_xphm}{0.78}{GW231123cg_xo4a}{0.83}{GW231123cg_tphm}{0.61}{GW231123cg_seob}{0.66}}}
\newcommand{\spinoneyminus}[1]{\IfEqCase{#1}{{GW231123cg_combined}{0.88}{GW231123cg_nrsur}{0.78}{GW231123cg_xphm}{0.59}{GW231123cg_xo4a}{0.90}{GW231123cg_tphm}{0.85}{GW231123cg_seob}{0.82}}}
\newcommand{\spinoneymed}[1]{\IfEqCase{#1}{{GW231123cg_combined}{0.14}{GW231123cg_nrsur}{0.10}{GW231123cg_xphm}{-0.14}{GW231123cg_xo4a}{0.47}{GW231123cg_tphm}{0.05}{GW231123cg_seob}{0.06}}}
\newcommand{\spinoneyplus}[1]{\IfEqCase{#1}{{GW231123cg_combined}{0.62}{GW231123cg_nrsur}{0.59}{GW231123cg_xphm}{0.82}{GW231123cg_xo4a}{0.31}{GW231123cg_tphm}{0.75}{GW231123cg_seob}{0.71}}}
\newcommand{\spinoneyzeropercent}[1]{\IfEqCase{#1}{{GW231123cg_combined}{-0.98}{GW231123cg_nrsur}{-0.98}{GW231123cg_xphm}{-0.97}{GW231123cg_xo4a}{-0.97}{GW231123cg_tphm}{-0.97}{GW231123cg_seob}{-0.98}}}
\newcommand{\spinoneyonepercent}[1]{\IfEqCase{#1}{{GW231123cg_combined}{-0.87}{GW231123cg_nrsur}{-0.85}{GW231123cg_xphm}{-0.85}{GW231123cg_xo4a}{-0.81}{GW231123cg_tphm}{-0.89}{GW231123cg_seob}{-0.88}}}
\newcommand{\spinoneyninetyninepercent}[1]{\IfEqCase{#1}{{GW231123cg_combined}{0.86}{GW231123cg_nrsur}{0.83}{GW231123cg_xphm}{0.82}{GW231123cg_xo4a}{0.86}{GW231123cg_tphm}{0.89}{GW231123cg_seob}{0.87}}}
\newcommand{\spinoneyfivepercent}[1]{\IfEqCase{#1}{{GW231123cg_combined}{-0.74}{GW231123cg_nrsur}{-0.68}{GW231123cg_xphm}{-0.73}{GW231123cg_xo4a}{-0.43}{GW231123cg_tphm}{-0.79}{GW231123cg_seob}{-0.76}}}
\newcommand{\spinoneyninetyfivepercent}[1]{\IfEqCase{#1}{{GW231123cg_combined}{0.77}{GW231123cg_nrsur}{0.69}{GW231123cg_xphm}{0.68}{GW231123cg_xo4a}{0.79}{GW231123cg_tphm}{0.80}{GW231123cg_seob}{0.77}}}
\newcommand{\spinoneyninetypercent}[1]{\IfEqCase{#1}{{GW231123cg_combined}{0.71}{GW231123cg_nrsur}{0.60}{GW231123cg_xphm}{0.57}{GW231123cg_xo4a}{0.75}{GW231123cg_tphm}{0.74}{GW231123cg_seob}{0.70}}}
\newcommand{\spinonezminus}[1]{\IfEqCase{#1}{{GW231123cg_combined}{0.75}{GW231123cg_nrsur}{0.72}{GW231123cg_xphm}{0.30}{GW231123cg_xo4a}{0.30}{GW231123cg_tphm}{0.28}{GW231123cg_seob}{0.37}}}
\newcommand{\spinonezmed}[1]{\IfEqCase{#1}{{GW231123cg_combined}{0.43}{GW231123cg_nrsur}{0.58}{GW231123cg_xphm}{-0.19}{GW231123cg_xo4a}{0.40}{GW231123cg_tphm}{0.51}{GW231123cg_seob}{0.53}}}
\newcommand{\spinonezplus}[1]{\IfEqCase{#1}{{GW231123cg_combined}{0.36}{GW231123cg_nrsur}{0.31}{GW231123cg_xphm}{0.30}{GW231123cg_xo4a}{0.21}{GW231123cg_tphm}{0.24}{GW231123cg_seob}{0.27}}}
\newcommand{\spinonezzeropercent}[1]{\IfEqCase{#1}{{GW231123cg_combined}{-0.87}{GW231123cg_nrsur}{-0.76}{GW231123cg_xphm}{-0.87}{GW231123cg_xo4a}{-0.74}{GW231123cg_tphm}{-0.57}{GW231123cg_seob}{-0.48}}}
\newcommand{\spinonezonepercent}[1]{\IfEqCase{#1}{{GW231123cg_combined}{-0.50}{GW231123cg_nrsur}{-0.38}{GW231123cg_xphm}{-0.64}{GW231123cg_xo4a}{-0.06}{GW231123cg_tphm}{0.07}{GW231123cg_seob}{-0.06}}}
\newcommand{\spinonezninetyninepercent}[1]{\IfEqCase{#1}{{GW231123cg_combined}{0.90}{GW231123cg_nrsur}{0.95}{GW231123cg_xphm}{0.56}{GW231123cg_xo4a}{0.66}{GW231123cg_tphm}{0.83}{GW231123cg_seob}{0.88}}}
\newcommand{\spinonezfivepercent}[1]{\IfEqCase{#1}{{GW231123cg_combined}{-0.32}{GW231123cg_nrsur}{-0.14}{GW231123cg_xphm}{-0.49}{GW231123cg_xo4a}{0.09}{GW231123cg_tphm}{0.23}{GW231123cg_seob}{0.16}}}
\newcommand{\spinonezninetyfivepercent}[1]{\IfEqCase{#1}{{GW231123cg_combined}{0.79}{GW231123cg_nrsur}{0.89}{GW231123cg_xphm}{0.11}{GW231123cg_xo4a}{0.61}{GW231123cg_tphm}{0.75}{GW231123cg_seob}{0.80}}}
\newcommand{\spinonezninetypercent}[1]{\IfEqCase{#1}{{GW231123cg_combined}{0.72}{GW231123cg_nrsur}{0.84}{GW231123cg_xphm}{0.03}{GW231123cg_xo4a}{0.58}{GW231123cg_tphm}{0.69}{GW231123cg_seob}{0.75}}}
\newcommand{\spintwoxminus}[1]{\IfEqCase{#1}{{GW231123cg_combined}{0.71}{GW231123cg_nrsur}{0.87}{GW231123cg_xphm}{0.58}{GW231123cg_xo4a}{0.55}{GW231123cg_tphm}{0.70}{GW231123cg_seob}{0.73}}}
\newcommand{\spintwoxmed}[1]{\IfEqCase{#1}{{GW231123cg_combined}{0.01}{GW231123cg_nrsur}{0.05}{GW231123cg_xphm}{-0.01}{GW231123cg_xo4a}{0.01}{GW231123cg_tphm}{-0.04}{GW231123cg_seob}{0.03}}}
\newcommand{\spintwoxplus}[1]{\IfEqCase{#1}{{GW231123cg_combined}{0.72}{GW231123cg_nrsur}{0.78}{GW231123cg_xphm}{0.56}{GW231123cg_xo4a}{0.59}{GW231123cg_tphm}{0.76}{GW231123cg_seob}{0.72}}}
\newcommand{\spintwoxzeropercent}[1]{\IfEqCase{#1}{{GW231123cg_combined}{-0.99}{GW231123cg_nrsur}{-0.99}{GW231123cg_xphm}{-0.97}{GW231123cg_xo4a}{-0.98}{GW231123cg_tphm}{-0.98}{GW231123cg_seob}{-0.98}}}
\newcommand{\spintwoxonepercent}[1]{\IfEqCase{#1}{{GW231123cg_combined}{-0.88}{GW231123cg_nrsur}{-0.92}{GW231123cg_xphm}{-0.79}{GW231123cg_xo4a}{-0.78}{GW231123cg_tphm}{-0.88}{GW231123cg_seob}{-0.87}}}
\newcommand{\spintwoxninetyninepercent}[1]{\IfEqCase{#1}{{GW231123cg_combined}{0.88}{GW231123cg_nrsur}{0.92}{GW231123cg_xphm}{0.76}{GW231123cg_xo4a}{0.79}{GW231123cg_tphm}{0.87}{GW231123cg_seob}{0.89}}}
\newcommand{\spintwoxfivepercent}[1]{\IfEqCase{#1}{{GW231123cg_combined}{-0.71}{GW231123cg_nrsur}{-0.82}{GW231123cg_xphm}{-0.59}{GW231123cg_xo4a}{-0.54}{GW231123cg_tphm}{-0.74}{GW231123cg_seob}{-0.70}}}
\newcommand{\spintwoxninetyfivepercent}[1]{\IfEqCase{#1}{{GW231123cg_combined}{0.72}{GW231123cg_nrsur}{0.83}{GW231123cg_xphm}{0.55}{GW231123cg_xo4a}{0.60}{GW231123cg_tphm}{0.72}{GW231123cg_seob}{0.75}}}
\newcommand{\spintwoxninetypercent}[1]{\IfEqCase{#1}{{GW231123cg_combined}{0.59}{GW231123cg_nrsur}{0.74}{GW231123cg_xphm}{0.42}{GW231123cg_xo4a}{0.46}{GW231123cg_tphm}{0.59}{GW231123cg_seob}{0.62}}}
\newcommand{\spintwoyminus}[1]{\IfEqCase{#1}{{GW231123cg_combined}{0.81}{GW231123cg_nrsur}{1.16}{GW231123cg_xphm}{0.63}{GW231123cg_xo4a}{0.57}{GW231123cg_tphm}{0.91}{GW231123cg_seob}{0.82}}}
\newcommand{\spintwoymed}[1]{\IfEqCase{#1}{{GW231123cg_combined}{0.02}{GW231123cg_nrsur}{0.29}{GW231123cg_xphm}{-0.04}{GW231123cg_xo4a}{0.00}{GW231123cg_tphm}{0.08}{GW231123cg_seob}{0.05}}}
\newcommand{\spintwoyplus}[1]{\IfEqCase{#1}{{GW231123cg_combined}{0.81}{GW231123cg_nrsur}{0.63}{GW231123cg_xphm}{0.63}{GW231123cg_xo4a}{0.59}{GW231123cg_tphm}{0.76}{GW231123cg_seob}{0.73}}}
\newcommand{\spintwoyzeropercent}[1]{\IfEqCase{#1}{{GW231123cg_combined}{-0.99}{GW231123cg_nrsur}{-0.99}{GW231123cg_xphm}{-0.98}{GW231123cg_xo4a}{-0.98}{GW231123cg_tphm}{-0.99}{GW231123cg_seob}{-0.99}}}
\newcommand{\spintwoyonepercent}[1]{\IfEqCase{#1}{{GW231123cg_combined}{-0.91}{GW231123cg_nrsur}{-0.95}{GW231123cg_xphm}{-0.84}{GW231123cg_xo4a}{-0.79}{GW231123cg_tphm}{-0.92}{GW231123cg_seob}{-0.88}}}
\newcommand{\spintwoyninetyninepercent}[1]{\IfEqCase{#1}{{GW231123cg_combined}{0.93}{GW231123cg_nrsur}{0.96}{GW231123cg_xphm}{0.79}{GW231123cg_xo4a}{0.82}{GW231123cg_tphm}{0.92}{GW231123cg_seob}{0.90}}}
\newcommand{\spintwoyfivepercent}[1]{\IfEqCase{#1}{{GW231123cg_combined}{-0.79}{GW231123cg_nrsur}{-0.87}{GW231123cg_xphm}{-0.67}{GW231123cg_xo4a}{-0.56}{GW231123cg_tphm}{-0.83}{GW231123cg_seob}{-0.77}}}
\newcommand{\spintwoyninetyfivepercent}[1]{\IfEqCase{#1}{{GW231123cg_combined}{0.82}{GW231123cg_nrsur}{0.91}{GW231123cg_xphm}{0.59}{GW231123cg_xo4a}{0.60}{GW231123cg_tphm}{0.83}{GW231123cg_seob}{0.78}}}
\newcommand{\spintwoyninetypercent}[1]{\IfEqCase{#1}{{GW231123cg_combined}{0.73}{GW231123cg_nrsur}{0.87}{GW231123cg_xphm}{0.44}{GW231123cg_xo4a}{0.46}{GW231123cg_tphm}{0.77}{GW231123cg_seob}{0.69}}}
\newcommand{\spintwozminus}[1]{\IfEqCase{#1}{{GW231123cg_combined}{0.50}{GW231123cg_nrsur}{0.40}{GW231123cg_xphm}{0.48}{GW231123cg_xo4a}{0.36}{GW231123cg_tphm}{0.37}{GW231123cg_seob}{0.44}}}
\newcommand{\spintwozmed}[1]{\IfEqCase{#1}{{GW231123cg_combined}{0.21}{GW231123cg_nrsur}{-0.06}{GW231123cg_xphm}{0.34}{GW231123cg_xo4a}{0.10}{GW231123cg_tphm}{0.33}{GW231123cg_seob}{0.32}}}
\newcommand{\spintwozplus}[1]{\IfEqCase{#1}{{GW231123cg_combined}{0.50}{GW231123cg_nrsur}{0.41}{GW231123cg_xphm}{0.47}{GW231123cg_xo4a}{0.55}{GW231123cg_tphm}{0.34}{GW231123cg_seob}{0.43}}}
\newcommand{\spintwozzeropercent}[1]{\IfEqCase{#1}{{GW231123cg_combined}{-0.94}{GW231123cg_nrsur}{-0.91}{GW231123cg_xphm}{-0.94}{GW231123cg_xo4a}{-0.80}{GW231123cg_tphm}{-0.87}{GW231123cg_seob}{-0.72}}}
\newcommand{\spintwozonepercent}[1]{\IfEqCase{#1}{{GW231123cg_combined}{-0.50}{GW231123cg_nrsur}{-0.61}{GW231123cg_xphm}{-0.41}{GW231123cg_xo4a}{-0.47}{GW231123cg_tphm}{-0.22}{GW231123cg_seob}{-0.31}}}
\newcommand{\spintwozninetyninepercent}[1]{\IfEqCase{#1}{{GW231123cg_combined}{0.85}{GW231123cg_nrsur}{0.53}{GW231123cg_xphm}{0.91}{GW231123cg_xo4a}{0.83}{GW231123cg_tphm}{0.79}{GW231123cg_seob}{0.87}}}
\newcommand{\spintwozfivepercent}[1]{\IfEqCase{#1}{{GW231123cg_combined}{-0.30}{GW231123cg_nrsur}{-0.47}{GW231123cg_xphm}{-0.14}{GW231123cg_xo4a}{-0.26}{GW231123cg_tphm}{-0.04}{GW231123cg_seob}{-0.13}}}
\newcommand{\spintwozninetyfivepercent}[1]{\IfEqCase{#1}{{GW231123cg_combined}{0.71}{GW231123cg_nrsur}{0.34}{GW231123cg_xphm}{0.81}{GW231123cg_xo4a}{0.65}{GW231123cg_tphm}{0.67}{GW231123cg_seob}{0.74}}}
\newcommand{\spintwozninetypercent}[1]{\IfEqCase{#1}{{GW231123cg_combined}{0.61}{GW231123cg_nrsur}{0.26}{GW231123cg_xphm}{0.74}{GW231123cg_xo4a}{0.53}{GW231123cg_tphm}{0.60}{GW231123cg_seob}{0.66}}}
\newcommand{\symmetricmassratiominus}[1]{\IfEqCase{#1}{{GW231123cg_combined}{0.05}{GW231123cg_nrsur}{0.01}{GW231123cg_xphm}{0.02}{GW231123cg_xo4a}{0.05}{GW231123cg_tphm}{0.01}{GW231123cg_seob}{0.01}}}
\newcommand{\symmetricmassratiomed}[1]{\IfEqCase{#1}{{GW231123cg_combined}{0.24}{GW231123cg_nrsur}{0.25}{GW231123cg_xphm}{0.24}{GW231123cg_xo4a}{0.20}{GW231123cg_tphm}{0.25}{GW231123cg_seob}{0.25}}}
\newcommand{\symmetricmassratioplus}[1]{\IfEqCase{#1}{{GW231123cg_combined}{0.01}{GW231123cg_nrsur}{0.00}{GW231123cg_xphm}{0.01}{GW231123cg_xo4a}{0.01}{GW231123cg_tphm}{0.00}{GW231123cg_seob}{0.00}}}
\newcommand{\symmetricmassratiozeropercent}[1]{\IfEqCase{#1}{{GW231123cg_combined}{0.12}{GW231123cg_nrsur}{0.17}{GW231123cg_xphm}{0.17}{GW231123cg_xo4a}{0.12}{GW231123cg_tphm}{0.20}{GW231123cg_seob}{0.17}}}
\newcommand{\symmetricmassratioonepercent}[1]{\IfEqCase{#1}{{GW231123cg_combined}{0.15}{GW231123cg_nrsur}{0.23}{GW231123cg_xphm}{0.21}{GW231123cg_xo4a}{0.14}{GW231123cg_tphm}{0.23}{GW231123cg_seob}{0.22}}}
\newcommand{\symmetricmassrationinetyninepercent}[1]{\IfEqCase{#1}{{GW231123cg_combined}{0.25}{GW231123cg_nrsur}{0.25}{GW231123cg_xphm}{0.25}{GW231123cg_xo4a}{0.23}{GW231123cg_tphm}{0.25}{GW231123cg_seob}{0.25}}}
\newcommand{\symmetricmassratiofivepercent}[1]{\IfEqCase{#1}{{GW231123cg_combined}{0.19}{GW231123cg_nrsur}{0.24}{GW231123cg_xphm}{0.22}{GW231123cg_xo4a}{0.15}{GW231123cg_tphm}{0.24}{GW231123cg_seob}{0.24}}}
\newcommand{\symmetricmassrationinetyfivepercent}[1]{\IfEqCase{#1}{{GW231123cg_combined}{0.25}{GW231123cg_nrsur}{0.25}{GW231123cg_xphm}{0.24}{GW231123cg_xo4a}{0.22}{GW231123cg_tphm}{0.25}{GW231123cg_seob}{0.25}}}
\newcommand{\symmetricmassrationinetypercent}[1]{\IfEqCase{#1}{{GW231123cg_combined}{0.25}{GW231123cg_nrsur}{0.25}{GW231123cg_xphm}{0.24}{GW231123cg_xo4a}{0.21}{GW231123cg_tphm}{0.25}{GW231123cg_seob}{0.25}}}
\newcommand{\thetajnminus}[1]{\IfEqCase{#1}{{GW231123cg_combined}{0.9}{GW231123cg_nrsur}{0.4}{GW231123cg_xphm}{0.4}{GW231123cg_xo4a}{0.3}{GW231123cg_tphm}{1.0}{GW231123cg_seob}{0.4}}}
\newcommand{\thetajnmed}[1]{\IfEqCase{#1}{{GW231123cg_combined}{1.3}{GW231123cg_nrsur}{1.3}{GW231123cg_xphm}{1.6}{GW231123cg_xo4a}{0.5}{GW231123cg_tphm}{1.9}{GW231123cg_seob}{1.2}}}
\newcommand{\thetajnplus}[1]{\IfEqCase{#1}{{GW231123cg_combined}{0.9}{GW231123cg_nrsur}{0.8}{GW231123cg_xphm}{0.4}{GW231123cg_xo4a}{2.1}{GW231123cg_tphm}{0.3}{GW231123cg_seob}{1.0}}}
\newcommand{\thetajnzeropercent}[1]{\IfEqCase{#1}{{GW231123cg_combined}{0.0}{GW231123cg_nrsur}{0.1}{GW231123cg_xphm}{0.1}{GW231123cg_xo4a}{0.0}{GW231123cg_tphm}{0.4}{GW231123cg_seob}{0.3}}}
\newcommand{\thetajnonepercent}[1]{\IfEqCase{#1}{{GW231123cg_combined}{0.2}{GW231123cg_nrsur}{0.6}{GW231123cg_xphm}{1.0}{GW231123cg_xo4a}{0.1}{GW231123cg_tphm}{0.8}{GW231123cg_seob}{0.7}}}
\newcommand{\thetajnninetyninepercent}[1]{\IfEqCase{#1}{{GW231123cg_combined}{2.6}{GW231123cg_nrsur}{2.2}{GW231123cg_xphm}{2.1}{GW231123cg_xo4a}{3.0}{GW231123cg_tphm}{2.3}{GW231123cg_seob}{2.4}}}
\newcommand{\thetajnfivepercent}[1]{\IfEqCase{#1}{{GW231123cg_combined}{0.4}{GW231123cg_nrsur}{0.9}{GW231123cg_xphm}{1.2}{GW231123cg_xo4a}{0.2}{GW231123cg_tphm}{0.9}{GW231123cg_seob}{0.9}}}
\newcommand{\thetajnninetyfivepercent}[1]{\IfEqCase{#1}{{GW231123cg_combined}{2.2}{GW231123cg_nrsur}{2.1}{GW231123cg_xphm}{2.0}{GW231123cg_xo4a}{2.6}{GW231123cg_tphm}{2.2}{GW231123cg_seob}{2.2}}}
\newcommand{\thetajnninetypercent}[1]{\IfEqCase{#1}{{GW231123cg_combined}{2.1}{GW231123cg_nrsur}{2.0}{GW231123cg_xphm}{1.9}{GW231123cg_xo4a}{0.9}{GW231123cg_tphm}{2.2}{GW231123cg_seob}{2.1}}}
\newcommand{\tiltoneminus}[1]{\IfEqCase{#1}{{GW231123cg_combined}{0.58}{GW231123cg_nrsur}{0.55}{GW231123cg_xphm}{0.41}{GW231123cg_xo4a}{0.25}{GW231123cg_tphm}{0.39}{GW231123cg_seob}{0.44}}}
\newcommand{\tiltonemed}[1]{\IfEqCase{#1}{{GW231123cg_combined}{1.07}{GW231123cg_nrsur}{0.83}{GW231123cg_xphm}{1.83}{GW231123cg_xo4a}{1.12}{GW231123cg_tphm}{0.97}{GW231123cg_seob}{0.92}}}
\newcommand{\tiltoneplus}[1]{\IfEqCase{#1}{{GW231123cg_combined}{0.93}{GW231123cg_nrsur}{0.91}{GW231123cg_xphm}{0.40}{GW231123cg_xo4a}{0.35}{GW231123cg_tphm}{0.35}{GW231123cg_seob}{0.46}}}
\newcommand{\tiltonezeropercent}[1]{\IfEqCase{#1}{{GW231123cg_combined}{0.01}{GW231123cg_nrsur}{0.01}{GW231123cg_xphm}{0.06}{GW231123cg_xo4a}{0.64}{GW231123cg_tphm}{0.02}{GW231123cg_seob}{0.02}}}
\newcommand{\tiltoneonepercent}[1]{\IfEqCase{#1}{{GW231123cg_combined}{0.25}{GW231123cg_nrsur}{0.13}{GW231123cg_xphm}{0.90}{GW231123cg_xo4a}{0.81}{GW231123cg_tphm}{0.40}{GW231123cg_seob}{0.28}}}
\newcommand{\tiltoneninetyninepercent}[1]{\IfEqCase{#1}{{GW231123cg_combined}{2.24}{GW231123cg_nrsur}{2.04}{GW231123cg_xphm}{2.42}{GW231123cg_xo4a}{1.64}{GW231123cg_tphm}{1.49}{GW231123cg_seob}{1.64}}}
\newcommand{\tiltonefivepercent}[1]{\IfEqCase{#1}{{GW231123cg_combined}{0.49}{GW231123cg_nrsur}{0.28}{GW231123cg_xphm}{1.42}{GW231123cg_xo4a}{0.87}{GW231123cg_tphm}{0.58}{GW231123cg_seob}{0.48}}}
\newcommand{\tiltoneninetyfivepercent}[1]{\IfEqCase{#1}{{GW231123cg_combined}{2.00}{GW231123cg_nrsur}{1.74}{GW231123cg_xphm}{2.23}{GW231123cg_xo4a}{1.46}{GW231123cg_tphm}{1.31}{GW231123cg_seob}{1.38}}}
\newcommand{\tiltoneninetypercent}[1]{\IfEqCase{#1}{{GW231123cg_combined}{1.85}{GW231123cg_nrsur}{1.56}{GW231123cg_xphm}{2.14}{GW231123cg_xo4a}{1.38}{GW231123cg_tphm}{1.24}{GW231123cg_seob}{1.27}}}
\newcommand{\tilttwominus}[1]{\IfEqCase{#1}{{GW231123cg_combined}{0.80}{GW231123cg_nrsur}{0.51}{GW231123cg_xphm}{0.66}{GW231123cg_xo4a}{0.89}{GW231123cg_tphm}{0.51}{GW231123cg_seob}{0.66}}}
\newcommand{\tilttwomed}[1]{\IfEqCase{#1}{{GW231123cg_combined}{1.23}{GW231123cg_nrsur}{1.64}{GW231123cg_xphm}{0.92}{GW231123cg_xo4a}{1.26}{GW231123cg_tphm}{1.15}{GW231123cg_seob}{1.12}}}
\newcommand{\tilttwoplus}[1]{\IfEqCase{#1}{{GW231123cg_combined}{0.84}{GW231123cg_nrsur}{0.50}{GW231123cg_xphm}{1.05}{GW231123cg_xo4a}{1.13}{GW231123cg_tphm}{0.47}{GW231123cg_seob}{0.66}}}
\newcommand{\tilttwozeropercent}[1]{\IfEqCase{#1}{{GW231123cg_combined}{0.01}{GW231123cg_nrsur}{0.04}{GW231123cg_xphm}{0.01}{GW231123cg_xo4a}{0.01}{GW231123cg_tphm}{0.06}{GW231123cg_seob}{0.03}}}
\newcommand{\tilttwoonepercent}[1]{\IfEqCase{#1}{{GW231123cg_combined}{0.19}{GW231123cg_nrsur}{0.85}{GW231123cg_xphm}{0.12}{GW231123cg_xo4a}{0.15}{GW231123cg_tphm}{0.42}{GW231123cg_seob}{0.23}}}
\newcommand{\tilttwoninetyninepercent}[1]{\IfEqCase{#1}{{GW231123cg_combined}{2.52}{GW231123cg_nrsur}{2.38}{GW231123cg_xphm}{2.49}{GW231123cg_xo4a}{2.80}{GW231123cg_tphm}{1.86}{GW231123cg_seob}{2.10}}}
\newcommand{\tilttwofivepercent}[1]{\IfEqCase{#1}{{GW231123cg_combined}{0.42}{GW231123cg_nrsur}{1.14}{GW231123cg_xphm}{0.26}{GW231123cg_xo4a}{0.37}{GW231123cg_tphm}{0.65}{GW231123cg_seob}{0.46}}}
\newcommand{\tilttwoninetyfivepercent}[1]{\IfEqCase{#1}{{GW231123cg_combined}{2.07}{GW231123cg_nrsur}{2.15}{GW231123cg_xphm}{1.97}{GW231123cg_xo4a}{2.39}{GW231123cg_tphm}{1.63}{GW231123cg_seob}{1.78}}}
\newcommand{\tilttwoninetypercent}[1]{\IfEqCase{#1}{{GW231123cg_combined}{1.87}{GW231123cg_nrsur}{2.03}{GW231123cg_xphm}{1.68}{GW231123cg_xo4a}{2.12}{GW231123cg_tphm}{1.52}{GW231123cg_seob}{1.62}}}
\newcommand{\totalmassminus}[1]{\IfEqCase{#1}{{GW231123cg_combined}{48}{GW231123cg_nrsur}{32}{GW231123cg_xphm}{24}{GW231123cg_xo4a}{16}{GW231123cg_tphm}{26}{GW231123cg_seob}{21}}}
\newcommand{\totalmassmed}[1]{\IfEqCase{#1}{{GW231123cg_combined}{322}{GW231123cg_nrsur}{319}{GW231123cg_xphm}{283}{GW231123cg_xo4a}{313}{GW231123cg_tphm}{356}{GW231123cg_seob}{341}}}
\newcommand{\totalmassplus}[1]{\IfEqCase{#1}{{GW231123cg_combined}{44}{GW231123cg_nrsur}{14}{GW231123cg_xphm}{25}{GW231123cg_xo4a}{16}{GW231123cg_tphm}{22}{GW231123cg_seob}{21}}}
\newcommand{\totalmasszeropercent}[1]{\IfEqCase{#1}{{GW231123cg_combined}{221}{GW231123cg_nrsur}{250}{GW231123cg_xphm}{221}{GW231123cg_xo4a}{228}{GW231123cg_tphm}{295}{GW231123cg_seob}{280}}}
\newcommand{\totalmassonepercent}[1]{\IfEqCase{#1}{{GW231123cg_combined}{259}{GW231123cg_nrsur}{277}{GW231123cg_xphm}{249}{GW231123cg_xo4a}{288}{GW231123cg_tphm}{319}{GW231123cg_seob}{310}}}
\newcommand{\totalmassninetyninepercent}[1]{\IfEqCase{#1}{{GW231123cg_combined}{378}{GW231123cg_nrsur}{338}{GW231123cg_xphm}{331}{GW231123cg_xo4a}{335}{GW231123cg_tphm}{385}{GW231123cg_seob}{370}}}
\newcommand{\totalmassfivepercent}[1]{\IfEqCase{#1}{{GW231123cg_combined}{274}{GW231123cg_nrsur}{287}{GW231123cg_xphm}{259}{GW231123cg_xo4a}{297}{GW231123cg_tphm}{330}{GW231123cg_seob}{319}}}
\newcommand{\totalmassninetyfivepercent}[1]{\IfEqCase{#1}{{GW231123cg_combined}{366}{GW231123cg_nrsur}{333}{GW231123cg_xphm}{308}{GW231123cg_xo4a}{329}{GW231123cg_tphm}{378}{GW231123cg_seob}{362}}}
\newcommand{\totalmassninetypercent}[1]{\IfEqCase{#1}{{GW231123cg_combined}{359}{GW231123cg_nrsur}{330}{GW231123cg_xphm}{300}{GW231123cg_xo4a}{326}{GW231123cg_tphm}{373}{GW231123cg_seob}{357}}}
\newcommand{\totalmasssourceminus}[1]{\IfEqCase{#1}{{GW231123cg_combined}{47}{GW231123cg_nrsur}{32}{GW231123cg_xphm}{28}{GW231123cg_xo4a}{18}{GW231123cg_tphm}{19}{GW231123cg_seob}{23}}}
\newcommand{\totalmasssourcemed}[1]{\IfEqCase{#1}{{GW231123cg_combined}{236}{GW231123cg_nrsur}{237}{GW231123cg_xphm}{241}{GW231123cg_xo4a}{198}{GW231123cg_tphm}{242}{GW231123cg_seob}{241}}}
\newcommand{\totalmasssourceplus}[1]{\IfEqCase{#1}{{GW231123cg_combined}{30}{GW231123cg_nrsur}{23}{GW231123cg_xphm}{30}{GW231123cg_xo4a}{30}{GW231123cg_tphm}{28}{GW231123cg_seob}{28}}}
\newcommand{\totalmasssourcezeropercent}[1]{\IfEqCase{#1}{{GW231123cg_combined}{157}{GW231123cg_nrsur}{157}{GW231123cg_xphm}{183}{GW231123cg_xo4a}{162}{GW231123cg_tphm}{191}{GW231123cg_seob}{190}}}
\newcommand{\totalmasssourceonepercent}[1]{\IfEqCase{#1}{{GW231123cg_combined}{180}{GW231123cg_nrsur}{185}{GW231123cg_xphm}{203}{GW231123cg_xo4a}{175}{GW231123cg_tphm}{216}{GW231123cg_seob}{209}}}
\newcommand{\totalmasssourceninetyninepercent}[1]{\IfEqCase{#1}{{GW231123cg_combined}{282}{GW231123cg_nrsur}{274}{GW231123cg_xphm}{288}{GW231123cg_xo4a}{238}{GW231123cg_tphm}{285}{GW231123cg_seob}{285}}}
\newcommand{\totalmasssourcefivepercent}[1]{\IfEqCase{#1}{{GW231123cg_combined}{189}{GW231123cg_nrsur}{204}{GW231123cg_xphm}{213}{GW231123cg_xo4a}{180}{GW231123cg_tphm}{223}{GW231123cg_seob}{218}}}
\newcommand{\totalmasssourceninetyfivepercent}[1]{\IfEqCase{#1}{{GW231123cg_combined}{266}{GW231123cg_nrsur}{259}{GW231123cg_xphm}{271}{GW231123cg_xo4a}{228}{GW231123cg_tphm}{270}{GW231123cg_seob}{269}}}
\newcommand{\totalmasssourceninetypercent}[1]{\IfEqCase{#1}{{GW231123cg_combined}{258}{GW231123cg_nrsur}{254}{GW231123cg_xphm}{263}{GW231123cg_xo4a}{221}{GW231123cg_tphm}{262}{GW231123cg_seob}{262}}}
\newcommand{\viewingangleminus}[1]{\IfEqCase{#1}{{GW231123cg_combined}{0.6}{GW231123cg_nrsur}{0.2}{GW231123cg_xphm}{0.2}{GW231123cg_xo4a}{0.3}{GW231123cg_tphm}{0.2}{GW231123cg_seob}{0.2}}}
\newcommand{\viewinganglemed}[1]{\IfEqCase{#1}{{GW231123cg_combined}{1.1}{GW231123cg_nrsur}{1.2}{GW231123cg_xphm}{1.4}{GW231123cg_xo4a}{0.5}{GW231123cg_tphm}{1.1}{GW231123cg_seob}{1.1}}}
\newcommand{\viewingangleplus}[1]{\IfEqCase{#1}{{GW231123cg_combined}{0.5}{GW231123cg_nrsur}{0.2}{GW231123cg_xphm}{0.2}{GW231123cg_xo4a}{0.3}{GW231123cg_tphm}{0.2}{GW231123cg_seob}{0.2}}}
\newcommand{\viewinganglezeropercent}[1]{\IfEqCase{#1}{{GW231123cg_combined}{0.0}{GW231123cg_nrsur}{0.1}{GW231123cg_xphm}{0.1}{GW231123cg_xo4a}{0.0}{GW231123cg_tphm}{0.4}{GW231123cg_seob}{0.3}}}
\newcommand{\viewingangleonepercent}[1]{\IfEqCase{#1}{{GW231123cg_combined}{0.2}{GW231123cg_nrsur}{0.6}{GW231123cg_xphm}{1.0}{GW231123cg_xo4a}{0.1}{GW231123cg_tphm}{0.8}{GW231123cg_seob}{0.7}}}
\newcommand{\viewingangleninetyninepercent}[1]{\IfEqCase{#1}{{GW231123cg_combined}{1.6}{GW231123cg_nrsur}{1.4}{GW231123cg_xphm}{1.6}{GW231123cg_xo4a}{1.1}{GW231123cg_tphm}{1.3}{GW231123cg_seob}{1.4}}}
\newcommand{\viewinganglefivepercent}[1]{\IfEqCase{#1}{{GW231123cg_combined}{0.3}{GW231123cg_nrsur}{0.9}{GW231123cg_xphm}{1.1}{GW231123cg_xo4a}{0.2}{GW231123cg_tphm}{0.9}{GW231123cg_seob}{0.8}}}
\newcommand{\viewingangleninetyfivepercent}[1]{\IfEqCase{#1}{{GW231123cg_combined}{1.5}{GW231123cg_nrsur}{1.4}{GW231123cg_xphm}{1.6}{GW231123cg_xo4a}{0.8}{GW231123cg_tphm}{1.2}{GW231123cg_seob}{1.3}}}
\newcommand{\viewingangleninetypercent}[1]{\IfEqCase{#1}{{GW231123cg_combined}{1.4}{GW231123cg_nrsur}{1.3}{GW231123cg_xphm}{1.5}{GW231123cg_xo4a}{0.7}{GW231123cg_tphm}{1.2}{GW231123cg_seob}{1.2}}}
\newcommand{\luminositydistancegpcminus}[1]{\IfEqCase{#1}{{GW231123cg_combined}{1.5}{GW231123cg_nrsur}{1.0}{GW231123cg_xphm}{0.3}{GW231123cg_xo4a}{1.4}{GW231123cg_tphm}{1.1}{GW231123cg_seob}{1.0}}}
\newcommand{\luminositydistancegpcmed}[1]{\IfEqCase{#1}{{GW231123cg_combined}{2.2}{GW231123cg_nrsur}{1.9}{GW231123cg_xphm}{0.9}{GW231123cg_xo4a}{3.5}{GW231123cg_tphm}{2.7}{GW231123cg_seob}{2.3}}}
\newcommand{\luminositydistancegpcplus}[1]{\IfEqCase{#1}{{GW231123cg_combined}{1.9}{GW231123cg_nrsur}{1.7}{GW231123cg_xphm}{0.4}{GW231123cg_xo4a}{1.2}{GW231123cg_tphm}{1.2}{GW231123cg_seob}{1.4}}}
\newcommand{\luminositydistancegpczeropercent}[1]{\IfEqCase{#1}{{GW231123cg_combined}{0.1}{GW231123cg_nrsur}{0.2}{GW231123cg_xphm}{0.1}{GW231123cg_xo4a}{0.7}{GW231123cg_tphm}{0.5}{GW231123cg_seob}{0.4}}}
\newcommand{\luminositydistancegpconepercent}[1]{\IfEqCase{#1}{{GW231123cg_combined}{0.5}{GW231123cg_nrsur}{0.6}{GW231123cg_xphm}{0.3}{GW231123cg_xo4a}{1.7}{GW231123cg_tphm}{1.2}{GW231123cg_seob}{0.9}}}
\newcommand{\luminositydistancegpcninetyninepercent}[1]{\IfEqCase{#1}{{GW231123cg_combined}{4.9}{GW231123cg_nrsur}{4.9}{GW231123cg_xphm}{1.5}{GW231123cg_xo4a}{5.2}{GW231123cg_tphm}{4.3}{GW231123cg_seob}{4.4}}}
\newcommand{\luminositydistancegpcfivepercent}[1]{\IfEqCase{#1}{{GW231123cg_combined}{0.7}{GW231123cg_nrsur}{0.8}{GW231123cg_xphm}{0.6}{GW231123cg_xo4a}{2.1}{GW231123cg_tphm}{1.6}{GW231123cg_seob}{1.3}}}
\newcommand{\luminositydistancegpcninetyfivepercent}[1]{\IfEqCase{#1}{{GW231123cg_combined}{4.2}{GW231123cg_nrsur}{3.5}{GW231123cg_xphm}{1.3}{GW231123cg_xo4a}{4.7}{GW231123cg_tphm}{3.8}{GW231123cg_seob}{3.6}}}
\newcommand{\luminositydistancegpcninetypercent}[1]{\IfEqCase{#1}{{GW231123cg_combined}{3.8}{GW231123cg_nrsur}{3.0}{GW231123cg_xphm}{1.2}{GW231123cg_xo4a}{4.5}{GW231123cg_tphm}{3.5}{GW231123cg_seob}{3.3}}}
\newcommand{\pmassgapmassonesourcelowerthansixty}[1]{\IfEqCase{#1}{{GW231123cg_combined}{0}{GW231123cg_nrsur}{0}{GW231123cg_xphm}{0}{GW231123cg_xo4a}{0}{GW231123cg_tphm}{0}{GW231123cg_seob}{0}}}
\newcommand{\pmassgapmassonesourceabovesixty}[1]{\IfEqCase{#1}{{GW231123cg_combined}{100}{GW231123cg_nrsur}{100}{GW231123cg_xphm}{100}{GW231123cg_xo4a}{100}{GW231123cg_tphm}{100}{GW231123cg_seob}{100}}}
\newcommand{\pmassgapmassonesourcelowerthanonehundredthirty}[1]{\IfEqCase{#1}{{GW231123cg_combined}{28}{GW231123cg_nrsur}{59}{GW231123cg_xphm}{0}{GW231123cg_xo4a}{9}{GW231123cg_tphm}{35}{GW231123cg_seob}{36}}}
\newcommand{\pmassgapmassonesourceaboveonehundredthirty}[1]{\IfEqCase{#1}{{GW231123cg_combined}{72}{GW231123cg_nrsur}{41}{GW231123cg_xphm}{100}{GW231123cg_xo4a}{91}{GW231123cg_tphm}{65}{GW231123cg_seob}{64}}}
\newcommand{\pmassgapmassonesourceaboveonehundredtwenty}[1]{\IfEqCase{#1}{{GW231123cg_combined}{94}{GW231123cg_nrsur}{80}{GW231123cg_xphm}{100}{GW231123cg_xo4a}{99}{GW231123cg_tphm}{96}{GW231123cg_seob}{93}}}
\newcommand{\pmassgapmassonesourcebetweensixtyandonehundredthirty}[1]{\IfEqCase{#1}{{GW231123cg_combined}{28}{GW231123cg_nrsur}{59}{GW231123cg_xphm}{0}{GW231123cg_xo4a}{9}{GW231123cg_tphm}{35}{GW231123cg_seob}{36}}}
\newcommand{\pmassgapmasstwosourcelowerthansixty}[1]{\IfEqCase{#1}{{GW231123cg_combined}{16}{GW231123cg_nrsur}{0}{GW231123cg_xphm}{1}{GW231123cg_xo4a}{77}{GW231123cg_tphm}{0}{GW231123cg_seob}{0}}}
\newcommand{\pmassgapmasstwosourceabovesixty}[1]{\IfEqCase{#1}{{GW231123cg_combined}{84}{GW231123cg_nrsur}{100}{GW231123cg_xphm}{99}{GW231123cg_xo4a}{23}{GW231123cg_tphm}{100}{GW231123cg_seob}{100}}}
\newcommand{\pmassgapmasstwosourcelowerthanonehundredthirty}[1]{\IfEqCase{#1}{{GW231123cg_combined}{99}{GW231123cg_nrsur}{99}{GW231123cg_xphm}{100}{GW231123cg_xo4a}{100}{GW231123cg_tphm}{98}{GW231123cg_seob}{98}}}
\newcommand{\pmassgapmasstwosourceaboveonehundredthirty}[1]{\IfEqCase{#1}{{GW231123cg_combined}{1}{GW231123cg_nrsur}{1}{GW231123cg_xphm}{0}{GW231123cg_xo4a}{0}{GW231123cg_tphm}{2}{GW231123cg_seob}{2}}}
\newcommand{\pmassgapmasstwosourceaboveonehundredtwenty}[1]{\IfEqCase{#1}{{GW231123cg_combined}{8}{GW231123cg_nrsur}{10}{GW231123cg_xphm}{2}{GW231123cg_xo4a}{0}{GW231123cg_tphm}{14}{GW231123cg_seob}{14}}}
\newcommand{\pmassgapmasstwosourcebetweensixtyandonehundredthirty}[1]{\IfEqCase{#1}{{GW231123cg_combined}{83}{GW231123cg_nrsur}{99}{GW231123cg_xphm}{99}{GW231123cg_xo4a}{23}{GW231123cg_tphm}{98}{GW231123cg_seob}{98}}}
\newcommand{\pchieffabovezero}[1]{\IfEqCase{#1}{{GW231123cg_combined}{89}{GW231123cg_nrsur}{90}{GW231123cg_xphm}{58}{GW231123cg_xo4a}{100}{GW231123cg_tphm}{100}{GW231123cg_seob}{100}}}
\newcommand{\pspinoneabovezeroseven}[1]{\IfEqCase{#1}{{GW231123cg_combined}{91}{GW231123cg_nrsur}{91}{GW231123cg_xphm}{73}{GW231123cg_xo4a}{100}{GW231123cg_tphm}{96}{GW231123cg_seob}{95}}}
\newcommand{\pspintwoabovezeroseven}[1]{\IfEqCase{#1}{{GW231123cg_combined}{63}{GW231123cg_nrsur}{89}{GW231123cg_xphm}{46}{GW231123cg_xo4a}{27}{GW231123cg_tphm}{83}{GW231123cg_seob}{70}}}
\newcommand{\pbothspinsabovezeroseven}[1]{\IfEqCase{#1}{{GW231123cg_combined}{57}{GW231123cg_nrsur}{81}{GW231123cg_xphm}{31}{GW231123cg_xo4a}{27}{GW231123cg_tphm}{79}{GW231123cg_seob}{65}}}
\newcommand{\medianabovepopulationmmax}[1]{\IfEqCase{#1}{{GW231123cg_combined}{98}{GW231123cg_nrsur}{96}{GW231123cg_xphm}{99}{GW231123cg_xo4a}{98}{GW231123cg_tphm}{97}{GW231123cg_seob}{97}}}
\newcommand{\finalkickminus}[1]{\IfEqCase{#1}{{GW231123cg_nrsur}{814}}}
\newcommand{\finalkickmed}[1]{\IfEqCase{#1}{{GW231123cg_nrsur}{884}}}
\newcommand{\finalkickplus}[1]{\IfEqCase{#1}{{GW231123cg_nrsur}{973}}}
\newcommand{\snrprecessionminus}[1]{\IfEqCase{#1}{{GW231123cg_combined}{1.2}{GW231123cg_nrsur}{1.4}{GW231123cg_xphm}{2.9}{GW231123cg_xo4a}{0.7}{GW231123cg_tphm}{0.6}{GW231123cg_seob}{0.9}}}
\newcommand{\snrprecessionmed}[1]{\IfEqCase{#1}{{GW231123cg_combined}{2.0}{GW231123cg_nrsur}{2.3}{GW231123cg_xphm}{5.2}{GW231123cg_xo4a}{1.1}{GW231123cg_tphm}{1.7}{GW231123cg_seob}{2.0}}}
\newcommand{\snrprecessionplus}[1]{\IfEqCase{#1}{{GW231123cg_combined}{5.2}{GW231123cg_nrsur}{4.0}{GW231123cg_xphm}{4.7}{GW231123cg_xo4a}{1.2}{GW231123cg_tphm}{1.1}{GW231123cg_seob}{1.4}}}
\newcommand{\snrtwoonemultipoleminus}[1]{\IfEqCase{#1}{{GW231123cg_combined}{0.4}{GW231123cg_nrsur}{0.5}{GW231123cg_xphm}{2.2}{GW231123cg_xo4a}{0.2}{GW231123cg_tphm}{0.2}{GW231123cg_seob}{0.3}}}
\newcommand{\snrtwoonemultipolemed}[1]{\IfEqCase{#1}{{GW231123cg_combined}{0.5}{GW231123cg_nrsur}{0.6}{GW231123cg_xphm}{4.5}{GW231123cg_xo4a}{0.4}{GW231123cg_tphm}{0.2}{GW231123cg_seob}{0.4}}}
\newcommand{\snrtwoonemultipoleplus}[1]{\IfEqCase{#1}{{GW231123cg_combined}{5.0}{GW231123cg_nrsur}{1.0}{GW231123cg_xphm}{2.7}{GW231123cg_xo4a}{0.4}{GW231123cg_tphm}{0.6}{GW231123cg_seob}{1.0}}}
\newcommand{\snrtwoonemultipolefivepercent}[1]{\IfEqCase{#1}{{GW231123cg_combined}{0.1}{GW231123cg_nrsur}{0.1}{GW231123cg_xphm}{2.3}{GW231123cg_xo4a}{0.2}{GW231123cg_tphm}{0.0}{GW231123cg_seob}{0.0}}}
\newcommand{\snrthreethreemultipoleminus}[1]{\IfEqCase{#1}{{GW231123cg_combined}{2.6}{GW231123cg_nrsur}{1.9}{GW231123cg_xphm}{4.1}{GW231123cg_xo4a}{2.1}{GW231123cg_tphm}{1.7}{GW231123cg_seob}{2.1}}}
\newcommand{\snrthreethreemultipolemed}[1]{\IfEqCase{#1}{{GW231123cg_combined}{3.3}{GW231123cg_nrsur}{2.6}{GW231123cg_xphm}{10.5}{GW231123cg_xo4a}{4.0}{GW231123cg_tphm}{2.1}{GW231123cg_seob}{2.6}}}
\newcommand{\snrthreethreemultipoleplus}[1]{\IfEqCase{#1}{{GW231123cg_combined}{9.2}{GW231123cg_nrsur}{2.7}{GW231123cg_xphm}{5.2}{GW231123cg_xo4a}{1.3}{GW231123cg_tphm}{1.9}{GW231123cg_seob}{2.7}}}
\newcommand{\snrthreethreemultipolefivepercent}[1]{\IfEqCase{#1}{{GW231123cg_combined}{0.7}{GW231123cg_nrsur}{0.6}{GW231123cg_xphm}{6.4}{GW231123cg_xo4a}{1.9}{GW231123cg_tphm}{0.4}{GW231123cg_seob}{0.5}}}
\newcommand{\snrfourfourmultipoleminus}[1]{\IfEqCase{#1}{{GW231123cg_combined}{2.4}{GW231123cg_nrsur}{2.2}{GW231123cg_xphm}{2.9}{GW231123cg_xo4a}{1.0}{GW231123cg_tphm}{1.3}{GW231123cg_seob}{2.0}}}
\newcommand{\snrfourfourmultipolemed}[1]{\IfEqCase{#1}{{GW231123cg_combined}{3.0}{GW231123cg_nrsur}{3.5}{GW231123cg_xphm}{7.3}{GW231123cg_xo4a}{1.2}{GW231123cg_tphm}{2.8}{GW231123cg_seob}{3.2}}}
\newcommand{\snrfourfourmultipoleplus}[1]{\IfEqCase{#1}{{GW231123cg_combined}{5.9}{GW231123cg_nrsur}{3.2}{GW231123cg_xphm}{4.3}{GW231123cg_xo4a}{1.1}{GW231123cg_tphm}{1.1}{GW231123cg_seob}{1.9}}}
\newcommand{\snrfourfourmultipolefivepercent}[1]{\IfEqCase{#1}{{GW231123cg_combined}{0.7}{GW231123cg_nrsur}{1.3}{GW231123cg_xphm}{4.4}{GW231123cg_xo4a}{0.2}{GW231123cg_tphm}{1.5}{GW231123cg_seob}{1.2}}}

\newcolumntype{Y}{>{\centering\arraybackslash}X}
\newcolumntype{Z}{>{\raggedright\arraybackslash}X}
\newcolumntype{K}{>{\raggedleft\arraybackslash}X}
\newcolumntype{U}{>{\hsize=1.01\hsize}Y}
\newcolumntype{V}{>{\hsize=1.2\hsize}Y}
\newcolumntype{W}{>{\hsize=0.71\hsize}Y}

\usepackage{makerobust}

\renewcommand{\today}{\number\day\space\ifcase\month\or
  January\or February\or March\or April\or May\or June\or
  July\or August\or September\or October\or November\or December\fi
  \space\number\year}

\definecolor{NOTECOLOR}{rgb}{0.4, 0.2, 0.1}
\definecolor{notecolor}{rgb}{0.4, 0.2, 0.1}
\definecolor{orangey}{rgb}{0.95, 0.42, 0.14}

\acrodef{LSC}[LSC]{LIGO Scientific Collaboration}
\acrodef{LVC}[LVC]{LIGO Scientific and Virgo Collaboration}
\acrodef{LVK}[LVK]{LIGO Scientific, Virgo and KAGRA Collaboration}
\acrodef{aLIGO}{Advanced Laser Interferometer Gravitational-Wave Observatory}
\acrodef{aVirgo}{Advanced Virgo}
\acrodef{LIGO}[LIGO]{Laser Interferometer Gravitational-Wave Observatory}
\acrodef{IFO}[IFO]{interferometer}
\acrodef{LHO}[LHO]{LIGO-Hanford}
\acrodef{LLO}[LLO]{LIGO-Livingston}
\acrodef{O2}[O2]{second observing run}
\acrodef{O1}[O1]{first observing run}
\acrodef{O3}[O3]{third observing run}
\acrodef{O3a}[O3a]{first half of the third observing run}
\acrodef{O3b}[O3b]{second half of the third observing run}
\acrodef{o4a}[O4a]{first part of the fourth observing run}

\acrodef{BH}[BH]{black hole}
\acrodefplural{BH}{black holes}
\acrodef{bbh}[BBH]{binary black hole}
\acrodef{BNS}[BNS]{binary neutron star}
\acrodef{IMBH}[IMBH]{intermediate-mass black hole}
\acrodef{NS}[NS]{neutron star}
\acrodef{BHNS}[BHNS]{black hole--neutron star binaries}
\acrodef{NSBH}[NSBH]{neutron star--black hole binary}
\acrodef{PBH}[PBH]{primordial black hole binaries}
\acrodef{CBC}[CBC]{compact binary coalescence}
\acrodef{gw}[GW]{gravitational wave}
\acrodef{GWH}[GW]{gravitational-wave}

\acrodef{cwb}[cWB]{coherent WaveBurst}

\acrodef{snr}[SNR]{signal-to-noise ratio}
\acrodef{far}[FAR]{false-alarm rate}
\acrodef{ifar}[IFAR]{inverse false-alarm rate}
\acrodef{FAP}[FAP]{false alarm probability}
\acrodef{PSD}[PSD]{power spectral density}

\acrodef{GR}[GR]{general relativity}
\acrodef{NR}[NR]{numerical relativity}
\acrodef{PN}[PN]{post-Newtonian}
\acrodef{EOB}[EOB]{effective-one-body}
\acrodef{ROM}[ROM]{reduced-order model}
\acrodef{IMR}[IMR]{inspiral--merger--ringdown}

\acrodef{PDF}[PDF]{probability density function}
\acrodef{PE}[PE]{parameter estimation}
\acrodef{CL}[CL]{credible level}

\acrodef{EOS}[EoS]{equation of state}

\acrodef{LAL}[LAL]{LIGO Algorithm Library}

\acrodef{KLD}[KLD]{Kullback--Leibler divergence}
\acrodef{JSD}[JSD]{Jensen--Shannon divergence}

\acrodef{PISN}[PISN]{pair-instability supernova}
\acrodefplural{PISN}[PISNe]{pair-instability supernovae}
\acrodef{AGN}[AGN]{active galactic nucleus}

\newcommand{\BH}[0]{\ac{BH}\xspace}

\newcommand{\gstlal}{GstLAL\xspace}

\newcommand{\cwb}{\ac{cwb}\xspace}
\newcommand{\pycbc}{PyCBC\xspace}
\newcommand{\mbta}{MBTA\xspace}

\graphicspath{{./}{figures/}}
\begin{document}

\newcommand{\Mc}{\ensuremath{\mathcal{M}}\xspace}
\newcommand{\Mtot}{\ensuremath{M}\xspace}
\newcommand{\Msun}{\ensuremath{\text{M}_\odot}\xspace}
\newcommand{\massone}{\ensuremath{m_1}\xspace}
\newcommand{\masstwo}{\ensuremath{m_2}\xspace}
\newcommand{\Mf}{\ensuremath{M_\mathrm{f}}\xspace}
\newcommand{\massratio}{\ensuremath{q}\xspace}
\newcommand{\masscomponent}{\ensuremath{m_i}\xspace}

\newcommand{\chieff}{\ensuremath{\chi_\mathrm{eff}}\xspace}
\newcommand{\chip}{\ensuremath{\chi_\mathrm{p}}\xspace}
\newcommand{\chif}{\ensuremath{\chi_\mathrm{f}}\xspace}
\newcommand{\spintilt}{\ensuremath{\theta_{{JS}}}\xspace}
\newcommand{\spinone}{\ensuremath{\chi_1}\xspace}
\newcommand{\vecspinone}{\ensuremath{\vec\chi_1}\xspace}
\newcommand{\spintwo}{\ensuremath{\chi_2}\xspace}
\newcommand{\vecspintwo}{\ensuremath{\vec\chi_2}\xspace}

\newcommand{\DL}{\ensuremath{D_\mathrm{L}}\xspace}
\newcommand{\DC}{\ensuremath{D_\mathrm{c}}\xspace}
\newcommand{\redshift}{\ensuremath{z}\xspace}

\newcommand{\skyareasymbol}{\ensuremath{\Delta \Omega}\xspace}

\newcommand\perMpcyr{\ensuremath{\text{Mpc}^{-3}\,\text{yr}^{-1}}}
\newcommand\Mpcyr{\ensuremath{\text{Mpc}^{3}\,\text{yr}}}
\newcommand\perGpcyr{\ensuremath{\text{Gpc}^{-3}\,\text{yr}^{-1}}}
\newcommand\Gpcyr{\ensuremath{\text{Gpc}^{3}\,\text{yr}}}
\newcommand{\DKLchip}{D_\mathrm{KL}^{\chi_\text{p}}}
\newcommand{\DKLchieff}{D_\mathrm{KL}^{\chi_\text{eff}}}
\newcommand{\VT}{\ensuremath{\langle VT \rangle}}
\newcommand{\pastro}{\ensuremath{p_{\mathrm{astro}}}}
\newcommand{\pterr}{\ensuremath{p_{\mathrm{terr}}}}
\newcommand{\pbbh}{\ensuremath{p_{\mathrm{BBH}}}}
\newcommand{\pbns}{\ensuremath{p_{\mathrm{BNS}}}}
\newcommand{\pnsbh}{\ensuremath{p_{\mathrm{NSBH}}}}
\newcommand{\comovingvt}{\ensuremath{\langle VT_\mathrm{c} \rangle}}
\newcommand{\Gpc}{\ensuremath{\text{Gpc}}\xspace}

\title{GW231123: a Binary Black Hole Merger with Total Mass 190-265\,$M_\odot$}

\author[0000-0003-4786-2698]{A.~G.~Abac}
\affiliation{Max Planck Institute for Gravitational Physics (Albert Einstein Institute), D-14476 Potsdam, Germany}
\author{I.~Abouelfettouh}
\affiliation{LIGO Hanford Observatory, Richland, WA 99352, USA}
\author{F.~Acernese}
\affiliation{Dipartimento di Farmacia, Universit\`a di Salerno, I-84084 Fisciano, Salerno, Italy}
\affiliation{INFN, Sezione di Napoli, I-80126 Napoli, Italy}
\author[0000-0002-8648-0767]{K.~Ackley}
\affiliation{University of Warwick, Coventry CV4 7AL, United Kingdom}
\author[0000-0001-5525-6255]{C.~Adamcewicz}
\affiliation{OzGrav, School of Physics \& Astronomy, Monash University, Clayton 3800, Victoria, Australia}
\author[0009-0004-2101-5428]{S.~Adhicary}
\affiliation{The Pennsylvania State University, University Park, PA 16802, USA}
\author{D.~Adhikari}
\affiliation{Max Planck Institute for Gravitational Physics (Albert Einstein Institute), D-30167 Hannover, Germany}
\affiliation{Leibniz Universit\"{a}t Hannover, D-30167 Hannover, Germany}
\author[0000-0002-4559-8427]{N.~Adhikari}
\affiliation{University of Wisconsin-Milwaukee, Milwaukee, WI 53201, USA}
\author[0000-0002-5731-5076]{R.~X.~Adhikari}
\affiliation{LIGO Laboratory, California Institute of Technology, Pasadena, CA 91125, USA}
\author{V.~K.~Adkins}
\affiliation{Louisiana State University, Baton Rouge, LA 70803, USA}
% ...


\date[\relax]{Compiled: \today}

\begin{abstract}

On 2023 November 23 the two  LIGO observatories both detected GW231123, a gravitational-wave signal 
consistent with the merger of two black holes with masses $\massonesourcemed{GW231123cg_combined}^{+\massonesourceplus{GW231123cg_combined}}_{-\massonesourceminus{GW231123cg_combined}}\, M_\odot$ and
$\masstwosourcemed{GW231123cg_combined}^{+\masstwosourceplus{GW231123cg_combined}}_{-\masstwosourceminus{GW231123cg_combined}}\, M_\odot$ (90\% credible intervals), at luminosity distance 0.7--4.1\,Gpc
and redshift of $\redshiftmed{GW231123cg_combined}^{+\redshiftplus{GW231123cg_combined}}_{-\redshiftminus{GW231123cg_combined}}$, and a network signal-to-noise ratio of $\sim$20.7. Both black holes exhibit high spins, $\aonemed{GW231123cg_combined}^{+\aoneplus{GW231123cg_combined}}_{-\aoneminus{GW231123cg_combined}}$ and
$\atwomed{GW231123cg_combined}^{+\atwoplus{GW231123cg_combined}}_{-\atwominus{GW231123cg_combined}}$ respectively.
A massive black hole remnant is supported by an independent ringdown analysis.
Some properties of GW231123 are subject to large systematic uncertainties, as indicated by differences in inferred parameters between signal models.
The primary black hole lies within or above the theorized mass gap where black holes between 60--130\,$M_\odot$ should be rare 
due to pair instability mechanisms, while the secondary spans the gap. The observation of GW231123 therefore suggests 
the formation of black holes from channels beyond standard stellar collapse, and that intermediate-mass black holes of mass 
$\sim$200\,$M_\odot$ form through gravitational-wave driven mergers.

\end{abstract}




\section{Introduction}\label{sec:intro}

From 2015 to 2020 the LIGO-Virgo-KAGRA Collaboration identified 69 gravitational-wave signals from binary black hole mergers with
false alarm rates below one per
year~\citep{LIGOScientific:2014pky,VIRGO:2014yos,abbott2020prospects,10.1093/ptep/ptaa125,LIGOScientific:2021vkt}.
 Of these, the most massive was the source of GW190521, with a merger remnant of
 $\sim$140\,$M_\odot$~\citep{LIGOScientific:2020iuh,LIGOScientific:2020ufj}.
The small number of observable cycles of GW190521 limits our ability to accurately infer the source's properties,
and subsequent studies have proposed a wide range of alternative interpretations, including
highly eccentric orbits, dynamical capture scenarios, exotic object mergers, and cosmic string
collapse~\citep{Gayathri:2020coq,Romero-Shaw:2020thy,Gamba:2021gap,CalderonBustillo:2020fyi,Aurrekoetxea:2023vtp}.
Here we present a yet more challenging signal: GW231123\_135430 (hereafter referred to as GW231123),
confidently observed through a coincident detection in both the LIGO Hanford and Livingston detectors during the first part of
their fourth observing run, O4a (2023 May 24 to 2024 January 16).
The combination of data from the two observatories was essential in making a confident detection.

GW231123 consists of $\sim$5 cycles over a frequency range of 30--80\,Hz, similar to GW190521.
We interpret GW231123 as a binary-black-hole merger and infer a
total mass between 190\,$M_\odot$ and 265\,$M_\odot$ and high component black-hole spins ($\sim$0.9 and $\sim$0.8).
While a few gravitational-wave candidates have been observed with similarly high total masses~\citep{LIGOScientific:2021vkt,Wadekar:2023gea},
none have false alarm rates less than 1 per year; in addition, GW231123 has both a large signal-to-noise ratio and high
statistical significance.
Such high masses and spins pose a challenge to our most accurate waveform models, leading to larger uncertainties in the
black-hole masses, and the binary orientation and distance, than any previous signal of comparable strength.

\Ac{PISN} and pulsational \acp{PISN} are
expected to preclude stellar collapse to black holes with masses $\approx60$--130$M_\odot$ \citep{2019ApJ...887...53F, Farmer:2020xne, Woosley:2021xba, Hendriks:2023yrw}, and
the majority of the astrophysical population of black holes inferred from gravitational-wave catalogs lies below this gap \citep{KAGRA:2021duu}.
The large measurement uncertainties in the source of GW231123 mean that the primary black hole may be within or \emph{beyond}
the mass gap, while 
the secondary mass spans the entire gap within the 90\% credible intervals of our analysis. 
It is also possible that the two black-hole  masses lie on either side of the gap.
The scenarios with the highest probability require a formation channel that populates the mass gap, such as prior stellar mergers (e.g., \citealt{DiCarlo:2019fcq, Renzo:2020smh, Kremer:2020wtp}), black-hole mergers (for a review, see \citealt{Gerosa:2021mno} and references therein), or accretion in a gaseous environment (e.g., \citealt{McKernan:2012rf}).
Though these channels can produce highly spinning black holes (e.g., a characteristic value of $\sim0.7$ for black-hole mergers), the inferred spins may be higher than typical of remnant black holes and those of remnants from mergers previously observed with gravitational waves.

In this paper we present the LIGO-Virgo-KAGRA analysis of GW231123.
In Section~\ref{sec:searches}, we establish GW231123 as a confident \ac{gw} detection.
In Section~\ref{sec:dq}, we discuss the data quality at the time of the observation.
In Section~\ref{sec:source properties}, we first discuss our treatment of waveform uncertainties, then present the source properties and
additional waveform consistency checks.
In Section~\ref{sec:ringdown}, we analyse the ringdown portion to test the consistency of a \ac{BH} remnant interpretation.
In Section~\ref{sec:astro} we present a range of potential astrophysical implications.
Although the \ac{bbh} merger scenario presented throughout this paper is the most plausible astrophysical explanation for the source of GW231123, alternative scenarios cannot be ruled out and we discuss a selection of these in Section~\ref{sec:alternatives}. We conclude in Section~\ref{sec:conclusion}, and provide additional material in support of our results in a series of appendices.

\begin{figure*}
	\centering
	\includegraphics[width=0.98\textwidth]{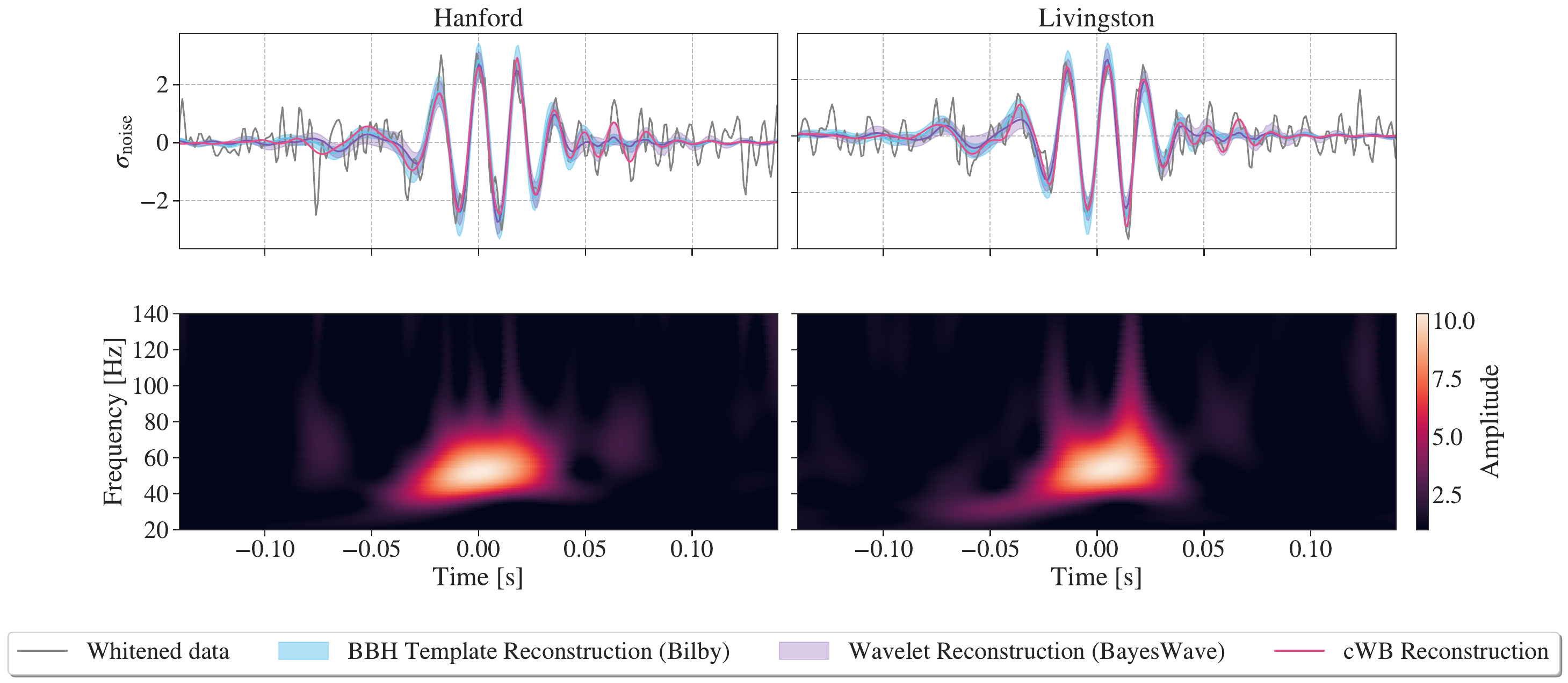}
	\caption{
		The \ac{gw} event GW231123 as observed by the LIGO Hanford (left panels) and LIGO Livingston (right panels) detectors.
		Time is measured relative to 2023 November 23 at 13:54:30.619 UTC.
		The top panels show the time-domain strain data (black), sampled at 1024\,Hz, whitened and then bandpass-filtered with a passband from 20\,Hz to 256\,Hz~\citep{LIGOScientific:2019hgc}.
		Also shown are the point-estimate whitened waveform from the cWB-BBH search (red), the 90\% credible interval of whitened waveforms inferred from a coherent Bayesian analysis using the combined samples from five \ac{bbh} waveform models (blue bands),
		and the 90\% credible interval inferred from BayesWave using a generic wavelet-based model (shaded purple). The vertical axis is in units of the noise standard deviation, $\sigma_{\rm noise}$.
		The bottom panels display the corresponding whitened time-frequency representations of the strain data, obtained using a continuous wavelet transform (CWT) with a Morlet–Gabor wavelet.
		The color scale is in units of the amplitude of the CWT coefficients.
	}
	\label{fig:data}
\end{figure*}

\section{Detection Significance} \label{sec:searches}

On 2023 November 23, at 13:54:30 UTC, the Advanced LIGO Hanford and Livingston detectors observed the \ac{gw} transient GW231123.
The Advanced Virgo and KAGRA detectors were not online at this time.
Despite its short duration ($\sim$0.1\,s) and limited bandwidth (Figure~\ref{fig:data}), coherent detection in both detectors allowed the 
signal to be identified in our analyses with high statistical significance, reported in terms of \ac{ifar}; see Table~\ref{table:search}. 
Without coincidence in two or more detectors, a high-mass \ac{bbh} signal like GW231123 would likely have been dismissed as a noise 
artefact (glitch). It was first detected by \pycbc Live, a matched-filter search for compact 
binaries~\citep{Allen:2004gu, Usman:2015kfa, Nitz:2017svb, DalCanton:2020vpm}. 
It was also reported by \cwb-BBH, a minimally modelled coherent excess power search~\citep{Mishra:2024zzs}. 
The \cwb-BBH search uses the \textit{WaveScan} time--frequency (TF) transformation~\citep{Klimenko:2022nji} and ranks 
identified triggers using a machine-learning classifier trained specifically on \ac{bbh} 
signals~\citep{Mishra:2021tmu, Mishra:2022ott, Mishra:2024zzs}.
In addition, the event was also detected by two model-independent low-latency \cwb searches, or burst searches, 
designed to identify generic GW transients: \cwb-2G and \cwb-XP. The former is based on the 
\textit{Wilson--Debauchies--Meyer} TF transformation~\citep{Klimenko:2008fu, Klimenko:2015ypf, Drago:2020kic}, 
while the latter uses the \textit{WaveScan} TF transformation. Both apply a machine-learning classifier (XGBoost), 
trained on generic white-noise-bursts to rank identified triggers~\citep{Szczepanczyk:2022urr}.
For further details on model-independent searches, see \cite{O4a:Allskyshort}.

Subsequent offline or archival reanalyses using improved background estimation and data quality information further increased the event’s statistical significance in both the \pycbc and \cwb pipelines (Table~\ref{table:search}).
Furthermore, two additional matched-filter searches, \gstlal~\citep{Messick:2016aqy, Sachdev:2019vvd, Hanna:2019ezx, Cannon:2020qnf, Sakon:2022ibh, ewing2023performance, Tsukada:2023edh, Joshi:2025nty}
and \mbta~\citep{Adams:2015ulm, Aubin:2020goo, Allene:2025saz}, which did not detect this event with significant confidence in low-latency (\ac{ifar} higher than 1 year),
recovered it in their offline analyses. These searches differ from \pycbc~\citep{Davies:2020tsx, Chandra:2021wbw, Davis:2022cmw, Kumar:2024bfe} in their implementation and use of signal--noise discriminators.
\gstlal's enhanced significance (the higher \ac{ifar} in Table.~\ref{table:search}) is primarily driven by a higher mass extension of the search with specific settings to compute the background for such higher mass mergers accurately. 
More details of the settings are provided in~\citep{GWTC:Methods, Joshi:2025nty}.
These changes improved the signal and noise models in this part of the parameter space in general, leading to a better recovery of this high mass signal. Additionally, \cwb-GMM, an entirely offline model-independent search, uses Gaussian Mixture Models~\citep{Gayathri:2020bly, Lopez:2021ikt, Smith:2024bsn} to rerank the triggers identified by \cwb-2G. 

\begin{table}[tb]
	\caption{Properties of the detection of GW231123 by various search pipelines.}
\begin{center}
\begin{tabular}{lccc}
\hline \hline
	\textbf{CBC pipelines} & Offline & Online      & Offline\\
       & SNR & IFAR (yr)  &  IFAR (yr) \\
\hline
PyCBC   & 19.9 & $> 100$    & 160 \\
GstLAL  & 20.1 & $2 \times 10^{-4}$ & $> 10000$ \\
MBTA    & 19.0   & --         & 60 \\
\cwb-BBH & 21.8 & $> 490$  & 9700 \\
\hline
	\textbf{\shortstack{Burst pipelines}} & & & \\
\hline
\cwb-2G  & 21.4 & $> 250$  & $> 490$ \\
\cwb-XP  & 21.1 & $> 240$  & $> 480$ \\
\cwb-GMM & 21.4 & --         & 100 \\
\hline \hline
\end{tabular}
\label{table:search}
\end{center}
\tablecomments{The significance is reported in terms of the inverse \ac{far} (IFAR) $= 1/$\ac{far} as measured by each search.}
\end{table}

The differences in the \acp{ifar} reported by the offline search pipelines—despite broadly consistent \acp{snr}—primarily reflect differences in their ability to separate GW231123-like signals from background noise in a comparable parameter range.
Similar discrepancies have been observed previously, particularly between matched-filter and minimally-modelled searches,
when searching for non-eccentric \ac{IMBH} binaries~\citep{CalderonBustillo:2017skv, Chandra:2020ccy, LIGOScientific:2020iuh, Chandra:2021xvs, Szczepanczyk:2020osv}.
These differences arise not only from how effectively each search separates signals from glitches, but also from the differing approaches used to estimate the noise background.

To assess whether the observed variation in statistical significance across pipelines is consistent with expectations, we conducted a dedicated injection campaign.
Using the \textsc{NRSur7dq4} (\textsc{NRSur}) waveform model~\citep{Varma:2019csw}, we simulated $\sim$8000 non-eccentric \ac{bbh} signals with intrinsic parameters consistent with those inferred for GW231123 (Section~\ref{sec:source properties}). We sampled the sky positions and binary orientations isotropically and drew redshifts uniformly in comoving volume up to $z_{\mathrm{max}}=1.5$, assuming a flat $\Lambda$CDM cosmology~\citep{Planck:2015fie}. We added these simulated signals uniformly over several days around the event and re-ran our offline search pipelines using the same configuration as applied to the real data.

We found that for simulated signals observed in both Advanced LIGO detectors, the CBC searches recovered the following fractions with a \ac{ifar} above 100 years,
\cwb-BBH 32\%, \pycbc 27\%, \gstlal 41\%, and \mbta 16\%. For the Burst searches, \cwb-2G and \cwb-XP each recovered 22\%, while \cwb-GMM recovered 10\%.
Since Burst searches identify coherent power across the detector network without relying on \ac{bbh} waveform models, their efficiencies are not directly comparable to CBC searches.
However, within each search category, detection pipelines reporting a higher \ac{ifar} for GW231123 consistently demonstrated higher recovery fractions for simulated signals with masses and
spins representative of those inferred for GW231123.

Given that all pipelines detected GW231123 with an \ac{ifar} above the typical threshold of 1 year used for population analyses, and a detailed background study for one pipeline (\cwb-BBH) identified GW231123 with an \ac{ifar} of 9700 years, we consider GW231123 to be a confident detection.

\section{Data quality} \label{sec:dq}

The event GW231123 was detected during the \ac{o4a}, a time when the LIGO Hanford and LIGO Livingston detectors were observing with a typical binary neutron star inspiral range of 152 Mpc and 160 Mpc \citep{Capote:2024rmo}. 
The detectors' data were calibrated in near real-time to produce the online dataset used for low-latency searches \citep{abbott2020guide,klimenko2016method,Tsukada:2023edh, Ewing:2023qqe,DalCanton:2020vpm,Chu:2020pjv, Aubin:2020goo} and parameter estimation \citep{Singer:2015ema, bilby_paper, Pankow:2015cra}. 
The calibration process subtracts linear spectral features from known instrumental sources, identified through auxiliary witness sensors, and intentionally injected calibration lines used to measure the instruments' response at various frequencies \citep{viets2018reconstructing, Sun:2020wke, Sun:2021qcg}.

Following the data-quality procedures established for \ac{o4a} \citep{LIGO:2024kkz}, including broadband noise subtraction \citep{Vajente:2019ycy} and data-quality report analysis routines \citep{LIGO:2021ppb}, detector data surrounding the event were evaluated for signs of non-Gaussian excess power (glitches) within the target time-frequency analysis window using a spectrogram-based glitch-identification tool~\citep{Vazsonyi:2022jul}. 
It was determined that glitches were present in each detector around, but not coincident with the event.

From spectrograms, we determined that a glitch was present in the LIGO Hanford data 1.7--1.1~s before the event, in a frequency range between 15--30~Hz. 
The glitch is possibly related to the LIGO Hanford differential arm control loop \citep{LIGOScientific:2014pky}. This control loop leads to nonstationary noise from the high root-mean-square drive applied to the electrostatic drive actuator. This issue has been fixed in the second part of the fourth observing run \citep{darm_improvement}. This glitch  was close to the event and within the time--frequency window used to infer the source properties, so  
BayesWave \citep{Cornish:2014kda, Cornish:2020dwh, Chatziioannou:2021ezd} was used to model simultaneously the compact binary signal and the glitch \citep{LIGO:2024kkz}. 
We removed this non-Gaussianity from the data by subtracting a phenomenological, wavelet-based model of the excess power noise~\citep{Hourihane:2022doe, Ghonge:2023ksb}. 
The glitch-subtracted data successfully passed the validation process, which compares the residual noise to Gaussian noise ~\citep{LIGO:2024kkz, Vazsonyi:2022jul}.
Additional broadband non-stationary noise was present in the Hanford detector in the hours of data surrounding GW231123, but we found no evidence that this impacted the analysis of GW231123. 

In LIGO Livingston data, a glitch was identified 3.0--2.0~s before the event, in a frequency range between 10--20~Hz. Given that LIGO Livingston had recurring low-frequency scattered light glitches \citep{LIGO:2024kkz}, this glitch was likely caused by scattered light.
We determined the time--frequency profile of the glitch to have no measurable effect on the GW231123 analysis, so the analyses from here on use the LIGO Livingston original data and the LIGO Hanford glitch-subtracted data.

\section{Source properties} \label{sec:source properties}

In the following, we describe the methods used to estimate the source properties (Section~\ref{sec:pe-methods}), and how we deal with the systematic differences in results from multiple signal models (Section~\ref{sec:systematics}).
Having discussed our methods and sources of error,
we present and discuss our estimates of the source properties in Section~\ref{sec:inference}, and finally our waveform consistency checks (Section~\ref{sec:waveformconsistency}).

\subsection{Methods}
\label{sec:pe-methods}

We report the properties of GW231123 using signal models for non-eccentric \ac{bbh} mergers in a coherent Bayesian analysis~\citep{LIGOScientific:2016vlm} of the LIGO Hanford and LIGO Livingston data around the time of GW231123. (We discuss potential eccentricity further in Section 6.)
We calculate the likelihood using
8\,s of data (6\,s before and 2\,s after the reported merger time of GW231123), and consider frequencies within the range 20--448~Hz. This range was chosen to contain the signal based on preliminary analyses at the time of the event, and to avoid loss of power at high frequencies due to low-pass filtering of the data~\citep{LIGOScientific:2025yae}. All analyses
employ standard priors used in previous analyses~\citep{LIGOScientific:2020ibl,LIGOScientific:2021usb,LIGOScientific:2021vkt}, and we use a Planck 2015 $\Lambda\mathrm{CDM}$ cosmology~\citep{Planck:2015fie}.
The \textsc{NRSur} analysis
employs a reduced prior mass range due to model constraints, mass ratios below 6:1. The other models employ a wider mass prior, mass ratios below $\sim$10:1, and no posterior support is found beyond the \textsc{NRSur} analysis.
We characterise the detector noise via the median estimate of different power spectral density (PSD) realizations calculated with BayesWave~\citep{Cornish:2014kda,Littenberg:2014oda}. As done previously~\citep{Chatziioannou:2019zvs,LIGOScientific:2021vkt}, we calculate the median PSD for data containing the trigger.
To sample the posterior distribution, we interface with the {\sc{dynesty}} nested sampling package~\citep{Speagle:2019ivv} via the {\sc{bilby}} library~\citep{bilby_paper,Romero-Shaw:2020owr}. We present results with 1000 live points, and verify that the results remain consistent when the number of live points is increased to 3000, as well as when we lower the frequency range to include data between 16--20~Hz.

\subsection{Waveform Systematics}
\label{sec:systematics}

The source properties of GW231123 lie in a challenging region of parameter space for current waveform models, to such an extent that
measurements using different models show significant disagreement, with multiple parameters failing to 
agree within 90\% credible intervals. 
(See Appendix~\ref{sec: systematics appendix} for examples.) For typical signals, our models are well within our observations' accuracy requirements, 
and the level of model disagreement for GW231123 has not been seen in any previous LVK \ac{gw} observation
with moderate \acp{snr} ($>$12). All models show strong support for spins $>$0.8, and since no theoretical signal model is 
calibrated to numerical-relativity (NR) waveforms from precessing binaries with spins above 0.8, waveform uncertainties are
one possible cause of the measurement differences. Hence, before presenting the source properties, we describe how we quantify 
waveform-model uncertainties. We do not study in detail the impact of Gaussian noise fluctuations or low-SNR glitches that are difficult to identify and mitigate using the methods presented in Section~\ref{sec:dq}.

We consider five state-of-the-art \ac{IMR} signal models, \textsc{NRSur7dq4}~\citep[\textsc{NRSur};][]{Varma:2019csw}, \textsc{SEOBNRv5PHM}~\citep[\textsc{v5PHM};][]{Ramos-Buades:2023ehm}, \textsc{IMRPhenomTPHM}~\citep[\textsc{TPHM};][]{Estelles:2021gvs}, \textsc{IMRPhenomXPHM}~\citep[\textsc{XPHM};][]{Colleoni:2024knd} and \textsc{IMRPhenomXO4a}~\citep[\textsc{XO4a};][]{Thompson:2023ase}. 
The first three model the signal in the time domain while the latter two natively employ the frequency domain. 
All models use information from numerical relativity to inform the merger-ringdown in the aligned-spin sector. 
However only \textsc{NRSur}; fully interpolates two-spin precessing systems in the precessing sector, while \textsc{XO4a} is calibrated to single-spin precessing systems; all other models employ results from post-Newtonian and perturbation theory through merger and ringdown.
(More details are given in Appendix~\ref{sec: systematics appendix}.)
The model papers referenced here include studies to assess the accuracy of these models across the \ac{bbh} 
parameter space, but here we focus on the likely region of parameter space for this observation; high total mass, $q = m_2/m_1 \geq1/3$, and moderate to high spins.

\begin{figure}[t]
   \centering
   \includegraphics[width=0.47\textwidth]{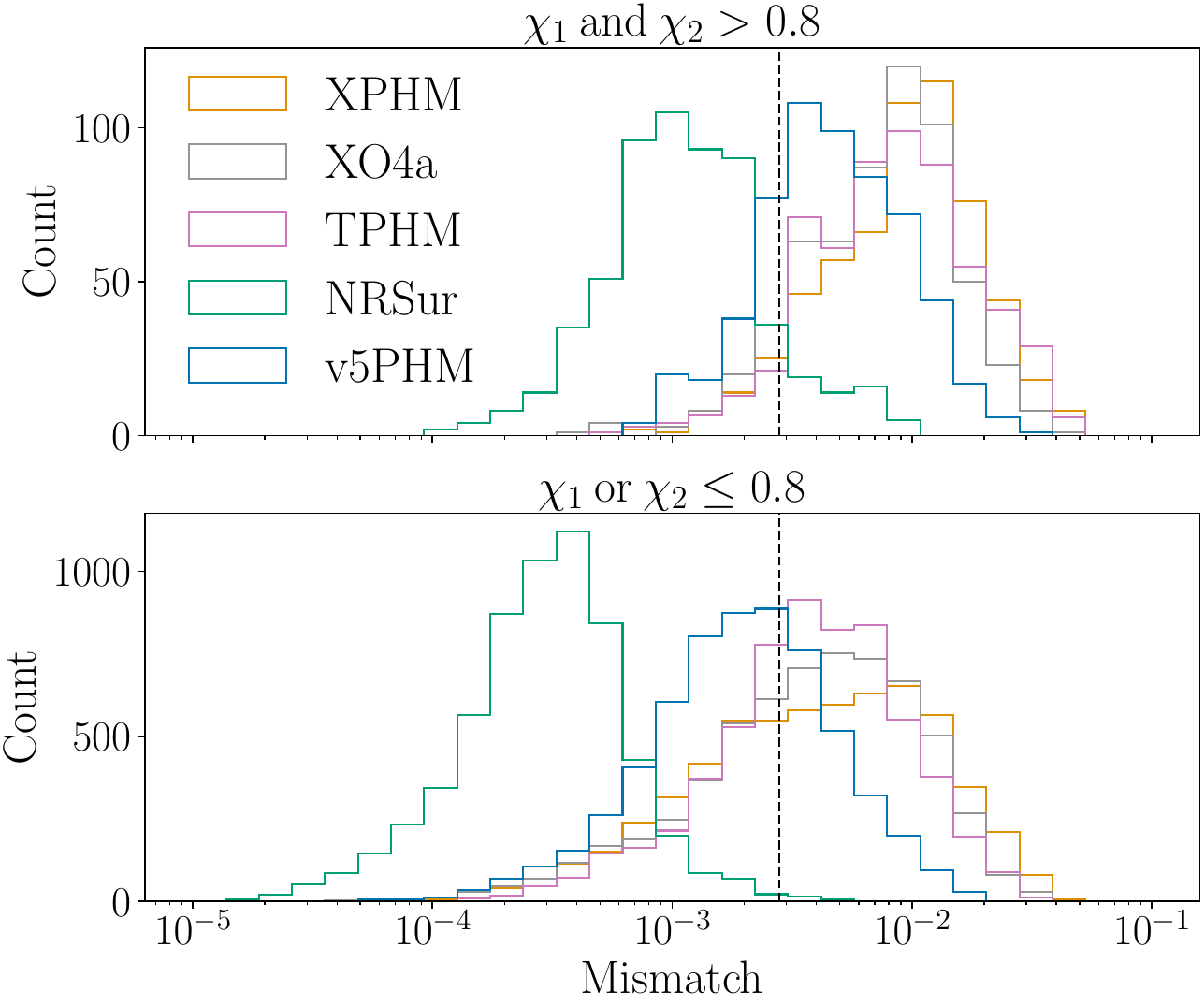} 
   \caption{Mismatch accuracy of the waveform models considered in this paper against 1123 NR simulations at a total mass of 300$M_\odot$ and a range of inclinations between $\iota=0$ and $\pi/2$. The vertical dashed line at a mismatch of 0.0028 shows
   the conservative criterion discussed in the text.}
   \label{fig: mismatch histogram}
\end{figure}

We quantify the models' accuracy against NR results, including a set of simulations that extend up to spins of 0.95~\citep{Boyle:2019kee, Hamilton:2023qkv, Scheel:2025jct}. 
A standard waveform accuracy measure is the mismatch between two waveforms~\citep{Cutler:1994ys},
where waveform uncertainties will not bias a parameter measurement if the model's mismatch uncertainty is less than
$\chi^2_k(1-p)/(2\rho^2)$~\citep{McWilliams:2010eq,Baird:2012cu}, where $\rho$ is the \ac{snr} and $\chi_k^2(1-p)$ is the chi-square value for $k$ degrees of freedom at probability $p$. 
For single-parameter measurements $k=1$ provides a lower bound~\citep{Thompson:2025ab}, so the mismatch criterion for the 90\% credible interval at $\rho = 22$
is $1.35/\rho^2 = 0.0028$. 
Figure~\ref{fig: mismatch histogram} reports the distribution of 
mismatches of each model against 1123 NR waveforms with $q \geq 1/3$, all scaled to the redshifted (detector-frame) total mass 
$(1+z)M = 300\,M_\odot$, at six equally spaced inclinations in $\cos\iota$ from $\iota=0$ to $\pi/2$ inclusive.
The mismatches are calculated for precessing systems~\citep{Schmidt:2014iyl, Harry:2016ijz} following the procedure described in~\cite{Hamilton:2021pkf}, maximising over time shifts, a global phase and template polarisation and optimising over in-plane spin rotations.
A subset of the simulations come from the third release of the SXS catalog \citep{Scheel:2025jct} and contain GW memory, which introduces a constant late-time offset that we handle for the mismatch calculations with a highpass filtering technique~\citep{Xu:2024ybt,Valencia:2024zhi,Chen:2024ieh} to mitigate possible artefacts in the Fourier domain.
We employ the same PSD for the LIGO Livingston detector as utilised in the coherent Bayesian analysis.
\textsc{NRSur} performs better than the other models (by roughly an order of magnitude for low-spin cases), and all other
models have comparable accuracy. However, \textsc{NRSur} does not meet the conservative accuracy criterion for all cases, 
and for spins greater than 0.8 the mismatches are higher in 10\% of 98 cases. 
Even if we apply a less conservative mismatch criterion from the literature (e.g., with $k=7$ for non-eccentric binaries and $p = 0.67$~\citep{Chatziioannou:2017tdw,Scheel:2025jct} we have
0.0072) there are configurations where \textsc{NRSur} exceeds the criterion (2\%).
The relative accuracy of models is also not uniform across all cases, e.g., 
we find cases in which other models 
show comparable or improved performance relative to \textsc{NRSur}.

To test whether these waveform uncertainties will result in biases, we performed our standard Bayesian parameter estimation analysis on a series of NR injections, as detailed in Appendix~\ref{sec: systematics appendix}. 
We observe that, while the five models considered here perform well for most signals, there are configurations where \emph{all models} 
may incur biases for massive high-spin signals. We also find that the relative performance of each model can change in the 
presence of Gaussian noise, although this requires more detailed study in future work. 
To properly correct for this, we would ideally marginalise over waveform 
uncertainties or incorporate model accuracy into Bayesian analyses ~\citep{Read:2023hkv,Khan:2024whs,Hoy:2024vpc,Pompili:2024yec,Kumar:2025nwb,Mezzasoma:2025moh}. 
Without access to a model of the waveform-model uncertainties, we follow what has been done previously~\citep{LIGOScientific:2016vlm} and combine the results from multiple models to marginalise over the model uncertainties. 
In choosing models in addition to \textsc{NRSur}, we note that all other models exhibit a comparable range of mismatches, and no model is 
clearly preferred in our injection studies, and so we include all five state-of-the-art models. We combine posterior results inferred 
from \textsc{NRSur}, \textsc{v5PHM}, \textsc{TPHM}, \textsc{XPHM}, \textsc{XO4a} with equal weight and report the combined 
samples throughout this paper. To illustrate the variation between the combined results and single models, in some figures we also 
show the \textsc{NRSur} results. In some analyses we expect the choice of model to have little impact, e.g., the detection significance 
study in Section~\ref{sec:searches}, and in these cases, we use only the \textsc{NRSur} samples.

\subsection{Inference}
\label{sec:inference}

Our Bayesian analysis indicates that
GW231123 was produced from a high-mass compact binary merger with highly spinning
components. We infer individual source component masses~$m_{1} = \massonesourcemed{GW231123cg_combined}^{+\massonesourceplus{GW231123cg_combined}}_{-\massonesourceminus{GW231123cg_combined}}\, M_{\odot}$ and~$m_{2} = \masstwosourcemed{GW231123cg_combined}^{+\masstwosourceplus{GW231123cg_combined}}_{-\masstwosourceminus{GW231123cg_combined}}\, M_{\odot}$ with spin magnitudes
$\chi_{1} = \aonemed{GW231123cg_combined}^{+\aoneplus{GW231123cg_combined}}_{-\aoneminus{GW231123cg_combined}}$ and
$\chi_{2} = \atwomed{GW231123cg_combined}^{+\atwoplus{GW231123cg_combined}}_{-\atwominus{GW231123cg_combined}}$.
We present a summary of the key source properties of GW231123 in Table~\ref{table:combined_pe}. Unless otherwise stated, we
report all mass measurements in the source frame,
and all measurements correspond to the median and 90\% symmetric credible level.

\begin{table}
\begin{ruledtabular}
    \caption{Source properties of GW231123.
    }
    \label{table:combined_pe}
    \renewcommand{\arraystretch}{1.2}
    \begin{center}
    \begin{tabular}{l l}
        \hline
        Primary mass $m_1 / \Msun$ & $\massonesourcemed{GW231123cg_combined}^{+\massonesourceplus{GW231123cg_combined}}_{-\massonesourceminus{GW231123cg_combined}}$ \\
        Secondary mass $m_2 / \Msun $ & $\masstwosourcemed{GW231123cg_combined}^{+\masstwosourceplus{GW231123cg_combined}}_{-\masstwosourceminus{GW231123cg_combined}}$ \\
        Mass ratio $q = m_2 / m_1$ & $\massratiomed{GW231123cg_combined}^{+\massratioplus{GW231123cg_combined}}_{-\massratiominus{GW231123cg_combined}}$ \\
        Total mass $M / \Msun$ & $\totalmasssourcemed{GW231123cg_combined}^{+\totalmasssourceplus{GW231123cg_combined}}_{-\totalmasssourceminus{GW231123cg_combined}}$ \\
        Final mass $M_{\rm{f}} / \Msun$ & $\finalmasssourcemed{GW231123cg_combined}^{+\finalmasssourceplus{GW231123cg_combined}}_{-\finalmasssourceminus{GW231123cg_combined}}$ \\
        Primary spin magnitude $\chi_1$ & $\aonemed{GW231123cg_combined}^{+\aoneplus{GW231123cg_combined}}_{-\aoneminus{GW231123cg_combined}}$ \\
        Secondary spin magnitude $\chi_2$ & $\atwomed{GW231123cg_combined}^{+\atwoplus{GW231123cg_combined}}_{-\atwominus{GW231123cg_combined}}$ \\
        Effective inspiral spin $\chi_{\rm eff}$ & $\chieffmed{GW231123cg_combined}^{+\chieffplus{GW231123cg_combined}}_{-\chieffminus{GW231123cg_combined}}$ \\
        Effective precessing spin $\chi_{\rm p}$ & $\chipmed{GW231123cg_combined}^{+\chipplus{GW231123cg_combined}}_{-\chipminus{GW231123cg_combined}}$ \\
        Final spin $\chi_{\rm f}$ & $\finalspinmed{GW231123cg_combined}^{+\finalspinplus{GW231123cg_combined}}_{-\finalspinminus{GW231123cg_combined}}$ \\
        Luminosity distance $D_{\rm L} / \rm{Gpc}$ & $\luminositydistancegpcmed{GW231123cg_combined}^{+\luminositydistancegpcplus{GW231123cg_combined}}_{-\luminositydistancegpcminus{GW231123cg_combined}}$ \\
        Inclination angle $\theta_{\mathrm{JN}} / \rm{rad}$ & $\thetajnmed{GW231123cg_combined}^{+\thetajnplus{GW231123cg_combined}}_{-\thetajnminus{GW231123cg_combined}}$ \\
        Source redshift $z$ & $\redshiftmed{GW231123cg_combined}^{+\redshiftplus{GW231123cg_combined}}_{-\redshiftminus{GW231123cg_combined}}$ \\
        Network matched filter SNR $\rho$ & $\networkmatchedfiltersnrmed{GW231123cg_combined}^{+\networkmatchedfiltersnrplus{GW231123cg_combined}}_{-\networkmatchedfiltersnrminus{GW231123cg_combined}}$ \\
    \end{tabular}
    \end{center}
	\tablecomments{We report combined results from five models that have been mixed with equal weight. In most cases we present the median value of the 1D marginalized posterior distribution and the 90\% symmetric credible intervals. For properties that have physical bounds, including the primary spin magnitude, secondary spin magnitude, mass ratio, effective precessing-spin, inclination angle and the final spin of the remnant, we report the median value as well as the 90\% highest posterior density (HPD) credible interval. The inclination of the binary is defined as the angle between the total angular momentum and the line of sight, $\theta_{\mathrm{JN}}$. All mass measurements are reported in the source frame. Our results are reported at a reference frequency of $10\, \mathrm{Hz}$. Results obtained with individual models can be found in Appendix~\ref{sec:pe_appendix}.}
\end{ruledtabular}
\end{table}

\begin{figure}[t!]
	\includegraphics[width=0.47\textwidth]{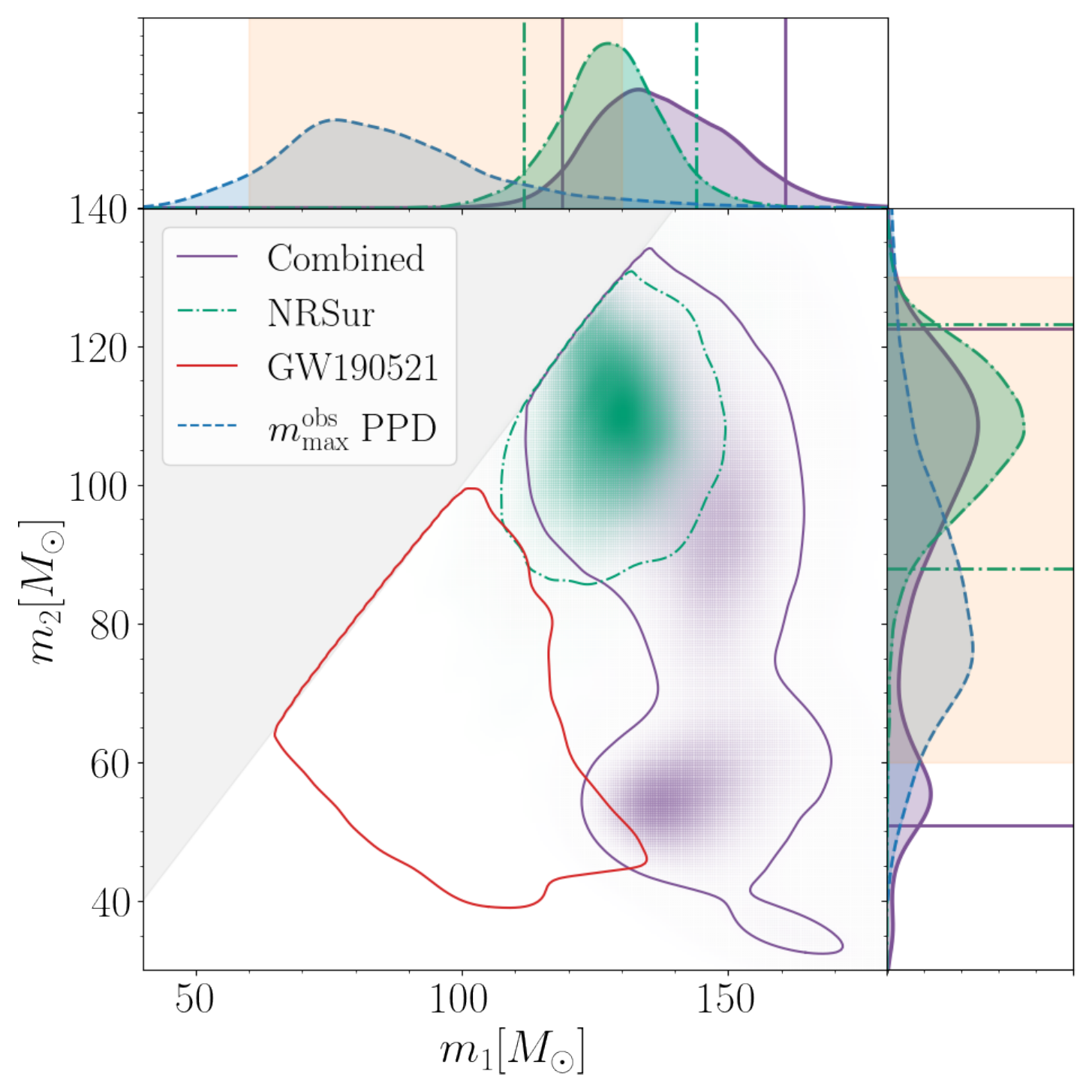}
	\caption{The posterior distribution of the primary and secondary source masses. We show the posterior distribution resulting from equally combining samples from five waveform models that include precession and higher-order multipoles (purple).
	We separately show the posterior distribution obtained with \textsc{NRSur} (green dash dot).
	We compare against estimates for the source frame masses of GW190521~\citep[red solid,][]{LIGOScientific:2020iuh,LIGOScientific:2020ufj,LIGOScientific:2021vkt}. Each contour, as well as the colored horizontal and vertical lines, shows the 90\% credible intervals.
	In blue dashed we show the posterior predictive distribution for the largest \ac{BH} mass $m_\mathrm{max}^\mathrm{obs}$ in mock catalogs similar to GWTC-3~\citep{LIGOScientific:2021vkt,KAGRA:2021duu}; see Section~\ref{sec:astro}. The solid orange bands show the putative mass gap from (pulsational) pair instability from 60--130\,$M_{\odot}$.
	}
    \label{fig:source_mass_inference}
\end{figure}

Although we observe differences depending on the model, the primary and secondary component
masses nevertheless have a significant probability of lying within the mass gap from (pulsational) \ac{PISN} processes, as shown in Figure~\ref{fig:source_mass_inference}, where we assume a nominal gap ranging from $\sim$60--130\,$M_{\odot}$
(see Section~\ref{sec:astro} for a detailed discussion). 
The binary's total mass is constrained to be within
\totalmasssourcefivepercent{GW231123cg_combined}--\totalmasssourceninetyfivepercent{GW231123cg_combined}\,$M_{\odot}$. This measurement
exceeds the 95th percentile of the inferred total mass from GW190521~\citep{LIGOScientific:2021vkt}. Assuming a FAR threshold of one per year, similar to~\citet{KAGRA:2021duu},
the source of GW231123 is the highest mass \ac{bbh} observed by the LVK to date; other lower significance high-mass observations have been discussed in~\citet{LIGOScientific:2021usb,Wadekar:2023gea,Williams:2024tna,Ruiz-Rocha:2025yno}.

\begin{figure}[t!]
	\includegraphics[width=0.47\textwidth]{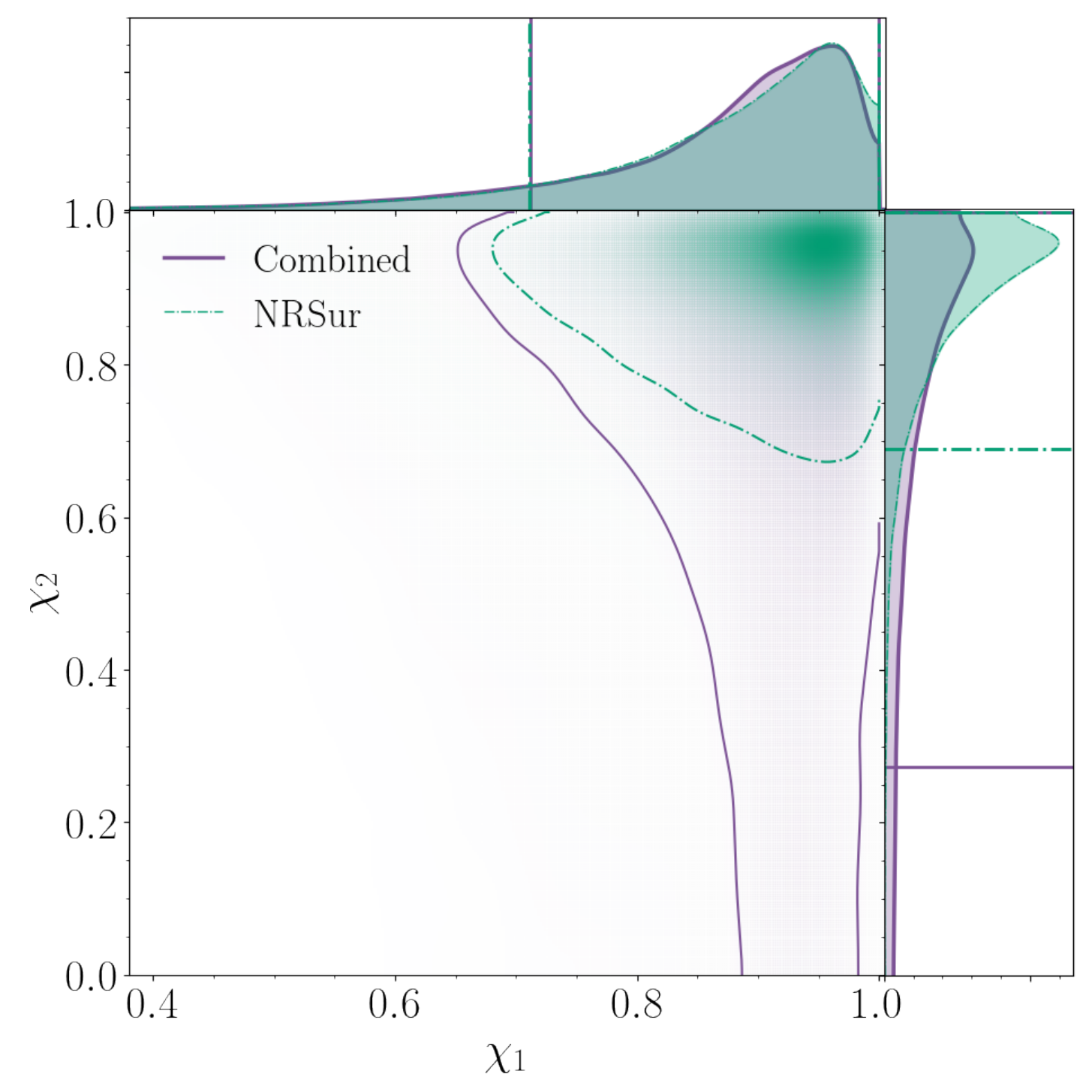}
	\caption{The posterior distribution of the primary and secondary spin magnitudes. We show the posterior distribution based on the combined samples (purple) and from the NRSur7dq4 waveform model (NRSur, green dash dot). Each contour, as well as the colored horizontal and vertical lines, shows the 90\% credible intervals.}
    \label{fig:source_spin_inference}
\end{figure}

We consistently infer that both \acp{BH} are highly spinning independent of the model we use. As shown in Figure~\ref{fig:source_spin_inference},
we infer that the primary spin magnitude $\chi_{1} \geq 0.7$ at $\pspinoneabovezeroseven{GW231123cg_combined}\%$ probability and the secondary spin magnitude $\chi_{2} \geq 0.7$ at $\pspintwoabovezeroseven{GW231123cg_combined}\%$ probability, see Sec.~\ref{sec:hierarchical} for details.
The primary component of GW231123 has one of the highest confidently measured \BH spins observed through \ac{gw}s~\citep[evidence for highly spinning \acp{BH} has also been presented in][]{Hannam:2021pit,Nitz:2020oeq,LIGOScientific:2021usb,LIGOScientific:2021vkt,Wadekar:2023gea,Williams:2024tna}.

We are unable to reliably infer the spin orientation of the binary; we infer polar angles between each spin vector and the orbital angular momentum that vary not only between models, but also when independently analysing data obtained by LIGO Livingston compared to LIGO Hanford, see Appendix~\ref{sec:pe_appendix}. In an attempt to understand these differences we carried out a series of analyses with different frequency ranges. We found that all models consistently infer greater support for spin components aligned with the orbital angular momentum, and no sign of systematics, when independently analysing LIGO Hanford data, and when excluding data from LIGO Livingston below 50\, Hz. However, as with the systematics issues discussed in Section~\ref{sec:systematics} and Appendix~\ref{sec: systematics appendix}, we were not able to conclusively reproduce this behaviour with injections of mock signals, and did not find any significant noise features below $50\, \mathrm{Hz}$ that could be the cause, although we did not perform a detailed study of Gaussian noise fluctuations or low-SNR glitches. When spin misalignment is inferred, we are unable to conclusively constrain the spin orientation away from aligned.

The uncertainty in the spin misalignment affects the inferred effective inspiral spin $\chi_{\mathrm{eff}}$, which parameterizes the spin aligned with the orbital angular momentum~\citep{Santamaria:2010yb,Ajith:2009bn}. Negative $\chi_{\mathrm{eff}}$ would imply that at least one spin is misaligned with the orbital angular momentum by more than ninety degrees. We cannot rule out $\chi_{\mathrm{eff}} < 0$, but there is an
$\pchieffabovezero{GW231123cg_combined}\%$ probability that $\chi_{\mathrm{eff}}$ is
positive.
The inferred effective precessing spin~\citep{Schmidt:2014iyl} is consistently measured between models and deviates from the prior,
$\chi_{\mathrm{p}} = \chipmed{GW231123cg_combined}^{+\chipplus{GW231123cg_combined}}_{-\chipminus{GW231123cg_combined}}$. Although we infer variation between models, we consistently obtain large Bayes factors ($10^{3}:1-10^{8}:1$) in favor of the precessing hypothesis compared to the spin-aligned hypothesis (spins aligned with the orbital angular momentum). Since the distribution of Bayes factors from noise alone is unknown, we additionally quantify the evidence for precession in GW231123 by computing the precession SNR, $\rho_\mathrm{p}$~\citep{Fairhurst:2019srr,Fairhurst:2019vut}. In the absence of any precession in the signal, we expect $\rho_{\mathrm{p}} < 2.1$ in 90\% of cases. We infer an SNR of $\rho_{\mathrm{p}} = \snrprecessionmed{GW231123cg_combined}^{+\snrprecessionplus{GW231123cg_combined}}_{-\snrprecessionminus{GW231123cg_combined}}$. Although the high SNR tail is consistent with the large Bayes factors~\citep{Green:2020ptm,Pratten:2020igi}, we infer non-negligible support below $\rho_{\mathrm{p}} = 2.1$. We are therefore unable to confidently claim precession in GW231123.
GW190521 was also found to exhibit mild evidence for spin-precession~\citep{LIGOScientific:2020iuh,LIGOScientific:2020ufj}.

We observe significant differences in the inferred luminosity distance and
inclination angle of GW231123's source, depending on the model, although we repeatedly infer
nearly symmetric distributions for the inclination angle around $\pi / 2\, \mathrm{rad}$ for all
models except {\textsc{XO4a}}.
We also infer substantial variation in the detector-frame quantities, despite seeing agreement between several models in the source-frame parameters. See Appendix~\ref{sec: systematics appendix} for a detailed discussion. 
Owing to disagreements in the inferred inclination angle of the binary, we similarly observe differences in the inferred SNRs in each higher-order multipole obtained by each model.
Following the methodology in \citet{LIGOScientific:2020stg,LIGOScientific:2020zkf}, where for each multipole the {\textsc{IMRPhenomXHM}} signal model~\citep{Garcia-Quiros:2020qpx} is used to remove any contribution parallel to the dominant multipole and to calculate the orthogonal optimal SNR~\citep{Mills:2020thr}, we nevertheless find that all models provide support for the
$(\ell, m) = (3, 3)$ multipole in GW231123. We infer an average orthogonal optimal SNR of $\snrthreethreemultipolemed{GW231123cg_combined}$ when combining the results from all models with equal weight.

The properties of the remnant \BH are estimated in different ways depending on the
model. We apply the
{\sc{NRSur7dq4Remnant}} model~\citep{Varma:2019csw} to the samples obtained by
{\sc{NRSur}},
and we average several fits calibrated to numerical relativity
simulations~\citep{Hofmann:2016yih,Healy:2016lce,Jimenez-Forteza:2016oae} for samples obtained by {\sc{v5PHM}}, {\sc{TPHM}}, {\sc{XPHM}}, {\sc{XO4a}}.
When combining the results with equal weight, we infer the final mass
and spin of the remnant \BH to be
 $M_{\mathrm{f}} = \finalmasssourcemed{GW231123cg_combined}^{+\finalmasssourceplus{GW231123cg_combined}}_{-\finalmasssourceminus{GW231123cg_combined}} \, M_{\odot}$ and $\chi_{\mathrm{f}} = \finalspinmed{GW231123cg_combined}^{+\finalspinplus{GW231123cg_combined}}_{-\finalspinminus{GW231123cg_combined}}$ respectively.
  For certain binary configurations, the remnant
 \BH may receive a recoil velocity that is enough to eject the remnant from its host
 galaxy~\citep{Merritt:2004xa}.
We infer a measurement of the remnant \BH's recoil velocity that differs from the effective prior distribution:
$v_{\mathrm{f}} = \finalkickmed{GW231123cg_nrsur}^{+\finalkickplus{GW231123cg_nrsur}}_{-\finalkickminus{GW231123cg_nrsur}} \,\mathrm{km}\,\mathrm{s}^{-1}$. This measurement is based on the {\sc{NRSur}} analysis and the {\sc{NRSur7dq4Remnant}} remnant model, the only fit providing recoil velocities estimates~\citep{Varma:2019csw,Varma:2020nbm}.

\subsection{Waveform Consistency Checks}
\label{sec:waveformconsistency}
To further assess whether a CBC signal with the inferred parameters in Section~\ref{sec:inference} adequately represents the data, 
we perform several consistency checks using a signal-agnostic approach that reconstructs coherent transient power, and through a model incorporating a modified wave dispersion relationship.
First, we compare the waveform of the maximum-likelihood sample from parameter estimation in Section~\ref{sec:pe-methods} to one obtained through minimally modelled analyses that make no assumptions about the source or morphology of the signal~\citep{Szczepanczyk:2020osv,Salemi:2019uea,Ghonge:2020suv}. Second, we conduct a residuals analysis, subtracting the best-fit waveform from the detector data and searching for coherent residual power~\citep{LIGOScientific:2021sio}. Discrepancies between the modelled and minimally modelled waveforms, or the presence of significant excess residual power, could indicate physical effects in addition to or alternative to those in our BBH signal models or unaccounted-for noise features ~\citep{Johnson-McDaniel:2021yge}.

For the waveform reconstruction comparisons, we use BayesWave~\citep{Cornish:2014kda, Cornish:2020dwh, Chatziioannou:2021ezd}, \cwb-2G~\citep{Klimenko:2008fu, Klimenko:2015ypf, Drago:2020kic}, and \cwb-BBH~\citep{Mishra:2024zzs} for the minimally modelled analysis. To evaluate the agreement between the signal as found by the modelled analysis and minimally modelled approach, we calculate the overlap between the maximum-likelihood sample from parameter estimation using the \textsc{NRSur} model and the median BayesWave or \cwb maximum-likelihood waveform. An overlap of 1 indicates perfect agreement, while an overlap of 0 indicates no similarity between waveforms.
To assess whether the overlaps are consistent with signals of comparable parameters and noise realizations, we perform a dedicated set of injections wherein we inject waveforms generated by draws from the posterior distribution of the source parameters into detector data surrounding the event. The BayesWave analysis performed 400 injections into approximately 8 hours of data surrounding the event, and the \cwb analysis injected about 2800 draws from the posterior distribution in an interval of two weeks around the event. 
We find good agreement between the minimally modelled and CBC waveform reconstructions.  The overlaps between the CBC maximum-likelihood waveform and BayesWave, \cwb-2G, and \cwb-BBH are 0.97, 0.96, and 0.98, respectively.  Compared to the distributions of overlaps from the injections, the $p$-values (defined as the fraction of injections with overlaps below that of the real event) are 0.74, 0.57, and 0.92 for BayesWave, \cwb-2G, and \cwb-BBH, respectively. Under the hypothesis that the overlaps between the BayesWave and cWB reconstructions with the maximum-likelihood waveform are drawn from the same distribution as the injection overlaps, the $p$-values should be distributed uniformly from 0 to 1, so these results indicate that the overlaps are consistent with expectations from systems similar to GW231123.

For the residuals test, we produce residual data by subtracting from the original data the maximum-likelihood waveform from the \textsc{NRSur} parameter estimation samples. If the signal has been modelled sufficiently, the residual data should be consistent with Gaussian noise.  We analyze the residual data with BayesWave, and calculate the 90\% credible upper limit on the recovered network SNR ($\mathrm{SNR}_{90}$).  To compare to expected values of $\mathrm{SNR}_{90}$, we also analyze segments of data (with no injected signal) selected randomly from 16384 s of data surrounding the event, and calculate the probability of obtaining an $\mathrm{SNR}_{90}$ higher than that of residual data. Details of this procedure can be found in~\citep{LIGOScientific:2021sio}.  We find no significant excess SNR in the residual data beyond what is expected from only Gaussian noise. 
Compared to the distribution of $\mathrm{SNR}_{90}$ from the noise-only runs, the $p$-value is 0.35.
This further confirms that minimally modelled tests do not flag any features in the data missed by the analyses described in Section~\ref{sec:pe-methods}.

We additionally search for post-ringdown echo signals~\citep{Tsang:2018uie,Tsang:2019zra}
with a BayesWave-based search, finding negative evidence for their presence (as quantified by the Bayes factor $\log_{10} \mathcal{B}^{\mathrm{signal}}_{\mathrm{noise}} < 0$), consistent with the above findings.
As a final consistency check, an analysis incorporating a modified wave dispersion relation due to non-zero graviton mass~\citep{LIGOScientific:2021sio} yields agreement with massless wave propagation when based on the \textsc{NRSur} or \textsc{TPHM} models.
Instead, when assuming the \textsc{XPHM} or \textsc{XO4a} templates, a statistically significant violation is found, suggesting missing signal components not captured by these models, which is consistent with the discussion of systematics in Section~\ref{sec:systematics}.

\section{Black-hole ringdown} \label{sec:ringdown}

Massive systems dominated by merger-ringdown, such as GW231123, are ideal to test the BH signal interpretation by applying BH spectroscopy techniques~\citep{Detweiler:1980gk,Dreyer:2003bv,Berti:2005ys, Berti:2009kk,Berti:2025hly}, yielding remnant properties under minimal assumptions on the remnant formation process.
We fit superpositions of damped sinusoids, aiming to associate them with characteristic quasi-normal modes (QNMs) of a BH, which drive its relaxation to equilibrium.
In principle, the resulting parameter estimates make it possible to robustly validate \ac{IMR} measurements, since a QNM description is generic to any BH remnant (e.g., a BH formed from an eccentric binary).

%
We truncate portions of data in the time domain at different analysis start times $t_{\rm start}$, and fit two sets of models.
The first (\texttt{DS-$N$}) is a superposition of $N$ damped-sinusoids
$ \sum_{j=1}^N A_j e^{i [ 2\pi f_j (t-t_{\rm start})+\phi_j ] } e^{- (t-t_{\rm start}) / \tau_j}$,
with constants $A_j, \phi_j, f_j, \tau_j$ as free parameters, assuming $f_j > 0$ (circularly polarized wave). 
In the second (\texttt{Kerr}), complex frequencies are identified with QNMs of a Kerr BH, $f_i = f_{\ell m n}(M^{\rm det}_{\rm f}, \chi_{\rm f})$ and $\tau_i = \tau_{\ell m n}(M^{\rm det}_{\rm f}, \chi_{\rm f})$, with detector frame (redshifted) mass $M^{\rm det}_{\rm f} $, spin $\chi_{\rm f}$, and ${\ell m n}$ the QNM angular ($\ell, m$) and overtone ($n$) indices.
In addition to the longest-lived ${\ell m n} = 220 $, we consider ${\ell m n} = \{221, 210, 200, 330, 320, 440 \}$, the linear QNMs with the largest predicted amplitudes for binary mergers~\citep{Kamaretsos:2012bs, London:2014cma, Cheung:2023vki, Zhu:2023fnf, Nobili:2025ydt, Carullo:2024smg}.
Here, we include both $\pm f_{\ell m n}$ contributions, accommodating generic signal polarizations, and $M^\mathrm{det}_{\rm f}$ enters the expression as mode amplitudes $A_{\ell m n}$ are degenerate with the source distance.

We use the \texttt{pyRing} pipeline~\citep{Carullo:2019flw} with standard analysis settings~\citep{Isi:2021iql, LIGOScientific:2021sio, Gennari:2023gmx}. We pre-condition the data by subtracting the $60$ Hz power line, which reduces the required analysis duration to $T=0.2 \,$s~\citep{Siegel:2024jqd}.
We sample the posterior distribution using the \texttt{CPNest} nested sampling algorithm~\citep{cpnest099}.
Times are relative to $t^{\rm pol}_{\rm peak} := \max_t [h_+^2(t) + h_{\times}^2(t)] = 1384782888.5998$s in the LIGO Hanford data, compatible with the median of the polarizations peak time from the \textsc{NRSur} reconstructed waveform, subject to an uncertainty $\mathcal{O}(0.01)$s.
The sky location is fixed to a value compatible with the \textsc{NRSur} maximum likelihood value, fixing the analysis start time in LIGO Livingston, as required by the truncated time-domain formulation of the analysis~\citep{Isi:2021iql}.
We have verified that repeating the analysis at different sky location values drawn from the \textsc{NRSur} posterior does not affect our conclusions.

%
Damped sinusoids are expected to be valid in the stationary regime of BH relaxation, typically $[10,20] G M^{\rm det}_{\rm f}  / c^3$ (namely $[14.7, 29.4] \, $ms assuming $M^{\rm det}_{\rm f} \simeq 298 M_{\odot}$) past the signal peak; fitting earlier may provide spurious support for additional modes~\citep{Berti:2025hly}.
However, a time-domain waveform reconstruction of GW231123 indicates a highly complex morphology, displaying a monotonic decay only after $t^{\rm strain}_{\rm peak} \simeq  t^{\rm pol}_{\rm peak} + 19$ ms.
This is significantly later than the nominal $t^{\rm pol}_{\rm peak}$, after which monotonic decay is expected for vanilla signal morphologies; see Appendix C.
Given this uncertainty, we explore a wide range of times $t_{\rm start} = t^{\rm pol}_{\rm peak} + [-7.4, 58.7] \, $ms in steps of $\approx 3 \, $ms.

\begin{figure}[t!]
	\includegraphics[width=0.5\textwidth]{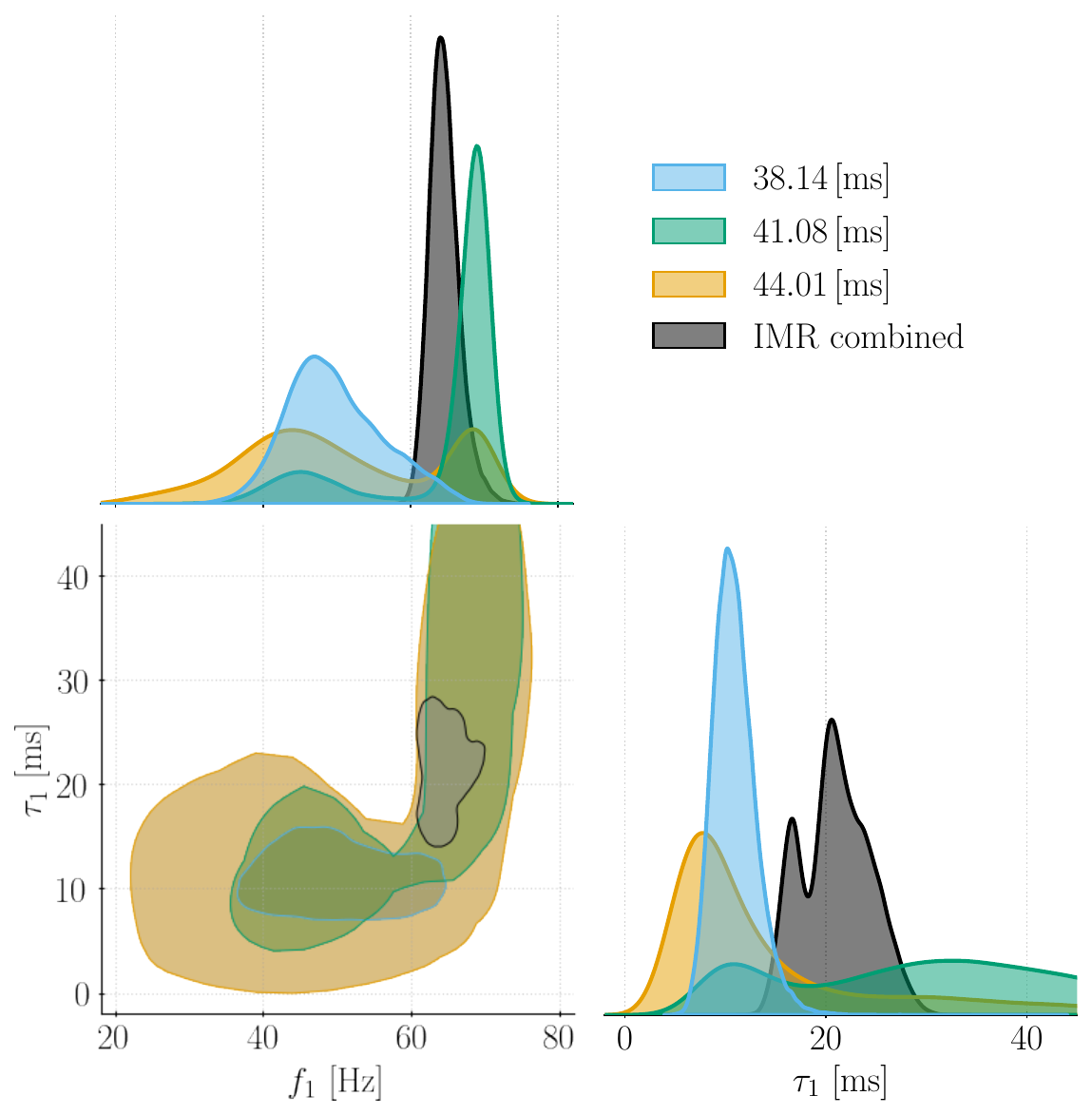}
	\caption{
	Two-dimensional frequency and damping time posterior distribution ($90\%$ credible levels), when starting the analysis at late times and assuming a single damped sinusoid with positive frequency; the combined IMR estimates for the longest-lived ${\ell m n} = {220}$ QNM are shown for comparison. For visualisation purposes, we display up to $\tau_1 = 45 $ms, while the posterior tail extends up to $\tau_1 = 80 $ms.
	}
    \label{fig:ringdown_properties}
\end{figure}

%
Fits that start at late times $t^{\rm start} \simeq t^{\rm pol}_{\rm peak} + 41 \, $ms, when we are confident on the validity of an exponential decay description, find preference for a single mode with both models (as quantified by Bayes factors $\mathcal{B}$).
A single damped sinusoid (\texttt{DS-1}) fit yields a multi-modal $f_1-\tau_1$ distribution, shown in Figure~\ref{fig:ringdown_properties}, unlike what is observed in previous events at such late times.
One peak with $f_1 \approx 68 \, $Hz and $A_1 \approx 2 \times 10^{-22}$ overlaps with the dominant Kerr $\ell m n = 220$ frequency $f_{220}$ as predicted by IMR models, supporting the BH hypothesis.
The damping time spans a broad range, overlapping with $\tau_{220}$.
A second peak is centred around $f_1\approx 45 \, $Hz, correlating with a larger amplitude value $A_1 \approx 6 \times 10^{-22}$.
Among linear Kerr QNMs predicted by the IMR models, this frequency peak shows the largest overlap with $f_{200}$.
The damping time is also bimodal: the peak associated to $f_1 \approx 45 \, $Hz is centred around $\tau_1 \approx 10 \, $ms, a smaller value compared to $\tau_{200} \approx 18 \, $ms, but overlapping the latter distribution.
Under a \texttt{Kerr} $2 2 0$ fit, the two-dimensional $M^{\rm det}_{\rm f}$ -- $\chi_{\rm f}$ distribution is also multi-modal: one peak overlaps with the IMR predictions, while a second prefers lower remnant spins.

%
Fitting at earlier times, Bayes factors indicate overwhelming preference ($\log_{10}\mathcal{B} > 6$) for two modes over one until $t_{\rm start} \simeq t^{\rm pol}_{\rm peak} + 32.3 \, $ms in both the \texttt{DS-$N$} and \texttt{Kerr} models, but in this range we may be fitting a complex merger signal, and a QNM superposition may not be valid.
%
%
At these early times, a \texttt{DS-2} fit yields two frequencies consistent with the two peaks observed at later times.
Amplitudes and damping times are comparable in magnitude between the two damped sinusoids, and both damping times are larger than the $\approx 10 \, $ms peak observed at later times.
In addition to the $220$ mode, in the time range explored we find the \texttt{Kerr} $320, 210, 200$ modes to be on average the most favoured by Bayes factors, while $330$ is preferred around $t_{\rm start} \simeq t^{\rm pol}_{\rm peak} + 23.5 \, $ms.
The \texttt{Kerr} two-modes combinations robustly yield a massive remnant, $M^{\rm det}_{\rm f} \gtrsim 200 M_{\odot}$, also at these earlier times.

These multi-mode combinations are in tension with IMR analyses.
The most favoured \texttt{Kerr} mode in addition to the $220$ according to Bayes factors, the $320$, implies a $M^{\rm det}_{\rm f}$ -- $\chi_{\rm f}$ distribution that does not overlap with IMR estimates.
Adding $210$ only results in partial overlap, while the $221$ overlaps to a larger degree.
However, the short-lived $221$ mode alone is not expected to give rise to the features observed in the signal until late times.
The $200$ mode addition results in the most significant overlap, with a $200$ amplitude comparable to the $220$, consistently with later times results.
While $\ell m n = 210, 200$ QNMs can be strongly excited in highly precessing systems with large mass asymmetry~\citep{OShaughnessy:2012iol, Zhu:2023fnf, Nobili:2025ydt}, the IMR analyses of GW231123 predict minimal power in the $200$ mode, as discussed in Appendix D.
Highly eccentric configurations can excite $m=0$ modes~\citep{Sperhake:2007gu}, but we lack sufficiently complete merger--ringdown models for eccentric-binary signals to reliably assess this possibility; see Section~\ref{sec:alternatives} for further discussion.
\texttt{DS-$N$} analyses with $N>3$ did not prefer more than two modes, but future investigations will be required to determine whether the observed features can be induced by a superposition of many overlapping modes.

%
In summary, the \texttt{Kerr} fits recover the remnant spin with large uncertainty, and they robustly predict $M^{\rm det}_{\rm f} \gtrsim 200 M_{\odot}$ at all times, supporting the interpretation of a massive BH remnant. 
Further investigations will be required to characterize the nature of the bi-modal features persistently observed in the signal and consistently interpret the multi-mode fits at earlier times.
Given these large uncertainties, we do not consider tests of the no-hair properties of BHs in general relativity that would be enabled by a confident two-mode identification~\citep{Detweiler:1980gk,Dreyer:2003bv,Berti:2005ys, Berti:2009kk,Gossan:2011ha,Brito:2018rfr,Carullo:2018sfu,Bhagwat:2019bwv,Isi:2021iql,Berti:2025hly}.

\section{Astrophysical implications}
\label{sec:astro}

Here, we discuss the astrophysical implications of the large masses and spins of GW231123's source and its possible origin given current understanding of (pulsational) \acp{PISN} and formation channels of merging \acp{bbh}.

\subsection{Single-event rate estimate}
\label{sec: Single-event rate estimate}

First, we quantify the merger rate of GW231123-like events following \citep{Kim:2002uw, LIGOScientific:2016kwr}.
The sensitive volume--time of the detectors to signals whose source properties are consistent with the posterior distribution of GW231123 is estimated using the \cwb-BBH results for the injection campaign in Section~\ref{sec:searches}. Assuming a constant merger rate $\mathcal{R}$ over comoving volume and source-frame time with prior $\propto1/\sqrt{\mathcal{R}}$ and a Poisson likelihood for the number of triggers,
we find $\mathcal{R} = 0.08_{-0.07}^{+0.19} \,\mathrm{Gpc}^{-3}\,\mathrm{yr}^{-1}$.
This is consistent with the rate of mergers like GW190521 ($0.13_{-0.11}^{+0.30}\,\mathrm{Gpc}^{-3}\,\mathrm{yr}^{-1}$; \citealt{LIGOScientific:2020iuh, LIGOScientific:2020ufj}) and upper limits of IMBH mergers (e.g., even the most stringent 90\% upper limit $<0.06\,\mathrm{Gpc}^{-3}\,\mathrm{yr}^{-1}$ from \citealt{LIGOScientific:2021tfm} and see Table~3 therein for constraints across source properties), but lower than the overall rate of BBHs with component masses $<100\,M_\odot$ inferred through GWTC-3 (16--61\,Gpc$^{-3}$\,yr$^{-1}$; \citealt{KAGRA:2021duu}).

\subsection{Relation to the previous inferred population}
\label{sec: population}

To further assess GW231123 in the context of the 69 \ac{bbh} mergers with $\mathrm{\acp{far}}<1\,\mathrm{yr}^{-1}$ through GWTC-3 \citep{LIGOScientific:2021vkt} and test if its masses and spins are surprising, we perform posterior predictive checks based on the \textit{fiducial} \ac{bbh} population fit from \citep[][Section~III\,C; we extend the prior on the maximum \ac{BH} mass up to $200\,M_\odot$ as there is support from GW190521 above the limit of 100\,$M_\odot$ imposed in the original GWTC-3 analysis]{KAGRA:2021duu}.
From the inferred population, we construct mock catalogs containing 69 detected events
and plot the distribution of their largest \ac{BH} mass $\approx83_{-26}^{+43}\,M_\odot$ in Figure~\ref{fig:source_mass_inference}.
Using the combined parameter estimates in Section~\ref{sec:inference}, the primary mass of GW231123 falls at the $98_{-5}^{+2}\,\%$ level of this distribution, indeed indicating that this event is an unlikely draw. However, due to large uncertainties in its masses, it is not conclusively an outlier as it may be less massive than the most massive mock events (equivalent comparisons for secondary and total mass are less significant).
Similarly, the largest BH spin in these catalogs is $0.78_{-0.14}^{+0.14}$, against which the primary and secondary spins of GW231123
can fall at any percentile and thus are not outliers.
Compared to its masses, the spins of GW231123's source are more consistent with the known population as it does not rule out large values.

\subsection{Possible formation channels}
\label{sec: Possible formation channel}

From theoretical predictions for the late-stage evolution of massive stars \citep{Fowler:1964zz, Barkat:1967zz, 1967ApJ...148..803R, 1968Ap&SS...2...96F, Bond:1984sn, Woosley:2002zz, Woosley:2016hmi}, contraction of the core leads to electron--positron pair production that reduces internal pressure support, causing further contraction that powers explosive nuclear burning and a rebounding shock.
For helium-core masses $\approx32$--$64\,M_\odot$,
multiple pulsational episodes can eject sufficient material to reduce the mass below the pair-instability regime, ending with a BH remnant.
A single pulse can entirely disrupt stars with larger helium cores, leaving behind no remnant in a \ac{PISN}.
At even larger helium-core masses $\gtrsim135\,M_\odot$, this is avoided as the high core temperature results in photodisintegration that accelerates gravitational collapse to a massive \ac{BH}.
This leads to the robust prediction from single-star evolution of the existence of a gap in the BH mass distribution.
Though this gap is broadly consistent with the range $\approx60$--$130\,M_\odot$, there are several theoretical uncertainties that affect
both the lower edge and total extent of the gap
\citep{Belczynski:2016jno, Stevenson:2019rcw, 2019ApJ...887...53F, Mapelli:2019ipt, Renzo:2020rzx, Marchant:2020haw, Woosley:2021xba, Hendriks:2023yrw}.
Uncertainties in nuclear reaction rates alone can shift the lower edge of the pair-instability mass gap from $\approx50\,M_\odot$ to $\approx100\,M_\odot$ \citep{Farmer:2020xne}.

Some stellar and binary evolution processes
are predicted to be able to
populate the
pair-instability mass gap.
Weaker stellar winds \citep{Mapelli:2019ipt} or core dredge-up episodes \citep{Costa:2020xbc} may allow a star to retain a hydrogen envelope and collapse to a BH with mass inside the gap \citep{Spera:2018wnw}. Short-period stellar binaries might avoid merging and produce binary BHs with large, equal masses $\sim100\,M_\odot$ from rapidly rotating metal-poor stars due to chemically homogeneous evolution \citep{deMink:2016vkw, Mandel:2015qlu, Marchant:2016wow}.
However, most models of isolated-binary formation predict small natal spins and at most one of the BHs spinning, due to tidal synchronization or accretion-induced spin up, and so binaries with masses and spins like those inferred from GW231123 are difficult to form \citep{Belczynski:2017gds, Qin:2018vaa, Fuller:2019sxi, Bavera:2020inc, Belczynski:2017gds, vanSon:2020zbk}.
In fact, the components BHs and especially the primary are so massive that they may have formed through core collapse above the pair-instability mass gap \citep{Ezquiaga:2020tns, Franciolini:2024vis}.

Alternatively, in hierarchical mergers, one or both of the binary components is the product of a previous \ac{bbh} merger, with characteristically large masses and spins \citep{Gerosa:2021mno}.
As seen in Figure~\ref{fig:source_spin_inference}, the spins inferred from GW231123 may be even larger than typically predicted from hierarchical \ac{BH} mergers \citep{Gerosa:2017kvu, Fishbach:2017dwv}, although the expected distribution for sources retained in their host environments may accommodate a wider range of spins \citep{Borchers:2025sid}.
Previous analyses have suggested evidence for hierarchical mergers in GW catalogs \citep{Kimball:2020qyd, Mould:2022ccw, Wang:2022gnx, Li:2023yyt, Pierra:2024fbl, Hussain:2024qzl, Antonini:2024het}, but the population of BH remnants receive gravitational recoils as high as $10^2$--$10^4\,\mathrm{km}\,\mathrm{s}^{-1}$ \citep{Doctor:2021qfn, Mahapatra:2021hme}, requiring environments with high escape speeds \citep{Antonini:2016gqe} such as dense stellar clusters \citep{Miller:2001ez, Antonini:2018auk, Rodriguez:2019huv, Fragione:2020nib, Mapelli:2020xeq, Sedda:2021abh, Kritos:2022non, Mahapatra:2022ngs} or \ac{AGN} disks \citep{Bartos:2016dgn, Stone:2016wzz, Mckernan:2017ssq, Yang:2019cbr, Tagawa:2019osr, McKernan:2019beu, Vaccaro:2023cwr, Sedda:2023big} to be retained.

This is in contrast to
stellar mergers, which receive smaller recoils from asymmetric mass loss \citep{Gaburov:2009kg, Glebbeek:2013kga} and therefore may be more efficient at producing \acp{BH} with large masses in dynamical environments. The large masses inferred from GW231123 may be explained by multiple such mergers in dense clusters or multiple systems \citep{Mapelli:2016vca, DiCarlo:2019fcq, Renzo:2020smh, Kremer:2020wtp, Gonzalez:2020xah, Rizzuto:2021atw, Costa:2022aka, ArcaSedda:2023mlv}, a scenario that could also describe the formation of massive central \acp{BH} \citep{PortegiesZwart:2004ggg, Greene:2019vlv}.

Besides mass transfer between the components of a stellar binary, \ac{BH} mass growth may also occur via accretion in other gaseous environments. \acp{BH} embedded in the disk of an \ac{AGN} may accrete material directly from the disk or from collisions with disk stars \citep{McKernan:2012rf}. Similarly, in dense clusters, \acp{BH} may accrete from stars after undoing dynamical interactions \citep{Giersz:2015mlk, Lopez:2018nkj, Kiroglu:2024xpc}. Furthermore, these accretion processes may also increase \ac{BH} spins.

A different possibility is that of primordial \acp{BH} being the binary components, which may exist across a range of mass scales, including within the pair-instability mass gap \citep{Bird:2016dcv, Bird:2022wvk,Clesse:2016vqa,Clesse:2020ghq}. However, there are remaining theoretical uncertainties, e.g., on whether primordial \acp{BH} could accrete sufficiently to spin up as rapidly as the \acp{BH} inferred from GW231123 \citep{Green:2020jor}.

Altogether, these theoretical predictions and their uncertainties make it difficult to determine whether or not the \acp{BH} in the
source of GW231123 have an astrophysical origin directly from stellar collapse.
We quantify this in more detail below.

\subsection{Stellar collapse}

\begin{figure}
	\includegraphics[width=\columnwidth]{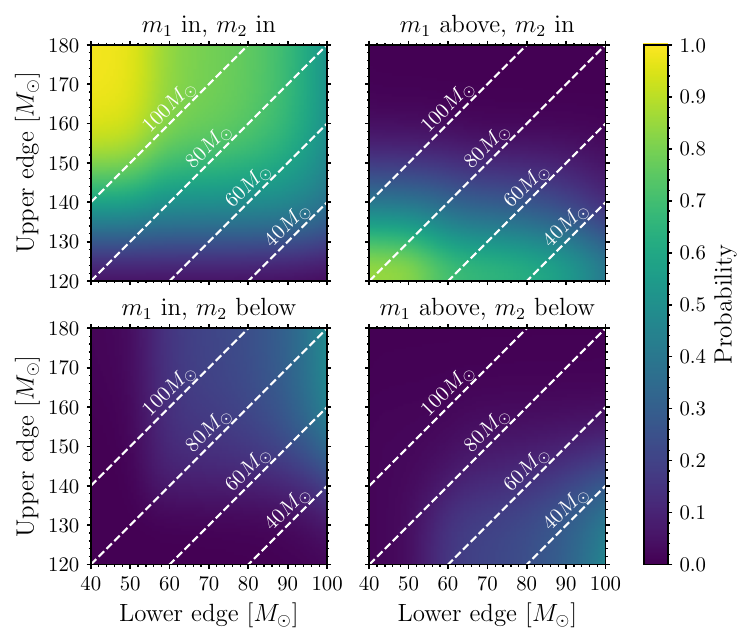}
	\caption{Probability using the combined parameter estimates for GW231123 that: both BHs are in the pair-instability mass gap (top left); the primary is above the gap and the secondary is within (top right); the primary is within and the secondary is below (bottom left); the primary is above and the secondary is below (bottom right). Probabilities are computed varying the lower and upper edges of the gap, while dashed lines mark constant gap widths.}
	\label{fig: gap-probability}
\end{figure}

To account for a range of possible locations for the pair-instability mass gap, in Figure~\ref{fig: gap-probability} we compute the probability that one or both component masses fall within the gap as a function of its lower edge from 40--100\,$M_\odot$ and upper edge from 120--180\,$M_\odot$, using the combined parameter estimates. In the following, we quote these probabilities specifically for the putative gap 60--130\,$M_\odot$. The probabilities that the secondary (primary) \ac{BH} lies in, above, and below this gap are 83\,\% (28\,\%), 1\,\% (72\,\%), and 16\,\% (0\,\%), respectively. Considering scenarios in which at least one of the components falls in this gap, the joint probability that: both \acp{BH} are in the gap (upper left panel of Figure~\ref{fig: gap-probability}) is 26\,\%; the primary is above while the secondary is within (upper right) is 57\,\%; and that the primary is within while the secondary is below (lower left) is 2\,\%. Alternatively, a scenario with neither \ac{BH} in the gap is possible in the case of a straddling binary \citep{Fishbach:2020qag}, with a primary \ac{BH} above the upper edge and secondary below the lower edge (lower right) having a probability of 14\,\%.

Overall, this implies that within the uncertainties on the combined parameter estimates (assuming our default prior) and the location of the pair-instability mass gap, scenarios with both \acp{BH} outside the gap have lower probability than those with at least one \ac{BH} in the gap.

\subsection{Hierarchical mergers}
\label{sec:hierarchical}

Given the high probability of at least one of the \acp{BH} lying inside the pair-instability mass gap, we consider the possibility that this is due to repeated \ac{bbh} mergers.
Assuming hierarchical origins, several works have inferred
the source properties of potential \acp{bbh} whose merger products are observed with GWs in a subsequent merger
\citep{Baibhav:2021qzw, Barrera:2022yfj, Alvarez:2024dpd, Mahapatra:2024qsy}.
We follow \citep{Alvarez:2024dpd} and use the \textsc{NRSur7dq4Remnant} surrogate model \citep{Varma:2019csw}
to find the distribution of \ac{bbh} source properties such that the corresponding distribution of \ac{BH} remnant properties reproduces the combined parameter estimates over mass and spin for the primary and secondary \ac{BH} inferred from GW231123.
As the primary-spin posterior favors large values $\gtrsim0.7$, this constrains the parent binary of the primary BH to have unequal masses $105_{-29}^{+24}\,M_\odot$ and $38_{-17}^{+33}\,M_\odot$; for more equal masses, both BH spins can reduce the total angular momentum if misaligned, whereas unequal-mass binaries are dominated by the single heavier BH. The parent binary of the primary BH may have had a large effective inspiral spin, with $\chieff=0.55_{-0.60}^{+0.25}$, but $\chieff\lesssim0$ is not ruled out. A similar picture holds for the secondary BH, with parent masses $73_{-37}^{+26}\,M_\odot$ and $25_{-15}^{+28}\,M_\odot$, but more uncertain effective inspiral spin $\chieff=0.29_{-0.88}^{+0.47}$ due to the larger uncertainty on the secondary spin in the source of GW231123.
These mergers would have imparted kicks of $749_{-630}^{+1320}\,\mathrm{km}\,\mathrm{s}^{-1}$ and $494_{-363}^{+1410}\,\mathrm{km}\,\mathrm{s}^{-1}$ in the case of the primary and secondary, respectively,
resulting in ejection from environments with escape speeds $\lesssim100\,\mathrm{km}\,\mathrm{s}^{-1}$, such as young star clusters or globular clusters
\citep{Antonini:2016gqe}.

The heavier of the two \acp{BH} in both parent binaries may also lie within the pair-instability mass gap, with probabilities 96\,\% and 71\,\% for the heavier parent of the primary and secondary, respectively, when taking a gap from 60--130\,$M_\odot$, as above. Therefore, if
either of the component \acp{BH} of GW231123's source
is interpreted as the product of a previous \ac{BH} merger, it may be the result of multiple previous mergers or require the components of the parent binary to have formed with larger masses via other astrophysical processes, such as stellar mergers or \ac{BH} accretion, as discussed in Section~\ref{sec: Possible formation channel}.

\section{Alternative interpretations}
\label{sec:alternatives}

All GW observations to date have been inferred to be from compact binaries consisting of \acp{BH} and/or neutron
stars~\citep{LIGOScientific:2016dsl,LIGOScientific:2020ibl,LIGOScientific:2021usb,LIGOScientific:2021vkt,Venumadhav:2019lyq,Olsen:2022pin,Mehta:2023zlk,Nitz:2021zwj, Wadekar:2023gea}, and we consider a BBH the most astrophysically plausible interpretation of GW231123, finding that a non-eccentric \ac{bbh} model
fits the signal with no significant residual.
Nonetheless, the low number of observable \ac{gw} cycles invites alternative interpretations. We discuss several here.

\subsection{Eccentricity}

Binaries formed in dense environments may retain residual eccentricity in the sensitive band of current \ac{gw} detectors~\citep{Antonini:2013tea, Samsing:2017xmd, Rodriguez:2017pec, Zevin:2018kzq, Chattopadhyay:2023pil, DallAmico:2023neb} or form with large eccentricities and merge promptly after due to a dynamical capture~\citep{Gold:2012tk, East:2012xq, Gamba:2021gap, Andrade:2023trh, Albanesi:2024xus},
but for high masses their \ac{gw} signals can be confused with those of non-eccentric mergers~\citep{Romero-Shaw:2020thy, CalderonBustillo:2020xms, Romero-Shaw:2022fbf}.
Our signal models assume a non-eccentric inspiral,
while state-of-the-art \ac{IMR} models that include eccentricity~\citep{Liu:2021pkr, Gamboa:2024hli, Paul:2024ujx, Albanesi:2025txj, Planas:2025feq} assume circularization in the merger--ringdown stages and would thus be unsuitable to infer the parameters of GW231123's source if it was eccentric when observed \citep{Ramos-Buades:2023yhy, Iglesias:2022xfc, Gupte:2024jfe, Planas:2025jny}.
Extensions of QNM amplitude models beyond eccentric non-spinning configurations~\citep{Carullo:2024smg} will be required to investigate the possible $m=0$ ringdown mode excitation hinted at in Section~\ref{sec:ringdown}.
Many studies have found that the merger-ringdown signal is robust with respect to moderate inspiral 
eccentricity~\citep{Hinder:2007qu, Huerta:2019oxn, Healy:2022wdn, Carullo:2023kvj, Nee:2025zdy}. 
Relaxing the non-eccentric assumption is not expected to significantly change our results unless the eccentricity is larger than 
$\sim0.6$ close to merger~\citep{Healy:2022wdn, Carullo:2023kvj}, which would be rare in the dynamical-capture scenarios above. 
For example, \cite{Chattopadhyay:2023pil} find an overall merger rate $<1\,\mathrm{Gpc}^{-3}\,\mathrm{yr}^{-1}$ in dense stellar clusters, 
$\sim$10\% with eccentricity $>0.1$ at a GW frequency of 10\,Hz and $\sim$10\% involving \acp{BH} with masses $>100\,M_\odot$, 
implying a rate of massive eccentric mergers $<0.01\,\mathrm{Gpc}^{-3}\,\mathrm{yr}^{-1}$, already at the lower limit of our 
constraint for GW231123 without considering the decline in the number of sources at increasing mass and eccentricity.
Although we do not explicitly rule out large eccentricity for the source of GW231123, we therefore consider it astrophysically unlikely.

\subsection{Gravitational lensing}

GW signals may be strongly lensed by galaxies or galaxy clusters, producing multiple copies of the original
signal~\citep{Hannuksela:2019kle,LIGOScientific:2021izm, LIGOScientific:2023bwz}.
However, no closely matching super-threshold counterpart candidates for GW231123 have been found from standard CBC searches.
GWs can also undergo wave-optics lensing~\citep{Takahashi:2003ix} when they encounter smaller objects ($\sim$$10^2$--$10^6\,M_\odot$
for signals in the LVK band).
GW231123 shows the strongest support for distorted lensed signals seen so far for both a point-mass model~\citep{Wright_2022}
and phenomenological analyses~\citep{Liu:2023ikc}, although preliminary background analyses suggest that some GW231123-like
signals may be mis-identified as lensed. More in-depth investigations are needed to assess the significance of the lensing hypothesis, and these will be presented in future work.

\subsection{Other scenarios}

Several possible burst-like sources \citep{Powell:2024bnp} of astrophysical and cosmological origin may produce signals of similar duration to GW231123, such as core-collapse supernovae, cosmic strings, and exotic compact objects.
For most supernova waveforms, the peak signal is expected at frequencies higher than observed in
GW231123~\citep{Abdikamalov:2020jzn,Mezzacappa:2024zph}.
The ringdown-dominated signals of high-mass \ac{bbh} mergers can be mimicked by waveforms from the collapse of cosmic strings \citep{LIGOScientific:2020ufj, Aurrekoetxea:2023vtp} and collisions of exotic compact objects (e.g. boson stars)~\citep{CalderonBustillo:2020fyi, Siemonsen:2023hko, Evstafyeva:2024qvp}.
Though we do not explicitly rule out these scenarios, the detection of GW231123 is consistent with the rates and properties of the currently understood population under the interpretation of a high-mass \ac{bbh} merger, which has higher astrophysical probability.

\section{Summary}
\label{sec:conclusion}

GW231123 is a short-duration \ac{gw} signal consisting of $\sim$5 observable cycles, most likely produced by a binary-black-hole
merger. On that basis, we infer a total mass between 190\,$M_\odot$ and 265\,$M_\odot$,
which is larger than any previously observed with high confidence in \ac{gw}s, and strong support for large spins on both black holes.
We report source property measurements with larger uncertainties than we would expect for a binary of this mass and
a signal with SNR $\sim$21, most likely due to uncertainties in current signal models at high spins.
A ringdown analysis also supports a massive remnant under minimal assumptions, consistent with full-signal estimates.
The measured masses
of GW231123's source lie at the edge of
the currently understood population of binary black holes.
The scenario with the highest probability is that at least one of the black hole sits in the pair-instability mass gap.
If either is interpreted as the product of a previous black-hole merger, at least one of the black holes in its parent binary
probably also lies in the mass gap.
Such a sequence of black-hole mergers would require an environment with high escape speed,
unless the black-hole masses are grown by other astrophysical processes,
such as stellar mergers.

Given the small number of observable \ac{gw} cycles, the large uncertainties in our measurements, and the limitations of
current signal models, we expect that there is much still to learn about GW231123 and its source. The feasibility of a wide range of other alternatives
to black-hole mergers remains to be investigated. Even within the binary-black-hole merger interpretation, we expect to learn more from detailed
studies of high-spin binaries, high-eccentricity mergers, hyperbolic encounters, and lensed signals.
Forthcoming analyses of the combined catalog of \ac{gw} events, alongside continued studies of pair-instability processes and the formation of intermediate-mass black holes, may help to reveal the origins of GW231123.
All studies will have to contend with the limited information
that can be extracted from short signals, but a clearer picture may emerge if a population of such signals is observed in future
observing runs.

Strain data from the LIGO detectors associated with GW231123 are available from the Gravitational Wave Open Science Center \footnote{https://doi.org/10.7935/anj7-6q40}. Samples from posterior distributions of the source parameters, additional materials, and notebooks for reproducing the figures are available on Zenodo \citep{zenodo}. The software packages used in our analyses are open-source.

\vspace{7pt}
%

\newif\ifcoreonly\coreonlytrue
\newif\ifkagra\kagratrue
\newif\ifheader\headerfalse
\ifheader
\begin{center}{\bf\Large
\ifkagra
Conference proceedings acknowledgements for \\ the LIGO Scientific Collaboration, the Virgo Collaboration and the KAGRA Collaboration
\else
Conference proceedings acknowledgements for \\ the LIGO Scientific Collaboration and the Virgo Collaboration
\fi
}\end{center}
\fi
This material is based upon work supported by NSF's LIGO Laboratory, which is a
major facility fully funded by the National Science Foundation.
The authors also gratefully acknowledge the support of
the Science and Technology Facilities Council (STFC) of the
United Kingdom, the Max-Planck-Society (MPS), and the State of
Niedersachsen/Germany for support of the construction of Advanced LIGO 
and construction and operation of the GEO\,600 detector. 
Additional support for Advanced LIGO was provided by the Australian Research Council.
The authors gratefully acknowledge the Italian Istituto Nazionale di Fisica Nucleare (INFN),  
the French Centre National de la Recherche Scientifique (CNRS) and
the Netherlands Organization for Scientific Research (NWO)
for the construction and operation of the Virgo detector
and the creation and support  of the EGO consortium. 
The authors also gratefully acknowledge research support from these agencies as well as by 
the Council of Scientific and Industrial Research of India, 
the Department of Science and Technology, India,
the Science \& Engineering Research Board (SERB), India,
the Ministry of Human Resource Development, India,
the Spanish Agencia Estatal de Investigaci\'on (AEI),
the Spanish Ministerio de Ciencia, Innovaci\'on y Universidades,
the European Union NextGenerationEU/PRTR (PRTR-C17.I1),
the ICSC - CentroNazionale di Ricerca in High Performance Computing, Big Data
and Quantum Computing, funded by the European Union NextGenerationEU,
the Comunitat Auton\`oma de les Illes Balears through the Conselleria d'Educaci\'o i Universitats,
the Conselleria d'Innovaci\'o, Universitats, Ci\`encia i Societat Digital de la Generalitat Valenciana and
the CERCA Programme Generalitat de Catalunya, Spain,
the Polish National Agency for Academic Exchange,
the National Science Centre of Poland and the European Union - European Regional
Development Fund;
the Foundation for Polish Science (FNP),
the Polish Ministry of Science and Higher Education,
the Swiss National Science Foundation (SNSF),
the Russian Science Foundation,
the European Commission,
the European Social Funds (ESF),
the European Regional Development Funds (ERDF),
the Royal Society, 
the Scottish Funding Council, 
the Scottish Universities Physics Alliance, 
the Hungarian Scientific Research Fund (OTKA),
the French Lyon Institute of Origins (LIO),
the Belgian Fonds de la Recherche Scientifique (FRS-FNRS), 
Actions de Recherche Concert\'ees (ARC) and
Fonds Wetenschappelijk Onderzoek - Vlaanderen (FWO), Belgium,
the Paris \^{I}le-de-France Region, 
the National Research, Development and Innovation Office of Hungary (NKFIH), 
the National Research Foundation of Korea,
the Natural Sciences and Engineering Research Council of Canada (NSERC),
the Canadian Foundation for Innovation (CFI),
the Brazilian Ministry of Science, Technology, and Innovations,
the International Center for Theoretical Physics South American Institute for Fundamental Research (ICTP-SAIFR), 
the Research Grants Council of Hong Kong,
the National Natural Science Foundation of China (NSFC),
the Israel Science Foundation (ISF),
the US-Israel Binational Science Fund (BSF),
the Leverhulme Trust, 
the Research Corporation,
the National Science and Technology Council (NSTC), Taiwan,
the United States Department of Energy,
and
the Kavli Foundation.
The authors gratefully acknowledge the support of the NSF, STFC, INFN and CNRS for provision of computational resources.

\ifcoreonly\else
{\bf For papers using O3b (and future) data, the following paragraph should be added for KAGRA.}\\
\fi
\ifkagra
This work was supported by MEXT,
the JSPS Leading-edge Research Infrastructure Program,
JSPS Grant-in-Aid for Specially Promoted Research 26000005,
JSPS Grant-in-Aid for Scientific Research on Innovative Areas 2402: 24103006,
24103005, and 2905: JP17H06358, JP17H06361 and JP17H06364,
JSPS Core-to-Core Program A.\ Advanced Research Networks,
JSPS Grants-in-Aid for Scientific Research (S) 17H06133 and 20H05639,
JSPS Grant-in-Aid for Transformative Research Areas (A) 20A203: JP20H05854,
the joint research program of the Institute for Cosmic Ray Research,
University of Tokyo,
the National Research Foundation (NRF),
the Computing Infrastructure Project of the Global Science experimental Data hub
Center (GSDC) at KISTI,
the Korea Astronomy and Space Science Institute (KASI),
the Ministry of Science and ICT (MSIT) in Korea,
Academia Sinica (AS),
the AS Grid Center (ASGC) and the National Science and Technology Council (NSTC)
in Taiwan under grants including the Science Vanguard Research Program,
the Advanced Technology Center (ATC) of NAOJ,
and the Mechanical Engineering Center of KEK.
\fi

Additional acknowledgements for support of individual authors may be found in the following document: \\
\texttt{https://dcc.ligo.org/LIGO-M2300033/public}.
For the purpose of open access, the authors have applied a Creative Commons Attribution (CC BY)
license to any Author Accepted Manuscript version arising.
We request that citations to this article use 'A. G. Abac {\it et al.} (LIGO-Virgo-KAGRA Collaboration), ...' or similar phrasing, depending on journal convention.

\ifcoreonly\else
{\bf For collaboration papers accepted after revisions made in response to referees, it is recommended that the 
following sentence be included: (with appropriate choice of singular or plural):}\\
We thank the anonymous journal referee(s) for helpful comments.
\fi

\ifcoreonly\else
{\bf For certain collaboration papers, it may be appropriate to acknowledge specific analysis software.
One template to consider is that used for the GW190425 discovery paper, for which the latex can be
found here:}\\
{\small https://git.ligo.org/publications/gw190425/gw190425-discovery/-/blob/master/gw190425-discovery.tex\#L259.}
\fi

\ifcoreonly\else
{\bf For collaboration papers released after March 2020, it may be appropriate to add the following special acknowledgement, BUT the journal may reject it (PRL rejected it). If one chooses, one can include such acknowledgements in arXiv/DCC versions only in that case:}\\
\fi
\ifcoreonly\else
{\it We would like to thank all of the essential workers who put their health at risk during the COVID-19 pandemic, without whom we would not have been able to complete this work.}
\fi

\software{
Calibration of the LIGO strain data was performed with a \gstlal-based calibration software pipeline~\citep{Viets:2017yvy}.
Data-quality products and event-validation results were computed using the {DMT}\xspace~\citep{DMTdocumentation}, {DQR}\xspace~\citep{DQRdocumentation}, {DQSEGDB}\xspace~\citep{Fisher:2020pnr}, {gwdetchar}\xspace~\citep{gwdetchar-software}, {hveto}\xspace~\citep{Smith:2011an}, iDQ\xspace~\citep{Essick:2020qpo}, {Omicron}\xspace~\citep{Robinet:2020lbf}, and {PythonVirgoTools}\xspace~\citep{pythonvirgotools} software packages and contributing software tools.
Analyses in this catalog relied on software from the LVK Algorithm Library Suite~\citep{lalsuite, swiglal}.
The detection of the signals and subsequent significance evaluations were performed with the \gstlal-based inspiral software pipeline~\citep{Messick:2016aqy,Sachdev:2019vvd,Hanna:2019ezx,Cannon:2020qnf}, with the \mbta pipeline~\citep{Adams:2015ulm,Aubin:2020goo}, with the \pycbc~\citep{Usman:2015kfa,Nitz:2017svb,Davies:2020tsx} packages, with \cwb-BBH pipeline ~\citep{Mishra:2024zzs}, \cwb-2G ~\citep{Klimenko:2008fu, Klimenko:2015ypf, Drago:2020kic} \cwb-XP ~\citep{Klimenko:2022nji}, \cwb-GMM ~\citep{Gayathri:2020bly, Lopez:2021ikt, Smith:2024bsn}.
Low-latency source localization was performed using BAYESTAR\xspace~\citep{Singer:2015ema}.
Estimates of the noise spectra and glitch models were obtained using BayesWave\xspace~\citep{Cornish:2014kda, Littenberg:2014oda,Cornish:2020dwh}.
Source-parameter estimation was primarily performed with the Bilby\xspace and {BilbyPipe}\xspace libraries~\citep{bilby_paper,Smith:2019ucc,Romero-Shaw:2020owr} using the \textsc{Dynesty}\xspace nested sampling package~\citep{Speagle:2019ivv}.
SEOBNRv5PHM waveforms used in parameter estimation were generated using pySEOBNR~\citep{Mihaylov:2023bkc}.
\textsc{PESummary} was used to postprocess and collate
parameter-estimation results~\citep{Hoy:2020vys}. Some of the parameter-estimation analyses
were managed with the Asimov library~\citep{Williams:2022pgn}.
Ringdown analyses were performed using the \texttt{pyRing}~\citep{pyRing} library, relying on the \texttt{CPNest} nested sampling algorithm~\citep{cpnest099}.
The manuscript content has been derived making use of additional publicly available software: 
\textsc{matplotlib}~\citep{matplotlib}, 
\textsc{numpy}~\citep{numpy}, 
\textsc{scipy}~\citep{scipy},
\textsc{seaborn}~\citep{seaborn},
\textsc{sxs}~\citep{Scheel:2025jct}.

}

\appendix
\section{Systematics studies}
\label{sec: systematics appendix}

For this analysis we consider the models \textsc{NRSur}, \textsc{v5PHM}, \textsc{TPHM}, \textsc{XPHM} and \textsc{XO4a}.
These models all describe precessing quasi-circular binaries and include higher multipole content.
The three model families \textsc{NRSur}, \textsc{SEOBNR} and \textsc{Phenom} use different approaches to model the waveforms~\citep{Chatziioannou:2024hju}. 
In short, \textsc{NRSur} interpolates between NR data~\citep{Field:2013cfa, Blackman:2015pia}, making it typically the most accurate of the models for high-mass signals, such as GW231123. 
The \textsc{SEOBNR} and \textsc{Phenom} families instead use a combination of analytical and numerical information to create a complete inspiral-merger-ringdown model applicable to systems at any total mass~\citep{Buonanno:1998gg, Buonanno:2000ef, Buonanno:2007pf, Ajith:2009bn}. 
The models \textsc{NRSur}, \textsc{v5PHM}, and \textsc{TPHM} calculate the signal in the time domain, while \textsc{XPHM} and \textsc{XO4a} model directly in the frequency domain.
These models comprise the five state-of-the-art models currently available for LVK analyses of observations in O4a.

\textsc{NRSur} is fully calibrated to numerical waveforms over the binary parameter space up to dimensionless spin magnitudes $\chi_1 = \chi_2 = 0.8$ and mass ratios $q=1/4$, and can be extrapolated up to dimensionless spin magnitudes $\chi_1 = \chi_2 = 1.0$ and mass ratios $q=1/6$. By construction, \textsc{NRSur} automatically includes all multipoles up to $\ell=4$ and characteristics of precession such as mode asymmetry~\citep{Varma:2019csw}.
By contrast, \textsc{v5PHM}, \textsc{TPHM}, and \textsc{XPHM} are calibrated to NR only in the aligned-spin sector~\citep{Pompili:2023tna, Estelles:2020twz, Pratten:2020fqn, Garcia-Quiros:2020qpx} and instead model precession either by extending post-Newtonian and effective-one-body results or by employing BH perturbation theory results through merger and ringdown. 
During the inspiral, \textsc{v5PHM}, \textsc{TPHM}, and \textsc{XPHM} implement precession dynamics by numerically evolving the spins~\citep{Khalil:2023kep, Estelles:2020osj,Colleoni:2024knd}.
\textsc{XO4a} uses closed-form, orbit-averaged expressions during the inspiral~\citep{Chatziioannou:2017tdw, Pratten:2020ceb} and phenomenological expressions calibrated to single-spin precessing simulations with $\chi_1<0.8$ through merger and ringdown~\citep{Hamilton:2021pkf}. Further, \textsc{XO4a} includes mode asymmetry of the dominant multipole~\citep{Ghosh:2023mhc}.

The different modeling approaches and treatments of the precession dynamics make these models relatively independent. In the presence of features in the data beyond the physical effects incorporated in the models (e.g., mismodelling in the high-spin regime, eccentricity, \ac{gw} memory, or noise artefacts) one might therefore expect the models to interact with these features differently and display model systematics, as are seen in the posteriors for this event. 
The accuracy of these models for typical signals has been comprehensively assessed through comparison to NR, both in the modelling papers themselves and elsewhere \citep[e.g., []{MacUilliam:2024oif}.
For GW231123 we have performed the accuracy analysis in Section~\ref{sec:systematics}, and a series of targeted NR injections, which we now describe. We hope that more can be learned in the future from improved models in the high-spin regime, and a detailed study of the behaviour of our models in Gaussian noise.

In order to investigate the likelihood of the presence of waveform systematics in the high total mass, comparable-mass ($q>1/3$), highly precessing region of parameter space, we perform a simulation study where we simulate a set of signals consisting of highly precessing NR waveforms from the SXS catalog~\citep{Boyle:2019kee, Scheel:2025jct} and recover with the five waveform models under consideration.
From several tens of simulations, we discuss here the results from two that span the range of observed results, from unbiased parameter estimation displaying no systematics to large systematic differences between models and clear biases in parameter recovery.

\begin{figure*}[t!]
	\begin{centering}
	\includegraphics[width=0.495\textwidth]{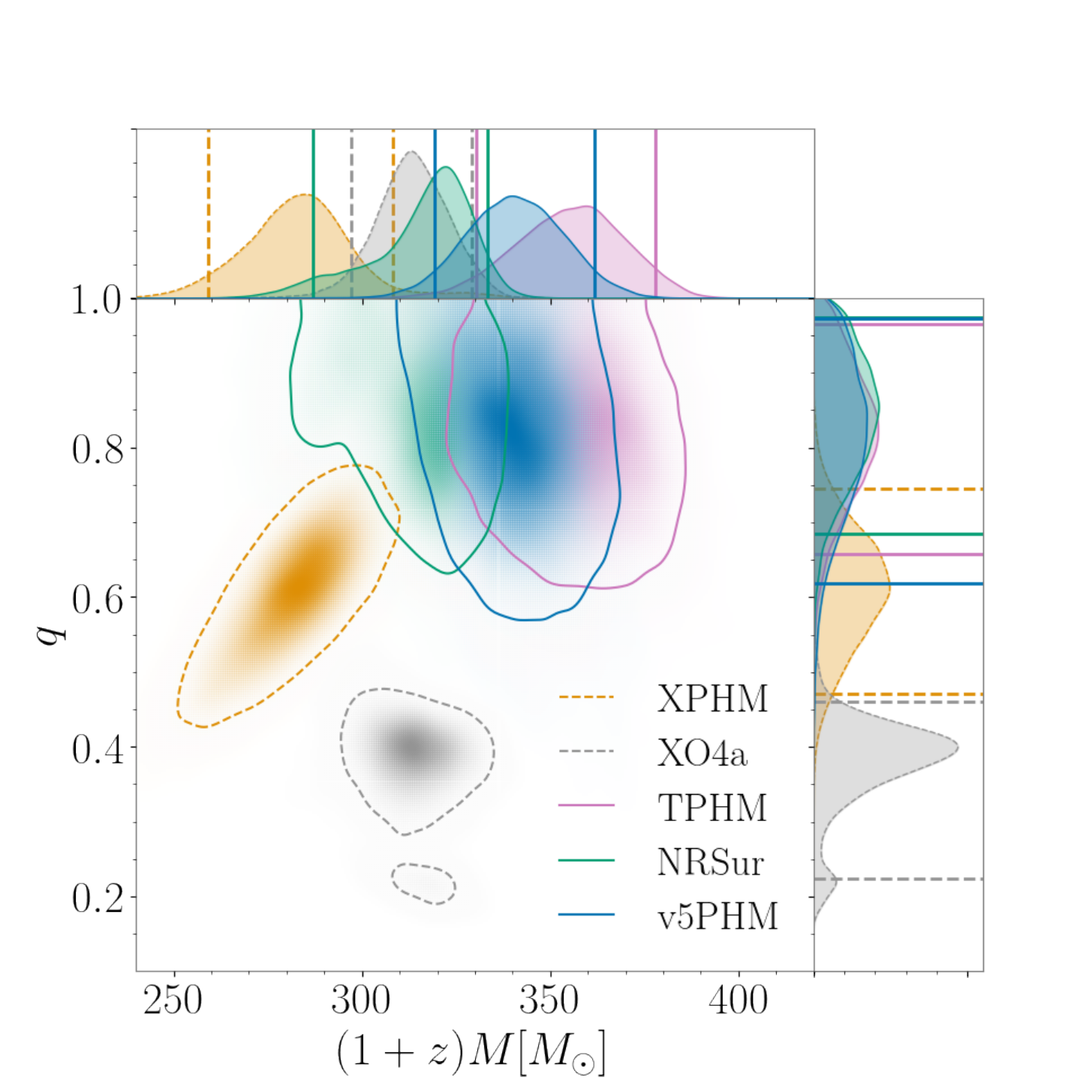}
	\end{centering}
	\includegraphics[width=0.495\textwidth]{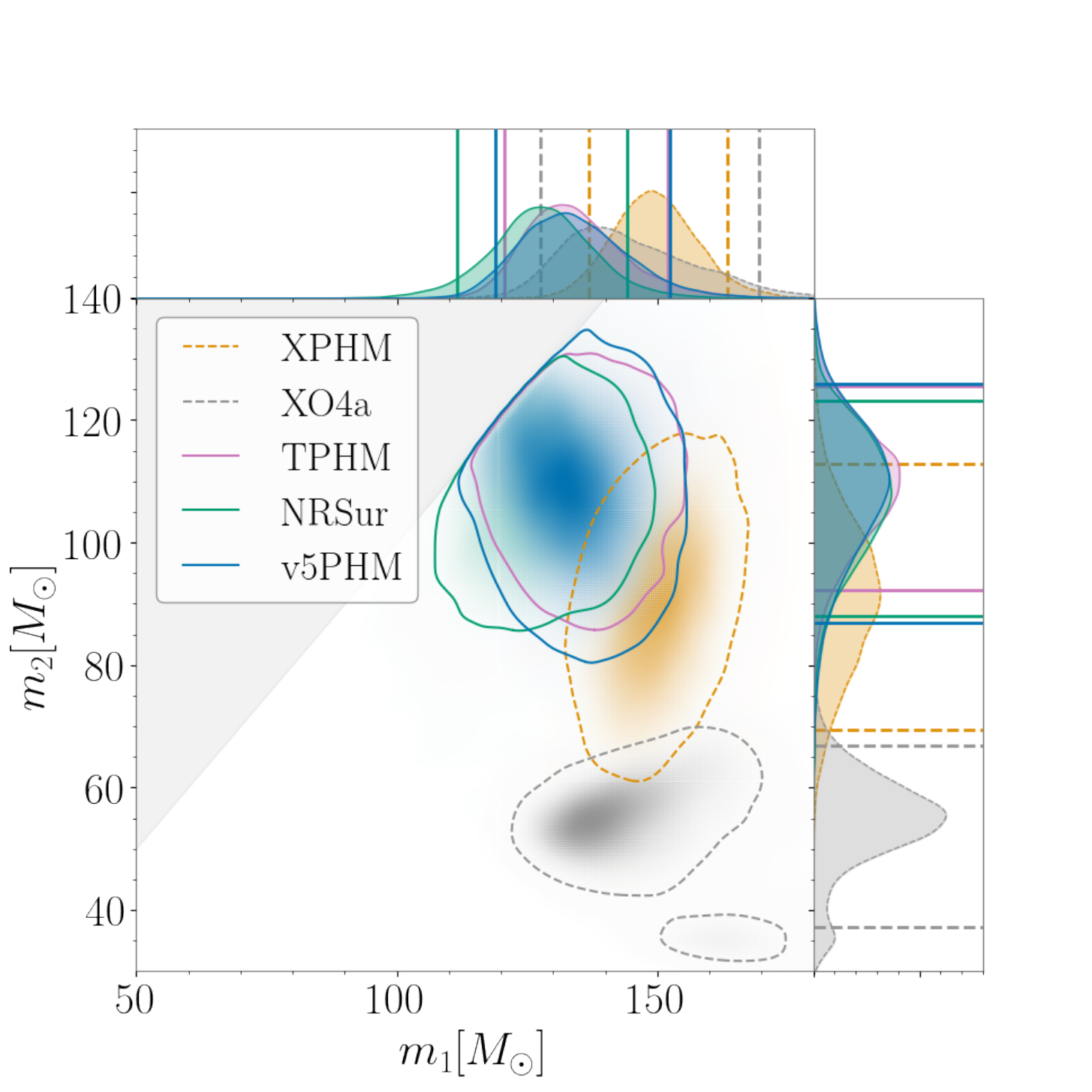}
	\caption{Marginalized posterior probability for the \emph{Left}: redshifted (detector-frame) total binary mass and the mass ratio, and \emph{Right}: primary and secondary source-frame masses inferred from GW231123 for each of the five models considered. Each contour, as well as the colored horizontal and vertical lines, shows the 90\% credible intervals. }
         \label{fig:model_total_mass_inference}
\end{figure*}

\begin{figure*}[t!]
	\begin{centering}
	\includegraphics[width=0.495\textwidth]{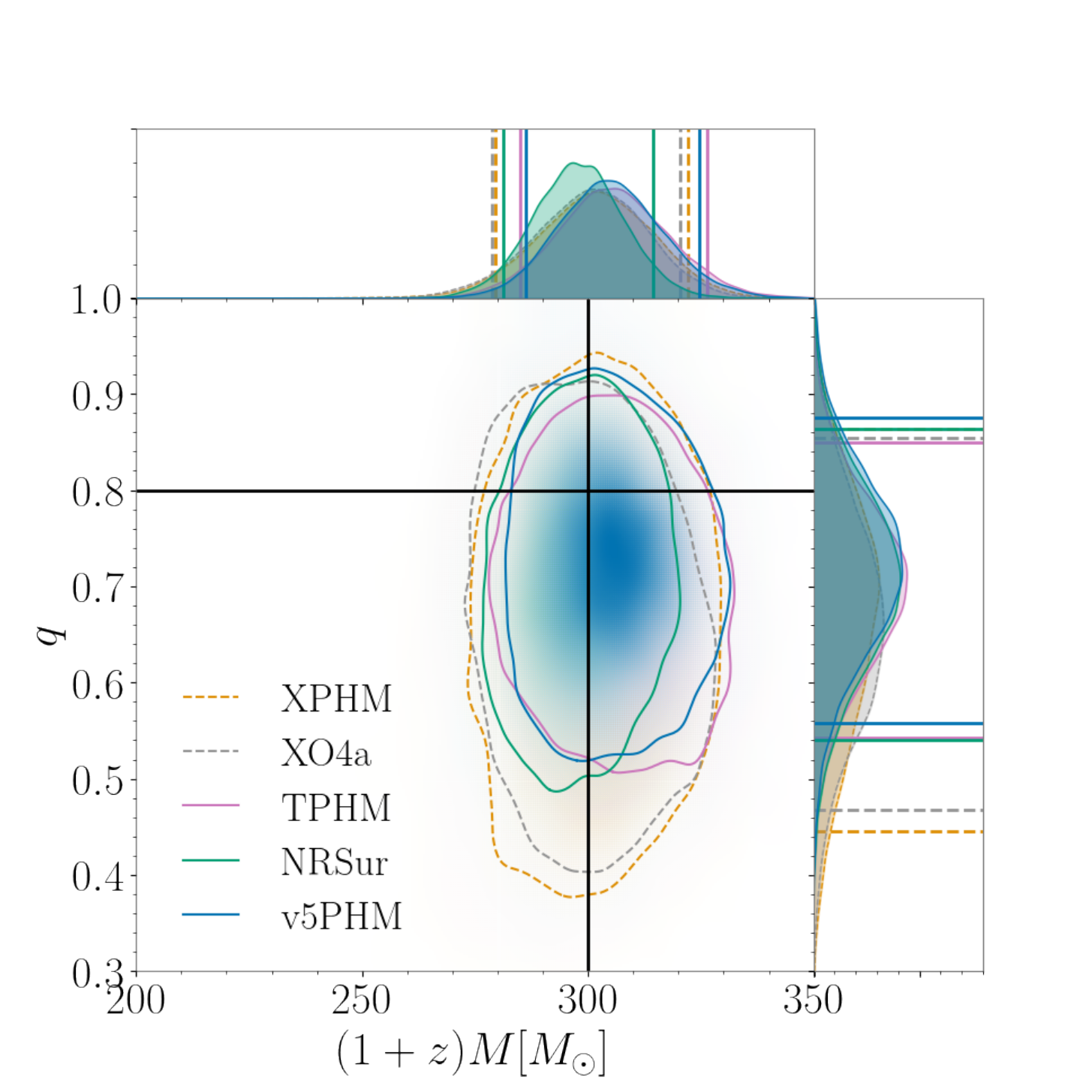}
	\includegraphics[width=0.495\textwidth]{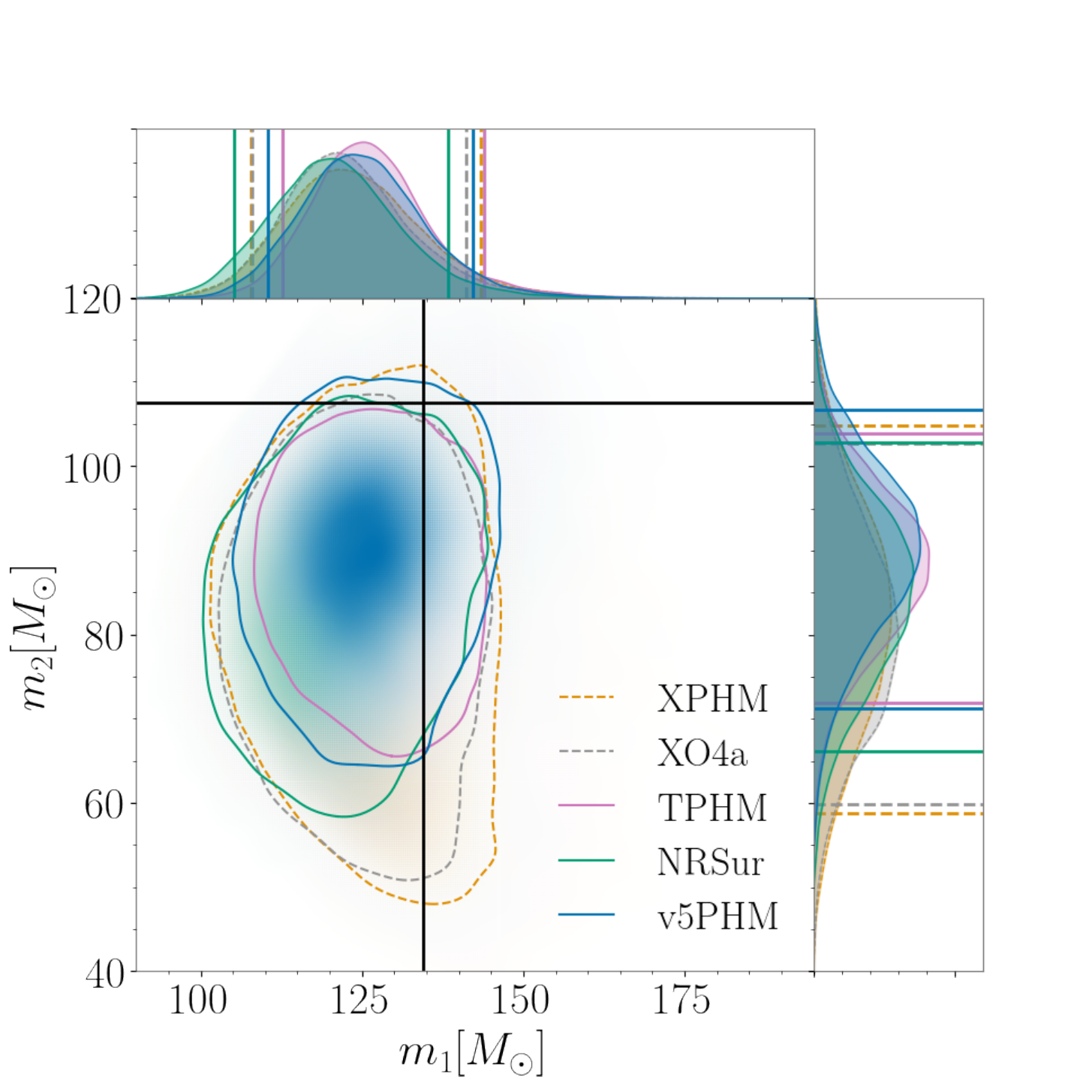}
	\includegraphics[width=0.495\textwidth]{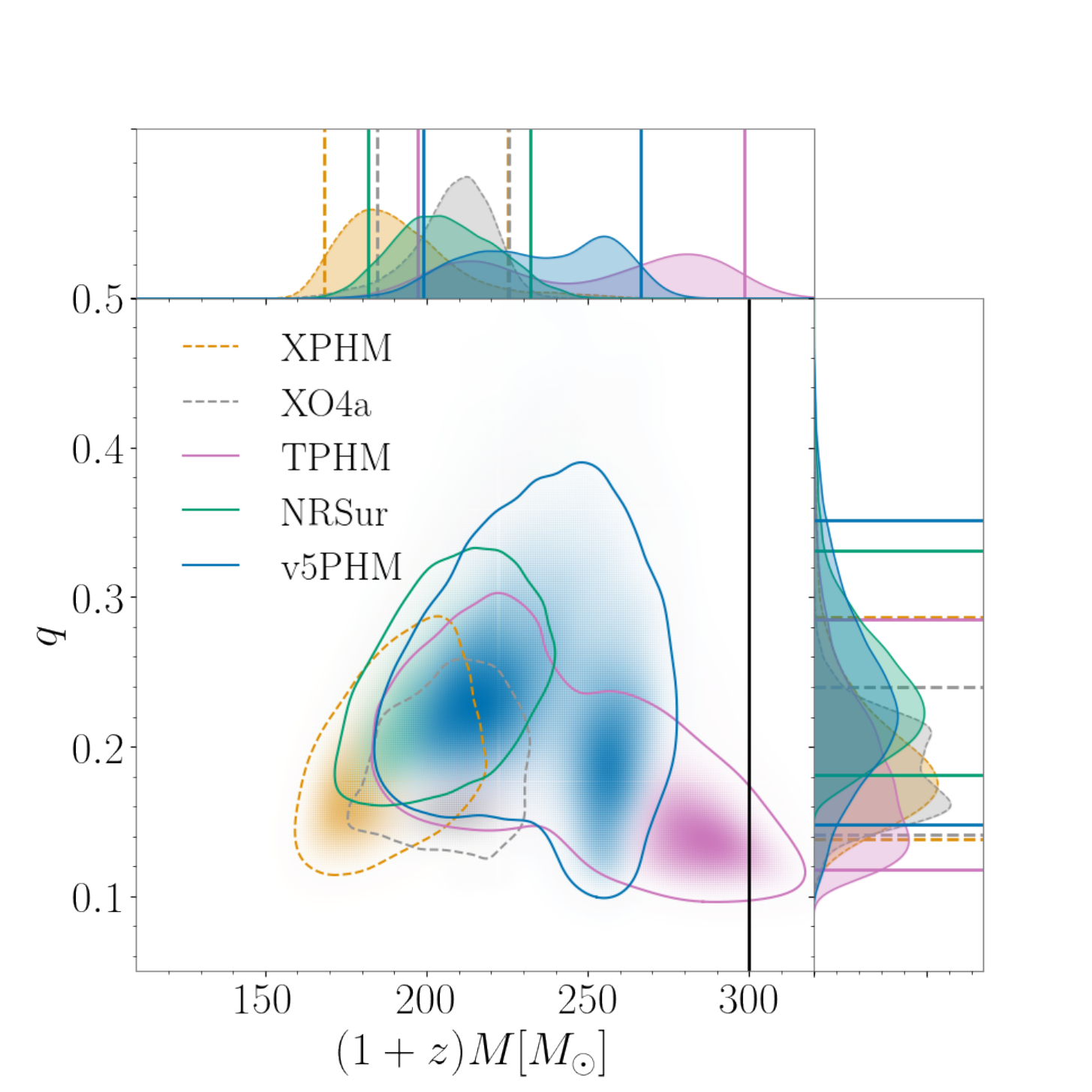}
	\includegraphics[width=0.495\textwidth]{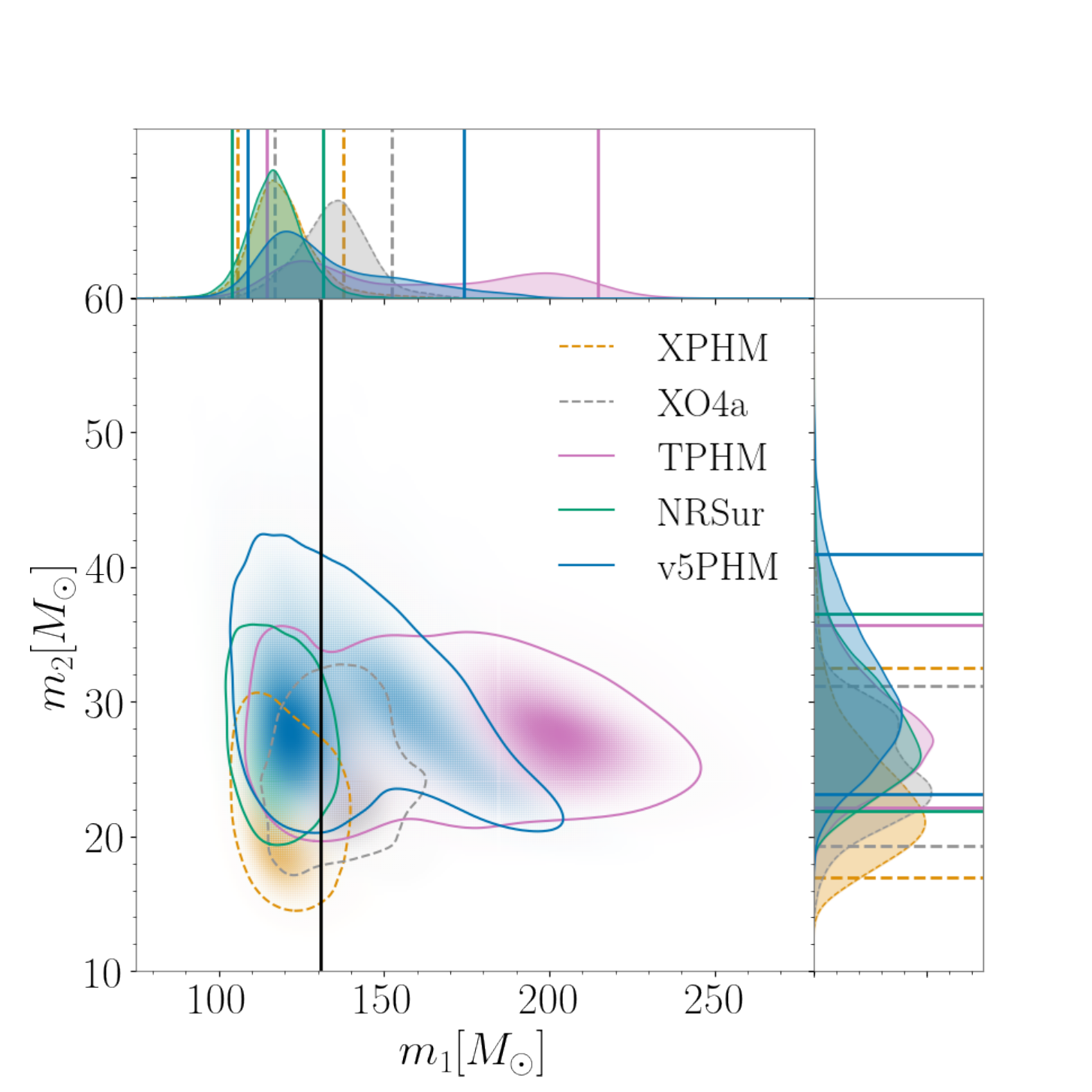}
	\end{centering}
	\caption{Marginalized posterior probability for (left column) redshifted (detector-frame) total binary mass and the mass ratio and (right column) primary and secondary source-frame masses inferred from two highly spinning precessing NR simulations with (detector-frame) total binary mass of $300\, M_{\odot}$ observed approximately edge on. The Top row shows the results for the {\textsc{SXS:BBH:0483}}~\citep{Boyle:2019kee} with masses $m_{1} \sim 135\, M_{\odot}$, $m_{2} \sim 110\, M_{\odot}$ and mass ratio $q = 0.8$.
The Bottom row shows the results for the {\textsc{SXS:BBH:4030}}~\citep{Scheel:2025jct} with masses $m_{1} = m_{2} \sim 130\, M_{\odot}$.
The 5 models used to analyse GW231123 were also used to analyse these simulations. Each contour, as well as the colored horizontal and vertical lines, shows the 90\% credible intervals. The black vertical and horizontal lines indicate the true source properties. In some panels, the true value is beyond the axis range of the figure.}
    \label{fig:model_total_mass_inference_injection}
\end{figure*}

For both configurations, we show the total mass and mass ratio as measured in the data (the detector frame). For high-mass binaries, we expect the total mass to be one of the most reliably measured quantities. 
The detector-frame masses are not the true source masses, but the \emph{redshifted} masses, and to calculate the true masses, we must also measure the redshift.
The relative accuracy of the detector-frame and source-frame masses may therefore differ, depending on the accuracy of the redshift. For this reason, we also show the individual masses $m_1$ and $m_2$ after 
correcting for the redshift.

The results for GW231123 are shown in Figure~\ref{fig:model_total_mass_inference}. In the left panel, we see clear evidence of systematics in the measurement of both the total mass and the mass ratio, with no overlap of the 90\% credible intervals for some models in both parameters. When we correct for the redshift, some of the differences appear to ``cancel out'', and we see agreement between several models in the source masses. This is likely coincidental; we expect
any model biases in the detector-frame masses and redshift to be independent. This expectation is borne out in the examples below.

In the majority of cases simulated, we were unable to reproduce this degree of systematics.
In both examples discussed here, we choose a large inclination angle as the mismatch performance is worst, and thus the associated expectations of evidence of systematics are greater, for systems with the greatest contribution from higher multipoles.
It should be noted, however, that since the orbital plane precesses, the inclination is not constant over the binary's evolution. 
An example of a typical recovery is shown in the top row of Figure~\ref{fig:model_total_mass_inference_injection}, where we consider the {\textsc{SXS:BBH:0483}} precessing NR simulation with total mass $M=300\, M_{\odot}$, mass ratio $q = 0.8$, and spin magnitudes $\chi_{1} = \chi_{2} = 0.80$ on both BHs. 
The simulation is added to zero-noise using the fiducial inclination angle $\iota = \pi/2\,\mathrm{rad}$ at 10\,Hz. 
For this configuration, the mismatch for \textsc{NRSur} was unambiguously below the conservative distinguishability criterion, with a value of $3.92\times10^{-4}$, while for the other models we see values $\mathcal{O}\left(10^{-3}\right)$.
In this case, the large differences in mismatch do not translate into noticeable differences in the accuracy of parameter recovery.
The posteriors from all models overlap and we can be confident in our recovered source properties. Note, however, that the 
source-frame $m_2$ is too low. This is a known bias for edge-on configurations: signals from face-on and face-off binaries are 
louder, meaning that larger distances (redshifts) are consistent with a fixed GW amplitude. This leads to a significantly larger prior 
volume, and thus a prior preference for smaller inclination angles~\citep{Usman:2018imj}, larger distances, and thus lower 
(redshifted) source masses.

Clear evidence of systematics was nevertheless seen in a limited number of simulations, as is demonstrated in the bottom row of Figure~\ref{fig:model_total_mass_inference_injection}.
We consider the {\textsc{SXS:BBH:4030}} precessing NR simulation with total mass $M=300\, M_{\odot}$, equal mass components ($q = 1$), and spin magnitudes $\chi_{1} = \chi_{2} = 0.95$ on both BHs. 
The simulation is added to zero-noise using the fiducial inclination angle $\iota = \pi/2\,\mathrm{rad}$ at 15\,Hz.
This injection was chosen from the set of cases with very high spins, with a mismatch between $2.36\times10^{-3}$ (\textsc{NRSur}) and $9.45\times10^{-3}$ (\textsc{TPHM}), mostly above the conservative indistinguishability criterion.
This \ac{NR} waveform also includes GW memory, which can require additional data processing for 
injection~\citep{Xu:2024ybt,Valencia:2024zhi,Chen:2024ieh}, but we find that our results are unchanged if we first subtract the memory
features before injection. 
We see unequivocal evidence of waveform systematics and biases in all models. Only the posterior of \textsc{TPHM} includes the true value of the detector-frame total mass, and all models exclude it at 90\% credibility.
No model recovers the true mass ratio ($q=1$). 
In the source-frame, the true value of $m_1$ lies in the 90\% credible region for all models, but $m_2$ is significantly biased from its true value of 100\,$M_\odot$.

\section{Source properties}
\label{sec:pe_appendix}

\begin{table}
\begin{ruledtabular}
    \caption{Individual source properties of GW231123 from each of the five models considered.}
    \label{table:individual_pe}
    \renewcommand{\arraystretch}{1.2}
    \begin{center}
    \begin{tabular}{l c c c c c}
        & XPHM & XO4a & TPHM & NRSur & v5PHM \\
        \hline
        Primary mass $m_1 / \Msun$ & $\massonesourcemed{GW231123cg_xphm}^{+\massonesourceplus{GW231123cg_xphm}}_{-\massonesourceminus{GW231123cg_xphm}}$ & $\massonesourcemed{GW231123cg_xo4a}^{+\massonesourceplus{GW231123cg_xo4a}}_{-\massonesourceminus{GW231123cg_xo4a}}$ & $\massonesourcemed{GW231123cg_tphm}^{+\massonesourceplus{GW231123cg_tphm}}_{-\massonesourceminus{GW231123cg_tphm}}$ & $\massonesourcemed{GW231123cg_nrsur}^{+\massonesourceplus{GW231123cg_nrsur}}_{-\massonesourceminus{GW231123cg_nrsur}}$ & $\massonesourcemed{GW231123cg_seob}^{+\massonesourceplus{GW231123cg_seob}}_{-\massonesourceminus{GW231123cg_seob}}$ \\
        Secondary mass $m_2 / \Msun $ & $\masstwosourcemed{GW231123cg_xphm}^{+\masstwosourceplus{GW231123cg_xphm}}_{-\masstwosourceminus{GW231123cg_xphm}}$ & $\masstwosourcemed{GW231123cg_xo4a}^{+\masstwosourceplus{GW231123cg_xo4a}}_{-\masstwosourceminus{GW231123cg_xo4a}}$ & $\masstwosourcemed{GW231123cg_tphm}^{+\masstwosourceplus{GW231123cg_tphm}}_{-\masstwosourceminus{GW231123cg_tphm}}$ & $\masstwosourcemed{GW231123cg_nrsur}^{+\masstwosourceplus{GW231123cg_nrsur}}_{-\masstwosourceminus{GW231123cg_nrsur}}$ & $\masstwosourcemed{GW231123cg_seob}^{+\masstwosourceplus{GW231123cg_seob}}_{-\masstwosourceminus{GW231123cg_seob}}$ \\
        Mass ratio $q = m_2 / m_1$ & $\massratiomed{GW231123cg_xphm}^{+\massratioplus{GW231123cg_xphm}}_{-\massratiominus{GW231123cg_xphm}}$ & $\massratiomed{GW231123cg_xo4a}^{+\massratioplus{GW231123cg_xo4a}}_{-\massratiominus{GW231123cg_xo4a}}$ & $\massratiomed{GW231123cg_tphm}^{+\massratioplus{GW231123cg_tphm}}_{-\massratiominus{GW231123cg_tphm}}$ & $\massratiomed{GW231123cg_nrsur}^{+\massratioplus{GW231123cg_nrsur}}_{-\massratiominus{GW231123cg_nrsur}}$ & $\massratiomed{GW231123cg_seob}^{+\massratioplus{GW231123cg_seob}}_{-\massratiominus{GW231123cg_seob}}$ \\
        Total mass $M / \Msun$ & $\totalmasssourcemed{GW231123cg_xphm}^{+\totalmasssourceplus{GW231123cg_xphm}}_{-\totalmasssourceminus{GW231123cg_xphm}}$ & $\totalmasssourcemed{GW231123cg_xo4a}^{+\totalmasssourceplus{GW231123cg_xo4a}}_{-\totalmasssourceminus{GW231123cg_xo4a}}$ & $\totalmasssourcemed{GW231123cg_tphm}^{+\totalmasssourceplus{GW231123cg_tphm}}_{-\totalmasssourceminus{GW231123cg_tphm}}$ & $\totalmasssourcemed{GW231123cg_nrsur}^{+\totalmasssourceplus{GW231123cg_nrsur}}_{-\totalmasssourceminus{GW231123cg_nrsur}}$ & $\totalmasssourcemed{GW231123cg_seob}^{+\totalmasssourceplus{GW231123cg_seob}}_{-\totalmasssourceminus{GW231123cg_seob}}$ \\
        Final mass $M_{\rm{f}} / \Msun$ & $\finalmasssourcemed{GW231123cg_xphm}^{+\finalmasssourceplus{GW231123cg_xphm}}_{-\finalmasssourceminus{GW231123cg_xphm}}$ & $\finalmasssourcemed{GW231123cg_xo4a}^{+\finalmasssourceplus{GW231123cg_xo4a}}_{-\finalmasssourceminus{GW231123cg_xo4a}}$ & $\finalmasssourcemed{GW231123cg_tphm}^{+\finalmasssourceplus{GW231123cg_tphm}}_{-\finalmasssourceminus{GW231123cg_tphm}}$ & $\finalmasssourcemed{GW231123cg_nrsur}^{+\finalmasssourceplus{GW231123cg_nrsur}}_{-\finalmasssourceminus{GW231123cg_nrsur}}$ & $\finalmasssourcemed{GW231123cg_seob}^{+\finalmasssourceplus{GW231123cg_seob}}_{-\finalmasssourceminus{GW231123cg_seob}}$ \\
        Primary spin magnitude $\chi_1$ & $\aonemed{GW231123cg_xphm}^{+\aoneplus{GW231123cg_xphm}}_{-\aoneminus{GW231123cg_xphm}}$ & $\aonemed{GW231123cg_xo4a}^{+\aoneplus{GW231123cg_xo4a}}_{-\aoneminus{GW231123cg_xo4a}}$ & $\aonemed{GW231123cg_tphm}^{+\aoneplus{GW231123cg_tphm}}_{-\aoneminus{GW231123cg_tphm}}$ & $\aonemed{GW231123cg_nrsur}^{+\aoneplus{GW231123cg_nrsur}}_{-\aoneminus{GW231123cg_nrsur}}$ & $\aonemed{GW231123cg_seob}^{+\aoneplus{GW231123cg_seob}}_{-\aoneminus{GW231123cg_seob}}$ \\
        Secondary spin magnitude $\chi_2$ & $\atwomed{GW231123cg_xphm}^{+\atwoplus{GW231123cg_xphm}}_{-\atwominus{GW231123cg_xphm}}$ & $\atwomed{GW231123cg_xo4a}^{+\atwoplus{GW231123cg_xo4a}}_{-\atwominus{GW231123cg_xo4a}}$ & $\atwomed{GW231123cg_tphm}^{+\atwoplus{GW231123cg_tphm}}_{-\atwominus{GW231123cg_tphm}}$ & $\atwomed{GW231123cg_nrsur}^{+\atwoplus{GW231123cg_nrsur}}_{-\atwominus{GW231123cg_nrsur}}$ & $\atwomed{GW231123cg_seob}^{+\atwoplus{GW231123cg_seob}}_{-\atwominus{GW231123cg_seob}}$ \\
        Effective inspiral spin $\chi_{\rm eff}$ & $\chieffmed{GW231123cg_xphm}^{+\chieffplus{GW231123cg_xphm}}_{-\chieffminus{GW231123cg_xphm}}$ & $\chieffmed{GW231123cg_xo4a}^{+\chieffplus{GW231123cg_xo4a}}_{-\chieffminus{GW231123cg_xo4a}}$ & $\chieffmed{GW231123cg_tphm}^{+\chieffplus{GW231123cg_tphm}}_{-\chieffminus{GW231123cg_tphm}}$ & $\chieffmed{GW231123cg_nrsur}^{+\chieffplus{GW231123cg_nrsur}}_{-\chieffminus{GW231123cg_nrsur}}$ & $\chieffmed{GW231123cg_seob}^{+\chieffplus{GW231123cg_seob}}_{-\chieffminus{GW231123cg_seob}}$ \\
        Effective precessing spin $\chi_{\rm p}$ & $\chipmed{GW231123cg_xphm}^{+\chipplus{GW231123cg_xphm}}_{-\chipminus{GW231123cg_xphm}}$ & $\chipmed{GW231123cg_xo4a}^{+\chipplus{GW231123cg_xo4a}}_{-\chipminus{GW231123cg_xo4a}}$ & $\chipmed{GW231123cg_tphm}^{+\chipplus{GW231123cg_tphm}}_{-\chipminus{GW231123cg_tphm}}$ & $\chipmed{GW231123cg_nrsur}^{+\chipplus{GW231123cg_nrsur}}_{-\chipminus{GW231123cg_nrsur}}$ & $\chipmed{GW231123cg_seob}^{+\chipplus{GW231123cg_seob}}_{-\chipminus{GW231123cg_seob}}$ \\
        Final spin $\chi_{\rm f}$ & $\finalspinmed{GW231123cg_xphm}^{+\finalspinplus{GW231123cg_xphm}}_{-\finalspinminus{GW231123cg_xphm}}$ & $\finalspinmed{GW231123cg_xo4a}^{+\finalspinplus{GW231123cg_xo4a}}_{-\finalspinminus{GW231123cg_xo4a}}$ & $\finalspinmed{GW231123cg_tphm}^{+\finalspinplus{GW231123cg_tphm}}_{-\finalspinminus{GW231123cg_tphm}}$ & $\finalspinmed{GW231123cg_nrsur}^{+\finalspinplus{GW231123cg_nrsur}}_{-\finalspinminus{GW231123cg_nrsur}}$ & $\finalspinmed{GW231123cg_seob}^{+\finalspinplus{GW231123cg_seob}}_{-\finalspinminus{GW231123cg_seob}}$ \\
        Luminosity distance $D_{\rm L} / \rm{Gpc}$ & $\luminositydistancegpcmed{GW231123cg_xphm}^{+\luminositydistancegpcplus{GW231123cg_xphm}}_{-\luminositydistancegpcminus{GW231123cg_xphm}}$ & $\luminositydistancegpcmed{GW231123cg_xo4a}^{+\luminositydistancegpcplus{GW231123cg_xo4a}}_{-\luminositydistancegpcminus{GW231123cg_xo4a}}$ & $\luminositydistancegpcmed{GW231123cg_tphm}^{+\luminositydistancegpcplus{GW231123cg_tphm}}_{-\luminositydistancegpcminus{GW231123cg_tphm}}$ & $\luminositydistancegpcmed{GW231123cg_nrsur}^{+\luminositydistancegpcplus{GW231123cg_nrsur}}_{-\luminositydistancegpcminus{GW231123cg_nrsur}}$ & $\luminositydistancegpcmed{GW231123cg_seob}^{+\luminositydistancegpcplus{GW231123cg_seob}}_{-\luminositydistancegpcminus{GW231123cg_seob}}$ \\
        Inclination angle $\theta_{\mathrm{JN}} / \rm{rad}$ & $\thetajnmed{GW231123cg_xphm}^{+\thetajnplus{GW231123cg_xphm}}_{-\thetajnminus{GW231123cg_xphm}}$ & $\thetajnmed{GW231123cg_xo4a}^{+\thetajnplus{GW231123cg_xo4a}}_{-\thetajnminus{GW231123cg_xo4a}}$ & $\thetajnmed{GW231123cg_tphm}^{+\thetajnplus{GW231123cg_tphm}}_{-\thetajnminus{GW231123cg_tphm}}$ & $\thetajnmed{GW231123cg_nrsur}^{+\thetajnplus{GW231123cg_nrsur}}_{-\thetajnminus{GW231123cg_nrsur}}$ & $\thetajnmed{GW231123cg_seob}^{+\thetajnplus{GW231123cg_seob}}_{-\thetajnminus{GW231123cg_seob}}$ \\
        Source redshift $z$ & $\redshiftmed{GW231123cg_xphm}^{+\redshiftplus{GW231123cg_xphm}}_{-\redshiftminus{GW231123cg_xphm}}$ & $\redshiftmed{GW231123cg_xo4a}^{+\redshiftplus{GW231123cg_xo4a}}_{-\redshiftminus{GW231123cg_xo4a}}$ & $\redshiftmed{GW231123cg_tphm}^{+\redshiftplus{GW231123cg_tphm}}_{-\redshiftminus{GW231123cg_tphm}}$ & $\redshiftmed{GW231123cg_nrsur}^{+\redshiftplus{GW231123cg_nrsur}}_{-\redshiftminus{GW231123cg_nrsur}}$ & $\redshiftmed{GW231123cg_seob}^{+\redshiftplus{GW231123cg_seob}}_{-\redshiftminus{GW231123cg_seob}}$ \\
        Network matched filter SNR $\rho$ & $\networkmatchedfiltersnrmed{GW231123cg_xphm}^{+\networkmatchedfiltersnrplus{GW231123cg_xphm}}_{-\networkmatchedfiltersnrminus{GW231123cg_xphm}}$ & $\networkmatchedfiltersnrmed{GW231123cg_xo4a}^{+\networkmatchedfiltersnrplus{GW231123cg_xo4a}}_{-\networkmatchedfiltersnrminus{GW231123cg_xo4a}}$ & $\networkmatchedfiltersnrmed{GW231123cg_tphm}^{+\networkmatchedfiltersnrplus{GW231123cg_tphm}}_{-\networkmatchedfiltersnrminus{GW231123cg_tphm}}$ & $\networkmatchedfiltersnrmed{GW231123cg_nrsur}^{+\networkmatchedfiltersnrplus{GW231123cg_nrsur}}_{-\networkmatchedfiltersnrminus{GW231123cg_nrsur}}$ & $\networkmatchedfiltersnrmed{GW231123cg_seob}^{+\networkmatchedfiltersnrplus{GW231123cg_seob}}_{-\networkmatchedfiltersnrminus{GW231123cg_seob}}$ \\
    \end{tabular}
    \end{center}
\tablecomments{As in Table~\ref{table:combined_pe} in most cases we present the median value of the 1D marginalized posterior distribution and the symmetric 90\% credible interval. For properties that have physical bounds we report the median value as well as the 90\% highest posterior density (HPD) credible interval. Our results are reported at a reference frequency of $10\, \mathrm{Hz}$.}
\end{ruledtabular}
\end{table}

In Table~\ref {table:individual_pe}, we present the individual source properties of GW231123 for each of the five models considered in the analysis of this event for those interested in a more detailed picture of the systematics.
As demonstrated in Appendix~\ref{sec: systematics appendix}, the source properties of this event lie in a challenging region of parameter space for all waveform models employed.
From the analysis performed here, we cannot guarantee that the results from any given model will be free from bias in this region of parameter space. 
We also find that different models fit the data better than others. All models except XPHM obtain a larger Bayesian evidence than the NRSur analysis, as reflected in the differing SNRs in Table~\ref{table:individual_pe}. For example, for some parameters XO4a yields significantly different results to many of the other models, yet it obtains a Bayes factor of at least 140:1 over NRSur. However, such differences are not necessarily indicative of one model being more accurate than another~\citep{Hoy:2022tst,Hoy:2024vpc}.
Consequently, we combine the posteriors from multiple models to achieve a conservative error estimate, which is reported throughout the main body of the paper.

\begin{figure*}[t!]
	\begin{center}
	\hspace{-2.4em}
	\includegraphics[width=0.36\textwidth]{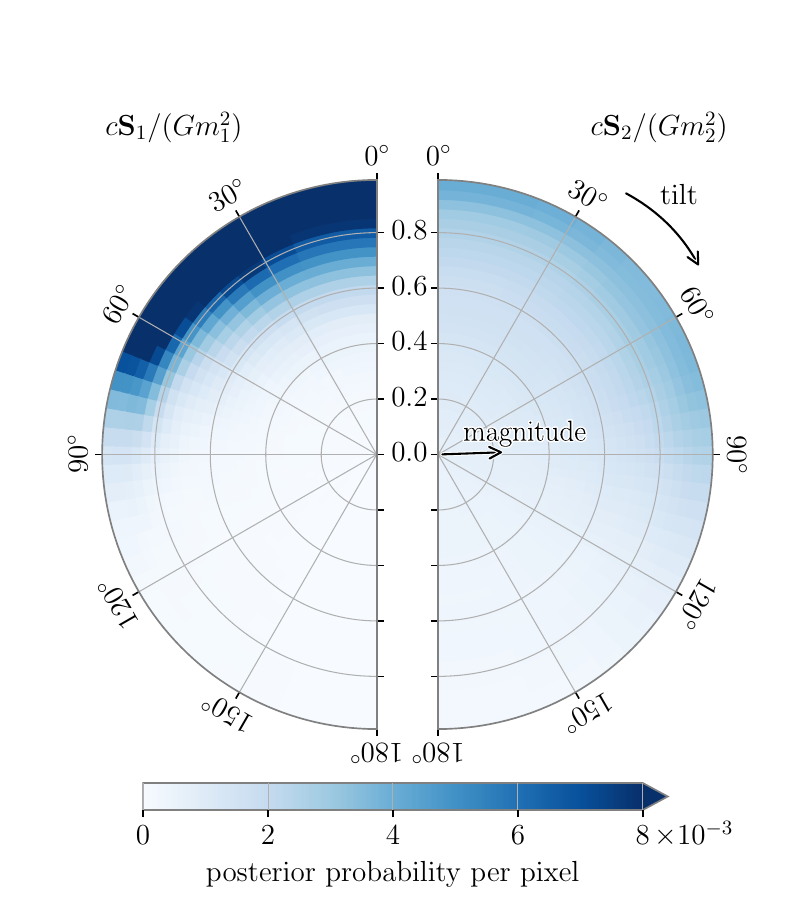}
	\hspace{-1.8em}
	\includegraphics[width=0.36\textwidth]{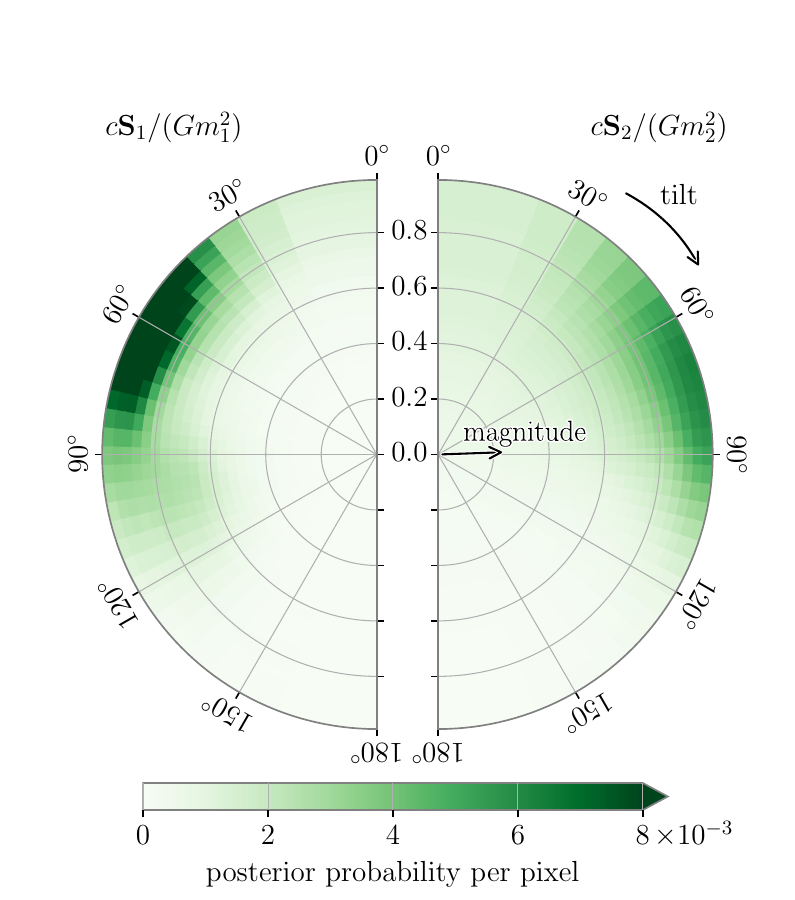}
	\hspace{-1.8em}
	\includegraphics[width=0.36\textwidth]{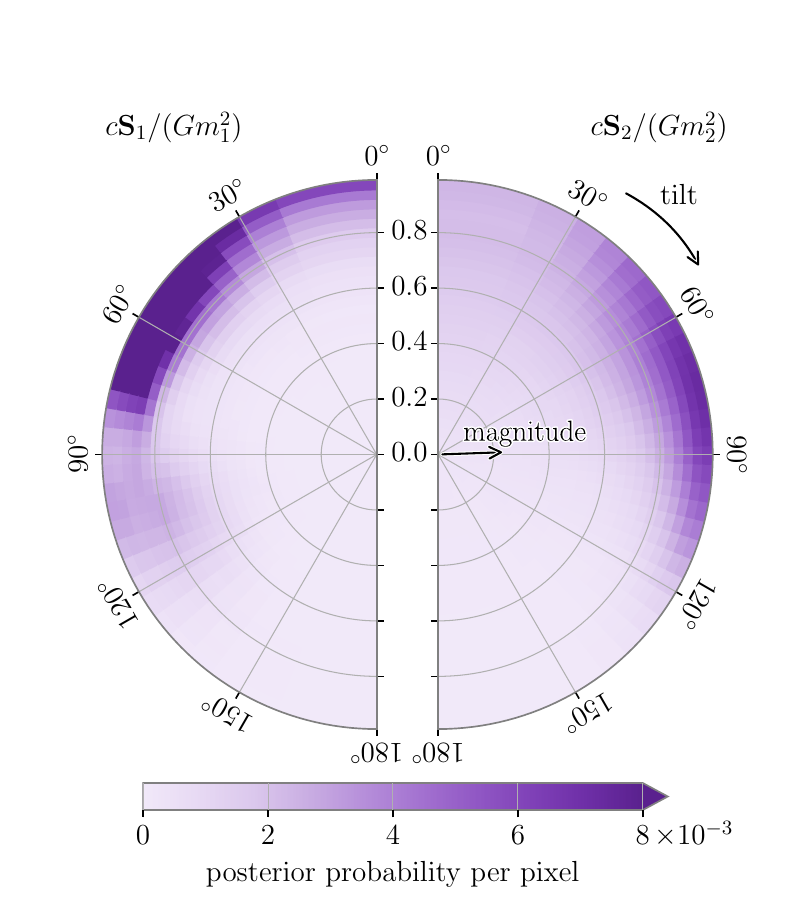}
	\end{center}
	\caption{Posterior probabilities for the dimensionless component spins, $c\boldsymbol{S}_{1}/(Gm^{2}_{1})$ and $c\boldsymbol{S}_{2}/(Gm^{2}_{1})$, relative to the orbital angular momentum axis $\hat{\boldsymbol{L}}$. From left to right, we compare the posterior probabilities obtained when analysing LIGO Hanford data only (blue), LIGO Livingston data only (green), and a coherent analysis of LIGO Hanford and LIGO Livingston data (purple). In all cases, we show the posterior distribution resulting from equally combining samples from five waveform models. The tilt angles are $0^\circ$ for spins aligned with the orbital angular momentum and $180^\circ$ for spins anti-aligned. Probabilities are marginalized over the azimuthal angles. The pixels have equal prior probability, being equally spaced in the spin magnitudes and the cosines of tilt angles. The spin orientations are defined at a fiducial GW frequency of 10 Hz.}
	\label{fig:ifo_spin_disk_inference}
\end{figure*}

We also illustrate in Figure~\ref{fig:ifo_spin_disk_inference} the differences in inferred spin orientation when considering the data from LIGO  
Hanford (left), LIGO Livingston (middle), and the full detector network (right). LIGO Hanford shows support for aligned-spin binaries, 
while LIGO Livingston has a clear preference for misalignment. The stronger signal in LIGO Livingston dominates the network results. 
The differences between the results in the two detectors could potentially be explained by lower signal power in LIGO Hanford (such that
precession is not measurable), but we have not been able to reproduce this discrepancy between detectors with injections in zero-noise, for example, of the \textsc{NRSur} 
waveform at its maximum-likelihood parameters. 

\section{Signal peak time}

\begin{figure*}
    \centering
    \includegraphics[width=0.95\linewidth]{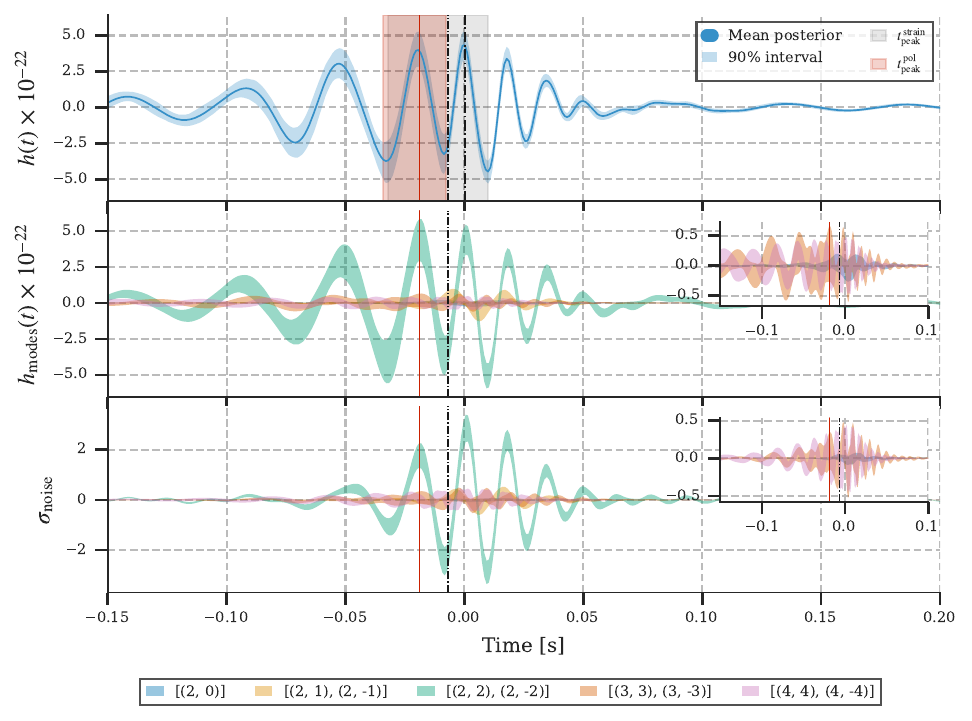}

    \caption{
    \textbf{Top panel:} 
    Posterior probability density functions of the \textsc{NRSur} waveform timeseries, obtained via \textsc{Bilby} using the \textsc{NRSur} waveform model in the LIGO Hanford detector. 
    The red band shows the uncertainty in the measurement of $t_\mathrm{peak}^\mathrm{pol}$, and the grey band shows the uncertainty in $t_\mathrm{peak}^\mathrm{strain}$.
    \textbf{Middle and bottom panel:} 
    Posterior probability density functions of the mode strain, $F_+ h_+ + F_\times h_\times$, with $h_+ - i h_\times = {}^{-2}Y_{\ell,m}h_{\ell,m} + {}^{-2}Y_{\ell,-m}h_{\ell,-m}$, shown for the LIGO Hanford detector. 
    The top part reports the unwhitened waveform and the bottom part the whitened one, showcasing the impact of whitening in visualising the signal morphology. 
    The inset focuses on the $(\ell, \pm m) = (2,0), (3,3), (4,4)$ modes. 
    The red line indicates the median of $t_\mathrm{peak}^\mathrm{pol}$, and the dashed-dotted black lines show the median of $t_\mathrm{peak}^{\mathrm{modes}}$.}
    \label{fig:hanford-recon-modes}
\end{figure*}

We require the time of peak GW emission to determine a valid starting time for ringdown analyses. 
The peak GW power is emitted at time $t^\mathrm{modes}_\mathrm{peak} = \max_t \sqrt{\sum_{\ell, m} \left| \dot{h}_{\ell m}(t) \right|^2}$, where $h_{\ell m}(t)$ are 
the multipoles in a spin-weighted spherical-harmonic decomposition of the signal. 
Alternatively, we can estimate the peak time using the peak of the polarisation $t^\mathrm{pol}_\mathrm{peak} = \underset{t}{\max} \left| h_+ - i h_\times \right|^2$.
This quantity depends on the binary's relative orientation to the detector and will be uncertain to within roughly one GW period, but it can be compared to an estimate computed through an unmodelled waveform reconstruction, allowing for a more agnostic analysis.
One can also conservatively estimate the onset of ringdown directly from the maximum value of the strain $t^{\rm strain}_{\rm peak}$, after which the signal displays a clear decay. 
Figure~\ref{fig:hanford-recon-modes} shows these times for GW231123, on top of the unwhitened \textsc{NRSur} strain reconstruction from which they were computed.
From this reconstruction, we find $t^{\rm strain}_{\rm peak}$ is $1384782888.6191^{+0.0098}_{-0.0322}$ s and $1384782888.6142^{+0.0107}_{-0.0195}$ s in the LIGO Hanford and Livingston detectors respectively. 
Instead, in the LIGO Hanford detector (chosen as reference for the ringdown analysis), we find $t^\mathrm{modes}_\mathrm{peak} - t^\mathrm{pol}_\mathrm{peak} \approx 6\,\mathrm{ms}$.

The differences among these estimates, together with the time-domain reconstruction shown in Figure~\ref{fig:hanford-recon-modes}, attest to the highly complex signal morphology and invite care when selecting a peak time definition to be used as reference in a ringdown analysis.
Hence, in the main text we repeat the analysis over a wide range of times, and plot results around a conservative $t^\mathrm{start} \approx t^{\rm strain}_{\rm peak} + 15 G M^{\rm det}_{\rm f} / c^3$ (assuming $M^{\rm det}_{\rm f} \simeq 298 M_{\odot}$), when we are confident on the validity of a QNM description.

\section{Higher-order radiation multipoles}

Given the support for large binary inclination from most IMR models, subdominant multipole moments (referred to as ``modes'' below) beyond the dominant $(\ell, m) = (2, \pm 2)$ spherical-harmonic multipole moment are expected to contribute appreciably to the observed signal~\citep{Blanchet:2013haa}. 
Here, we investigate in detail their contribution throughout the signal.
Using \textsc{NRSur} posterior samples, we estimate optimal \ac{snr} values of $2.27_{-1.05}^{+1.45}$ for the $(3,\pm3)$ mode, $2.92_{-0.89}^{+0.70}$ for the $(4,\pm4)$ mode, $0.68_{-0.48}^{+3.29}$ for the $(2,\pm1)$ mode and $0.25_{-0.16}^{+0.65}$ for the $(2,0)$ mode. 
Unlike the $(3,3)$ and $(4,4)$ modes, the inferred distribution for the $(2,1), (2,0)$ modes are consistent with expectations from random Gaussian noise fluctuations, implying a lack of statistically significant support for their presence in the data. 
Relevant to the ringdown analysis, the strain contribution from the $(2,0)$ mode remains significantly subdominant compared to the more prominent $(3,\pm3)$ and $(4,\pm4)$ modes throughout the signal duration, as illustrated in Figure~\ref{fig:hanford-recon-modes} (showing LIGO Hanford, with similar conclusions obtained for LIGO Livingston). 
As seen in the bottom panel, the whitening process suppresses the lower-frequency content of the signal, causing the peak amplitude to appear quieter relative to the higher-frequency ringdown. 
This filtering effect also reduces the visibility of subdominant modes such as $(2, \pm 1)$ and $(2, 0)$, which fall largely outside the detector's sensitive band. In contrast, the $(3, \pm 3)$ and $(4, \pm 4)$ modes remain visible post-merger due to their higher frequency content, making them detectable. This result is consistent with the \ac{snr} estimates above.
The IMR modes are defined with respect to the binary's total angular momentum at a given reference time during the inspiral, while the ringdown modes are defined with respect to the remnant spin at asymptotically late times. The direction between these two vectors may be offset by a few degrees~\citep{Hamilton:2021pkf}, but we do not expect that would be sufficient to increase the power in $(2,1)$ or $(2,0)$ to a level measurable in Gaussian noise, i.e., above an SNR of $\sim$2.1.

In summary, we conclude that a significant excitation of the $(2,0,0)$ or $(2,1,0)$ ringdown modes, suggested by the overlap of damped sinusoids fitting parameters with the remnant properties inferred by \textsc{NRSur}, is in tension with \textsc{NRSur} multipole moments content.

\clearpage

\bibliography{references}{}
\bibliographystyle{aasjournal}

\end{document}